\documentclass[longbibliography,nofootinbib,twocolumn,showpacs,amsmath,amstex,amssymb,mathfonts,superscriptaddress,prb]{revtex4-1}

\usepackage{graphicx}
\usepackage{color}
\usepackage{bm}
\usepackage{amsmath}
\usepackage{amssymb}
\usepackage{amsthm}
\usepackage{amsfonts}
\usepackage[normalem]{ulem}
\usepackage[colorlinks,bookmarks=true,citecolor=blue,linkcolor=red,urlcolor=blue]{hyperref}

\usepackage{bbm}

\newcommand*{\id}{{\normalfont\hbox{1\kern-0.15em \vrule width .8pt depth-.5pt}}}

\newcommand{\be}{\begin{equation}}
\newcommand{\ee}{\end{equation}}
\newcommand{\bq}{\begin{eqnarray}}
\newcommand{\eq}{\end{eqnarray}}

\newcommand{\bs}[1]{\boldsymbol{#1}}

\theoremstyle{theorem}

\theoremstyle{theorem}

\theoremstyle{definition}

\theoremstyle{definition}

\theoremstyle{remark}

\theoremstyle{theorem}

\makeatletter
\def\@fnsymbol#1{\ensuremath{\ifcase#1\or \dagger\or \ddagger\or
   \mathsection\or \mathparagraph\or \|\or **\or \dagger\dagger
   \or \ddagger\ddagger \else\@ctrerr\fi}}
    \makeatother

\begin{document}
\title{Equivalence between vortices, twists and chiral gauge fields in Kitaev's honeycomb lattice model}

\author{Matthew D. Horner}
\thanks{M.D.H. and A.F. contributed equally to this work.}
\affiliation{School of Physics and Astronomy, University of Leeds, Leeds, LS2 9JT, United Kingdom}

\author{Ashk Farjami}
\thanks{M.D.H. and A.F. contributed equally to this work.}
\affiliation{School of Physics and Astronomy, University of Leeds, Leeds, LS2 9JT, United Kingdom}

\author{Jiannis K. Pachos}
\affiliation{School of Physics and Astronomy, University of Leeds, Leeds, LS2 9JT, United Kingdom}

\date{\today }

\begin{abstract}

We demonstrate that $\mathbb{Z}_2$ gauge transformations and lattice deformations in Kitaev's honeycomb lattice model can have the same description in the continuum limit in terms of a chiral gauge field. The chiral gauge field is coupled to the Majorana fermions that satisfy the Dirac dispersion relation in the non-Abelian sector of the model. For particular values, the effective chiral gauge field becomes equivalent to the $\mathbb{Z}_2$ gauge field, enabling us to associate effective fluxes to lattice deformations. Motivated by this equivalence, we consider Majorana-bounding $\pi$ vortices and Majorana-bounding lattice twists and demonstrate that they are adiabatically connected to each other. This equivalence opens the possibility for novel encoding of Majorana-bounding defects that might be easier to realise in experiments. 

\end{abstract}

\maketitle

\section{Introduction}

Identifying the effective quantum field theory description of condensed-matter systems offers a simple and powerful way to understand their properties and predict their behaviour. For example, two-dimensional lattice models, such as graphene~\cite{Neto,DiVincenzo,Semenoff}, with a low-energy description in terms of Dirac fermions can be understood in terms of the powerful formalism of relativistic physics. Such an effective description of a model determines the main properties of its ground state and it can reveal the nature of its low-lying excitations. 
Similar to graphene, Kitaev's honeycomb lattice model~\cite{Kitaev} (KHLM) has a low-energy limit described by the Majorana version of the Dirac equation~\cite{Farjami}. 

The main interest in the KHLM is that vortices imprinted in the system trap localised Majorana zero modes that behave as non-Abelian anyons~\cite{Kitaev,Kitaev2,Pachos,Vidal3,Ville1,Self,Otten,Vidal4,Jackiw,Hou,Ville4,Ville5}. This property, together with the possibility of realising this model in the laboratory with crystallised materials~\cite{Chaloupka,Choi,Jackeli,Banerjee}, makes KHLM of interest to anyonic quantum computation~\cite{Kitaev2,DasSarma,Nayak} as well as to the investigation of fundamental physics of materials that support non-Abelian anyons. 

Although materials which display the properties of a pure Kitaev model are far off, there have been many studies introducing strains and defects in candidate materials such as ruthenium chloride~\cite{Lampen-Kelley,Do,Winter_2017,Kim}. Recently, it has been shown that not only vortices but twists in the form of lattice deformations can trap Majorana zero modes~\cite{Petrova,Petrova2,Willans} exhibiting the same non-Abelian statistics. This generalises the ways we have in realising Majorana anyons and the possible ways we can use to manipulate them. Nevertheless, lattice twists and vortices do not appear to have any connection between them apart from their common characteristic of trapping Majorana zero modes. 

Field theory provides an analytically tractable means to study lattice models and reveals the underlying relativistic and geometric description. Recently, these techniques have been applied to the KHLM~\cite{Farjami}, topological superconductors~\cite{Golan,Nissinen_2020} and Weyl superfluids~\cite{Nissinen_2020}, revealing the Riemann-Cartan (RC)~\cite{Hehl} nature of the continuum limit. In this paper, we propose to build upon these studies by considering \textit{chirality} and chiral gauge fields, which is a rather exotic concept of high energy physics that permeates to condensed matter systems. Massless fermions in $3+1$ dimensions can be described by spinors which are reducible into a pair of Weyl fermions of opposite chirality. This chirality, either left-handed or right-handed, signals how these objects transform under Lorentz transformations. The weak interaction of the Standard Model is chiral in nature as its interactions treat left- and right-handed particles differently~\cite{Maggiore}. Chirality also arises naturally in lattice gauge theories~\cite{Creutz} and condensed matter systems such as Weyl semimetals whose low-energy excitations are described by Weyl fermions. There is an intimate relationship between chiral gauge fields and torsion in the continuum limit which allows one to produce strain-induced gauge fields by inserting deformations to the lattice~\cite{laurila2020torsional,Cortijo,Sumiyoshi,Landsteiner,Grushin_2016,Pikulin_2016,Gorbar_2017,Ferreiros_2019}. Upon coupling to gauge fields, these systems can exhibit the chiral anomaly~\cite{Gian,Bertlmann,laurila2020torsional,Landsteiner,Pikulin_2016}, where chiral symmetry is broken resulting in a non-conserved current and a generalised quantum Hall effect~\cite{Liu}. Chirality has also been discussed in the context of graphene~\cite{Jackiw2}, phase transitions~\cite{Ville3}, and Landau levels~\cite{Rachel,laurila2020torsional,Grushin_2016}.

As Majorana fermions are charge-neutral they cannot couple to a $U(1)$ electromagnetic gauge field, however they can interact with a $U(1)_A$ {\em chiral} gauge field. These chiral gauge fields naturally generalise the $\mathbb{Z}_2$ gauge field that can be present only at the lattice level of the KHLM to the continuum limit. Indeed, we apply techniques from lattice gauge theory to demonstrate the equivalence between $\mathbb{Z}_2$ gauge fields on the lattice and $U(1)_A$ chiral gauge fields in the continuum, generalising the results of the $U(1)$ lattice gauge theory description of graphene~\cite{Giuliani,Neto,Gusynin}. Moreover, we show these chiral gauge fields also provide a faithful encoding of lattice deformations such as dislocations and twists in the continuum level, while preserving the relativistic description of the model. Hence, we are able to demonstrate that in the continuum limit of the model the lattice twists are equivalent to $\mathbb{Z}_2$ gauge transformations. 

This opens up the exciting possibility that localised $\mathbb{Z}_2$ gauge fields and localised twists that can trap Majorana zero modes are physically equivalent. To verify this, we show that Majorana zero modes trapped in $\mathbb{Z}_2$ vortices are {\em adiabatically} connected to Majorana zero modes trapped by twists. As a result, any lattice realisation of the chiral gauge field like twists, vortices or a hybrid of the two can trap Majorana zero modes. This opens up the possibility to experimentally realise Majorana zero modes with a wide variety of defects.

This paper is organised as follows. In Sec. \ref{sec:KHLM}, we review the KHLM and its corresponding relativistic continuum limit in the form of a Dirac Hamiltonian. In Sec. \ref{sec:chiral_continuum}, we discuss a possible generalisation of the Dirac action by upgrading its $U(1)_A$ chiral symmetry to a local symmetry with the introduction of chiral gauge fields. We then provide a general discussion of the relationship between gauge fields and Fermi points of lattice models, specifically how the insertion of a gauge field has the effect of shifting the Fermi points. In Sec. \ref{sec:gauge}, we apply this interpretation to the KHLM with a $\mathbb{Z}_2$ gauge field and two types of twists in the honeycomb lattice, and identify the corresponding continuum limit chiral gauge fields for each case. In particular, we identify that the continuum limit of a global $\mathbb{Z}_2$ gauge field and a particular type of twist in the lattice yields the same continuum limit. Finally, in Sec. \ref{sec:zero_modes} we demonstrate that when the $\mathbb{Z}_2$ gauge field and twists are inserted locally, they produce identical zero modes. We end the paper with a conclusion and Appendices containing further discussions of material in the paper for the interested reader.

\section{The Kitaev honeycomb lattice model}
\label{sec:KHLM}
In this section we shall provide a brief introduction to the KHLM and its continuum limit.

\subsection{Fermionisation}

KHLM is an exactly solvable model describing spin-$\frac{1}{2}$ particles residing on the vertices of a honeycomb lattice~\cite{Kitaev}. These spins are coupled via two- and three-body interactions with respective coupling constants $\{ J_x,J_y,J_z \}$ and $K$. 
By employing an appropriate fermionisation procedure, the spin Hamiltonian can be re-expressed as a tight-binding Hamiltonian describing Majorana fermions $c_i$ hopping on the vertices $i$ of a honeycomb lattice coupled to a $\mathbb{Z}_2$ gauge field $u_{ij}$ which lives on the links $(i,j)$, see (\ref{eq:KHLM_many_body_ham}). In the Majorana picture, the two- and three-body interactions become nearest and next-to-nearest-neighbour hopping terms, with corresponding hopping amplitudes $\{ J_x,J_y,J_z \}$ and $K$, respectively. The $\mathbb{Z}_2$ gauge field has the interesting property that its vortices trap Majorana zero modes that behave as non-Abelian anyons. 

We define the \textit{no-vortex sector} of the model as the case where the $\mathbb{Z}_2$ gauge field takes the trivial configuration of $u_{ij} = +1$ for all links. In this case, the system is translationally invariant with respect to a unit cell consisting of two neighbouring vertices, say $a$ and $b$, that form the triangular sub-lattices $A$ and $B$, respectively, of the honeycomb lattice. The Hamiltonian of the system $H = \sum_{\langle i,j\rangle} h_{ij} c^a_i c^b_j$ can be diagonalised via a Fourier transform to yield 
\begin{equation}
H = \frac{1}{4} \int \mathrm{d}^2 q \psi^\dagger_{\boldsymbol{q}} h(\boldsymbol{ q}) \psi_{\boldsymbol{q}}, 
\end{equation}
where 
\begin{equation}
h(\boldsymbol{q}) = 
\begin{pmatrix}
\Delta(\boldsymbol{q}) & -f(\boldsymbol{q}) \\
-f^*(\boldsymbol{q}) & - \Delta(\boldsymbol{q})
\end{pmatrix} \label{eq:KHLM_sp_ham}
\end{equation}
is the single-particle Hamiltonian and $\psi_{\boldsymbol{ q}} = (c^a_{\boldsymbol{ q}} \ i c^b_{\boldsymbol{ q}})^\mathrm{T}$, with $c^a_{\boldsymbol{ q}}$ and $c^b_{\boldsymbol{ q}}$ being the momentum space Majorana fermions residing on the corresponding sub-lattice. The entries of $h(\boldsymbol{q})$ are given by
$f(\boldsymbol{q}) = 2(J^x e^{i \boldsymbol{q} \cdot \boldsymbol{n}_1} + J^y e^{i \boldsymbol{q} \cdot \boldsymbol{n}_2} + J^z)$, 
and
$\Delta(\boldsymbol{q})  = 2K [  - \sin(\boldsymbol{q} \cdot \boldsymbol{n}_1) + \sin(\boldsymbol{q} \cdot \boldsymbol{n}_2) 
 + \sin( \boldsymbol{p} \cdot ( \boldsymbol{n}_1 - \boldsymbol{n}_2) ) ]$, where $\boldsymbol{n}_1 = (\sqrt{3}/2,3/2)$ and $\boldsymbol{n}_2 = (-\sqrt{3}/2,3/2)$ are the honeycomb lattice basis vectors, and the corresponding dispersion relation is given by
\begin{equation}
E(\boldsymbol{q}) = \pm \sqrt{|f(\boldsymbol{q})|^2 + \Delta^2(\boldsymbol{q})}. \label{eq:dispersion}
\end{equation}
The single-particle Hamiltonian Eq. (\ref{eq:KHLM_sp_ham}) has the symmetries $\sigma^x h(\boldsymbol{q}) \sigma^x = h(-\boldsymbol{q})$ and $\sigma^y h^*(\boldsymbol{q}) \sigma^y = -h(\boldsymbol{q})$, which are parity and particle-hole symmetries, respectively. The first symmetry imposes the constraint $E(\boldsymbol{q})= E(- \boldsymbol{q})$ on the dispersion and the second symmetry tells us that the upper and lower bands come in $\pm$ pairs, which is seen explicitly in (\ref{eq:dispersion}).

\subsection{Continuum Limit}

We are interested in investigating the properties of the ground state or low-lying excited states of the model that can reveal the phase of the system as well as its possible anyonic excitations. 
Similar to graphene, this model has two independent, isolated Fermi points $\boldsymbol{q} = \boldsymbol{P}_{\pm}$ in the Brillouin zone for which the dispersion $E(\boldsymbol{q})$ takes its minimum value. Around these points, the dispersion is linear in momentum so describes relativistic excitations. For the case of $K=0$, the Fermi points satisfy $E(\boldsymbol{P}_{\pm})=0$.  For the \textit{isotropic case} with $J_x = J_y = J_z \equiv J$, the Fermi points are given by
\begin{equation}
\boldsymbol{P}_\pm = \pm \left( \frac{4 \pi}{3 \sqrt{3}},0 \right). \label{eq:iso_fermi_point}
\end{equation} 
A non-zero $K$ simply opens a gap in the dispersion at the Fermi points. The parity symmetry of the Hamiltonian implies that the Fermi points always come in $\pm$ pairs, i.e., $\boldsymbol{P}_+ = - \boldsymbol{P}_-$. 

The effective description of the model about the ground state, where all negative energy states (valence band) are occupied, is obtained by restricting momenta to lie in a small neighbourhood of the two Fermi points as $\boldsymbol{q} = \boldsymbol{P}_\pm + \boldsymbol{p}$. For each Fermi point, we define the two-component Weyl spinors $\psi_\pm(\boldsymbol{p}) \equiv (c^a_\pm(\boldsymbol{p}) \ ic^b_\pm(\boldsymbol{p}))$, where $c^{a/b}_\pm(\boldsymbol{p}) \equiv c^{a/b}_{\boldsymbol{P}_\pm + \boldsymbol{p}}$, and the corresponding low-energy Hamiltonians $h_\pm(\boldsymbol{p}) \equiv h(\boldsymbol{P}_\pm + \boldsymbol{p})$, to first order in $\boldsymbol{p}$. 

One can consider both Fermi points simultaneously by regarding excitations about the two Fermi points as two \textit{chiral} degrees of freedom. We achieve this by combining the pair of two-component Weyl spinors $\psi_\pm$ into a single four-component Dirac-like spinor with the definition $\Psi(\boldsymbol{p}) = (\psi_+ , \sigma^x \psi_-)^\mathrm{T} = (c^a_+ , ic^b_+ , ic^b_- , c^a_-)^\mathrm{T}$. We then take the direct sum of $h_+(\boldsymbol{p})$ and $h_-(\boldsymbol{p})$  in their respective bases defined by $\Psi(\boldsymbol{p})$ to yield the total $4 \times 4$ Hamiltonian,
\begin{equation}
h_\text{KHLM}(\boldsymbol{p})  = 3J \gamma^0(  \gamma^1 p_x -  \gamma^2 p_y ) - i 3\sqrt{3}K \gamma^1 \gamma^2,
\label{eq:DiracKit}
\end{equation}
which takes the form of a massless Dirac Hamiltonian defined on a $(2+1)$-dimensional Minkowksi space-time with torsion~\cite{Farjami}. The continuum limit has provided us with a representation of the gamma matrices given by
\begin{equation}
\gamma^0 =
\begin{pmatrix}
0 &\mathbb{I} \\
\mathbb{I} & 0
\end{pmatrix}
= \sigma^x\otimes\mathbb{I}, \quad
\boldsymbol{\gamma}=
\begin{pmatrix}
0 & -\boldsymbol{\sigma} \\
\boldsymbol{\sigma} & 0
\end{pmatrix}
= -i\sigma^y\otimes \boldsymbol{\sigma},\label{eq:gamma_matrices}
\end{equation}
where $\boldsymbol{\sigma} =(\sigma^x,\sigma^y,\sigma^z)$ are the Pauli matrices and $\mathbb{I}$ is the two-dimensional identity, which obey the $(3+1)$-dimensional Clifford algebra $\{ \gamma^A, \gamma^B \} = 2 \eta^{A B}$, where $\eta^{A B} = \mathrm{diag}(1,-1,-1,-1)$ is the Minkowski metric. We also define the fifth gamma matrix,
\begin{align}
\gamma^5 & =i\gamma^0\gamma^1\gamma^2\gamma^3=
\begin{pmatrix}
\mathbb{I} & 0 \\
0 & -\mathbb{I}
\end{pmatrix} 
=\sigma^z\otimes\mathbb{I}, \label{eq:gamma_5}
\end{align}
which obeys $\{ \gamma^5 , \gamma^A \} = 0$ for all gamma matrices. This particular representation of the gamma matrices is known as the \textit{chiral} representation. Note that, despite working on a $(2+1)$-dimensional space, we are able to define a $(3+1)$-dimensional representation as we are working with $4 \times 4$ matrices, however at this stage $\gamma^3$ is redundant. In this paper, we use the notation that early upper-case Latin indices $A,B,\ldots$ range over $0,1,2,3$, while early lower-case Latin indices $a,b,\ldots$ range over $0,1,2$. These are orthonormal frame indices and we refer to any gamma matrices with such indices as \textit{flat space} gamma matrices to contrast with the curved space gamma matrices to be defined later. Moreover, the single-particle Hamiltonian $h(\boldsymbol{p})$ has charge-conjugation symmetry. More details about the derivation of the continuous limit of the KHLM can be found in Appendix~\ref{app:contlimit}. 

An important observation to make is that the four-dimensional spinor $\Psi(\boldsymbol{ p})= (\psi_+ , \sigma^x \psi_-)^\mathrm{T}$ is a Majorana spinor, i.e., charge neutral~\cite{Maggiore}. This is due to the fact that the two-component Weyl spinors $\psi_\pm$ about each Fermi point $\boldsymbol{P}_\pm$ are not independent. In general, charge conjugation in momentum space is defined as $\Psi^{(c)}(\boldsymbol{p}) = C\Psi^\dagger(-\boldsymbol{p})$, where $C$ is the unitary charge conjugation matrix obeying $C^\dagger \gamma^A C = - (\gamma^A)^*$ for all gamma matrices and $\dagger$ denotes taking the Hermitian conjugate of each component without taking the transpose of the spinor. In our chiral representation Eqs. (\ref{eq:gamma_matrices}), the charge conjugation matrix is given by $C=- \sigma^y \otimes \sigma^y = -i \gamma^2$. We observe that the spinor $\Psi(\boldsymbol{ p}) $ is a Majorana spinor, i.e. $\Psi^{(c)}(\boldsymbol{p}) = \Psi(\boldsymbol{p})$, which is shown using the fact that in momentum space Majorana modes obey $c^\dagger_\pm(\boldsymbol{p}) = c_\mp(-\boldsymbol{p})$. 

\section{Chiral gauge fields in the continuum}
\label{sec:chiral_continuum}

The continuum limit of the isotropic KHLM is described by the Majorana version of the Dirac Hamiltonian given by Eq. (\ref{eq:DiracKit}). While Majorana fermions do not couple to $U(1)$ gauge fields, they can be coupled to a $U(1)_A$ \textit{chiral} gauge field. 
In this section, we investigate how one could realise chiral gauge fields in the continuum limit of a lattice model. 

\subsection{The Dirac action formalism}

The most general continuum limit of the KHLM is not only relativistic but is defined on a space-time with both curvature and torsion~\cite{Farjami}. Such general space-times are called \textit{Riemann-Cartan} space-times which are characterised by a nontrivial metric $g_{\mu \nu}$ and affine connection $\Gamma^\rho_{\ \mu \nu}$~\cite{Hehl}. For the purposes of defining spinors on a Riemann-Cartan space, we translate to the equivalent language of dreibein $e_a^{\ \mu}$ and spin connection $\omega^a_{\mu b}$ whose Latin indices are with respect to a local orthonormal frame. For brevity, we present only the relevant material in this paper and point the reader to a self-contained review of Riemann-Cartan theory applied to the KHLM in Ref.~\onlinecite{Farjami}. We use the notation that Greek letters $\mu, \nu,\ldots$ represent $(2+1)$-dimensional general coordinate indices, whilst later lower-case Latin indices $i,j,\ldots$ represent the spatial coordinate indices only. 

The action for a spin-$\frac{1}{2}$ particle $\psi$ of mass $m$ on a static $(2+1)$-dimensional Riemann-Cartan space-time $M= \mathbb{R} \times \Sigma$ is given by~\cite{Farjami}
\begin{equation}
S_\text{RC} = \int_M \mathrm{d}^{2+1}x \bar{\Psi} \left( i \gamma^\mu \partial_\mu - \frac{i}{8} \phi \gamma^0 \gamma^1 \gamma^2 + \frac{i}{2} \partial_i \gamma^i -m \right) \Psi. \label{eq:dirac action 2}
\end{equation}
where $\{ \gamma^\mu = e^\mu_{\ a} \gamma^a \}$ are the curved space gamma matrices, $\phi$ is the \textit{torsion pseudoscalar} related to the torsion of the space-time by $T_{abc} = \frac{\phi}{3!} \epsilon_{abc}$, $\Psi = \sqrt{|e|} \psi$ is the spinor density obeying flat-space anti-commutation relations, $|e| = \mathrm{det}[e^a_{\ \mu}]$ is the determinant of the dreibein and $\bar{\Psi} = \Psi^\dagger \gamma^0$ is the Dirac adjoint. The Hamiltonian density corresponding to the action Eq. (\ref{eq:dirac action 2}) is given by $\mathcal{H} = \Psi^\dagger h \Psi$, where $h$ is the \textit{single-particle} Hamiltonian given by
\begin{equation}
h(\boldsymbol{p}) = e^{\ i}_a \gamma^0 \gamma^a p_i + {i\over 8} \phi \gamma^1\gamma^2 - \frac{i}{2} \partial_i e^{\ i}_a \gamma^0 \gamma^a + m \gamma^0, 
\label{eq:RC Ham} 
\end{equation}
which is given explicitly in terms of the dreibein and the flat-space gamma matrices. A comparison of Eq. (\ref{eq:RC Ham}) with Eq. (\ref{eq:DiracKit}) reveals that the continuum limit of the isotropic and homogeneous case is described by a massless Dirac Hamiltonian on a Minkowski space-time with torsion. Further discussion of the dreibein of more general continuum limits is provided in Appendix \ref{app:contlimit}.

\subsection{Gauging the chiral symmetry}

The continuum limit of the KHML has provided us with four-component \textit{Majorana} spinors. A $U(1)$ transformation is not compatible with a Majorana spinor $\Psi$ because it does not preserve the Majorana reality condition $\Psi^{(c)} = \Psi$, i.e., if $\Psi$ is a Majorana spinor, then $\Psi' = e^{i \alpha} \Psi$ is not a Majorana spinor. For this reason, we cannot couple Majorana spinors to a $U(1)$ gauge field and therefore these particles are electrically neutral. However, the \textit{massless} action (\ref{eq:dirac action 2}) has a global and internal $U(1)_A$ chiral symmetry~\cite{Maggiore} which \textit{is} compatible with Majorana spinors, where the subscript $A$ stands for axial. A $U(1)_A$ transformation is defined by  
\begin{equation}
\Psi(x) \rightarrow e^{i \alpha \gamma^5} \Psi(x), \quad \bar{\Psi}(x) \rightarrow \bar{\Psi}(x)e^{i \alpha \gamma^5},  \label{eq: chiral transformation}
\end{equation}
where $ \alpha \in \mathbb{R}$. This chiral transformation preserves the reality condition, i.e., if $\Psi$ is a Majorana spinor, then $\Psi' = e^{i \alpha \gamma^5} \Psi$ is also a Majorana spinor. In the chiral representation of the gamma matrices Eqs. (\ref{eq:eq:gamma_matrices}), we see that a chiral transformation simply corresponds to two opposite $U(1)$ transformations of each Weyl spinor component of $\Psi$. Note that the names ``chiral" and ``axial" are used interchangeably in the literature. The term chiral in our context refers to anything associated with $\gamma^5$.

We upgrade this chiral symmetry to a local symmetry by introducing the gauge field $A_\mu$ with a corresponding gauge-covariant derivative,
\begin{equation}
D^A_\mu \Psi = \partial_\mu \Psi +iA_\mu \gamma^5 \Psi,
\end{equation}
which transforms as $D^A_\mu  \Psi \rightarrow  e^{i \alpha \gamma^5} D^A_\mu \Psi$ under the simultaneous transformation $\Psi \rightarrow e^{i \alpha \gamma^5} \Psi$ and $A_\mu \rightarrow A_\mu - \partial_\mu \alpha$, for a space-dependent parameter $\alpha(x)$. Replacing the partial derivatives in the massless version of the action Eq. (\ref{eq:dirac action 2}) with covariant derivatives yields the single-particle Hamiltonian
\begin{equation}
h(\boldsymbol{p}) = e^{\ i}_a \gamma^0 \gamma^a \left( p_i + A_i \gamma^5 \right) + A_0 \gamma^5 + {i\over 8} \phi \gamma^1\gamma^2 - \frac{i}{2} \partial_i e^{\ i}_a \gamma^0 \gamma^a.
\label{eq:HamChiral}
\end{equation}
It is worth noticing that the temporal component of the chiral gauge field  $A_0\gamma^5$ commutes with all the other terms in the Hamiltonian Eq. (\ref{eq:HamChiral}). Hence, its presence does not influence any of the physical observables and can be neglected. The temporal component also has no dreibein coefficient as the only non-zero temporal dreibein on a static space-time is given by $e_0^{\ t} = 1$.

Appendix~\ref{app:gen_actions} presents all possible terms one can add to the Majorana version of the Dirac Hamiltonian to generalise it, including the chiral term presented here.

\subsection{Gauge fields and Fermi points}
\label{sec:Fermichiral}

In lattice gauge theory, there is a general approach for minimally coupling a matter field living on the vertices $i$ of a lattice to a gauge field living on the links $(i,j)$. This is achieved by multiplying the tunnelling couplings of the matter field in the many-body Hamiltonian by Wilson lines of the form $u_{ij} = \exp( i e\int_i^j \mathrm{d} \boldsymbol{l} \cdot  \boldsymbol{A} )$, where $u_{ij}$ is an element of a Lie group, $\boldsymbol{A}$ is an element of the corresponding Lie algebra and $e$ is the charge of the matter field~\cite{Aidelsburger,Gusynin,Rothe,Munster,Giuliani}. This is sometimes called a \textit{Peierls substitution}. When taking the continuum limit of the lattice model, the Peierls substitution becomes equivalent to the usual minimal coupling prescription of substituting $\boldsymbol{p} \rightarrow \boldsymbol{p} + e\boldsymbol{A}$. Hence, for lattice models like graphene that give rise to a Dirac equation in the continuum limit, the minimal coupling is manifested by a shift of the model's Fermi points in a {\em parallel} fashion by $-e \boldsymbol{A}$~\cite{Volovik}.

The KHLM is comprised of Hermitian Majorana modes $c_i $ that are charge neutral, c.f. Eq. \ref{eq:KHLM_many_body_ham}. Hence, they can only couple to a gauge field that has real-valued Wilson line elements, e.g., $u_{ij}\in \mathbb{Z}_2 = \{ 1,-1 \}$. However, due to the parity symmetry of the KHLM, these real-valued Wilson lines will cause the Fermi points of the model to shift in an \textit{anti-parallel} fashion, resulting in an emergent $U(1)_A$ chiral gauge field in the continuum limit. To see this, consider the single-particle Hamiltonian Eq. (\ref{eq:KHLM_sp_ham}) of the KHLM. When taking the continuum limit, we Taylor expand about the Fermi points of the model. In lattice models these Fermi points always come in pairs~\cite{Nielsen}, which is seen explicitly in the KHLM as we have two inequivalent Fermi points $\boldsymbol{P}_\pm$ in the Brillouin zone. We define the effective continuum limit Hamiltonians about each Fermi point by restricting the momenta $\boldsymbol{q}$ to take values $\boldsymbol{q} = \boldsymbol{P}_\pm + \boldsymbol{p}$, for small $\boldsymbol{p}$, as
\begin{equation}
h_\pm(\boldsymbol{p}) \equiv h(\boldsymbol{P}_\pm + \boldsymbol{p}) = \boldsymbol{p} \cdot \boldsymbol{\nabla} h(\boldsymbol{P}_\pm ) + O(p^2).  \label{eq:h_taylor}
\end{equation}
Modifications to the model that preserve the form of the Hamiltonian Eq. (\ref{eq:KHLM_sp_ham}), such as varying the strength of the couplings $\{ J_i \}$, inserting a $\mathbb{Z}_2$ gauge field or adding in extra couplings, will have the effect of modifying the single-particle Hamiltonian as $h(\boldsymbol{q}) \rightarrow {h'}(\boldsymbol{q})$. In general, the new Fermi points $\boldsymbol{P}'_\pm$ will be different giving rise to a shift
\begin{equation}
\Delta \boldsymbol{P}_\pm = {\boldsymbol{P}}'_\pm - \boldsymbol{P}_\pm.
\end{equation}
By restricting momenta to take small values about the new Fermi points as $\boldsymbol{q} = {\boldsymbol{P}}'_\pm + {\boldsymbol{p}}'$, the continuum limit Hamiltonians about the new points are given by
\begin{equation}
{h}'_\pm({\boldsymbol{p}'}) \equiv {h'}({\boldsymbol{P}}'_\pm + {\boldsymbol{p}}') = {\boldsymbol{p}}' \cdot \boldsymbol{\nabla} {h}'({\boldsymbol{P}}'_\pm) + O({p'}^2). \label{eq:h_taylor_prime}
\end{equation}
In general, $ \boldsymbol{p} \neq {\boldsymbol{p}'}$, so direct comparison of the continuum limits Eqs. (\ref{eq:h_taylor}) and (\ref{eq:h_taylor_prime}) cannot be done. Nevertheless, employing the relation ${\boldsymbol{p}}' = \boldsymbol{p} - \Delta \boldsymbol{P}_\pm$ the expansion Eq. (\ref{eq:h_taylor_prime}) becomes
\begin{equation}
{h}'_\pm(\boldsymbol{p}) = (\boldsymbol{p} - \Delta \boldsymbol{P}_\pm ) \cdot \boldsymbol{\nabla} {h}'(\boldsymbol{P}'_\pm) + O(p'^2). 
\label{eqn:trans}
\end{equation}
Now that both Hamiltonians Eqs. (\ref{eq:h_taylor}) and (\ref{eqn:trans}) are written down in the same coordinate system, one can compare them. We see that the shift in the Fermi points $\Delta \boldsymbol{P}_\pm$ appears in the Hamiltonian in the same way that a gauge field would appear if we were to apply the minimal coupling prescription.
 
As the Fermi points of the KHLM are always $\pm$ symmetric due to parity symmetry, the Fermi points shift oppositely as $\Delta \boldsymbol{P}_+ = - \Delta \boldsymbol{P}_-$, which means the gauge field about each Fermi point is given by $\boldsymbol{A}_\pm =  -\Delta \boldsymbol{P}_\pm$ (take the charge $e=1$). We see the gauge field couples \textit{chirally} to each Fermi point, i.e., with a sign depending upon the Fermi point, so when $h'_+(\boldsymbol{p})$ and $h'_-(\boldsymbol{p})$ are combined to give a $4 \times 4$ Hamiltonian, the generated gauge field is a chiral gauge field of the form $\boldsymbol{A}\gamma^5 $, where
\begin{equation}
\boldsymbol{A} = - \Delta \boldsymbol{P}_+. \label{eq:gauge_field}
\end{equation}
In the following we will consider particular modifications in the couplings of KHLM and determine the resulting chiral gauge fields.

\section{Chiral gauge fields from the lattice model}
\label{sec:gauge}

We now modify the lattice model to obtain a chiral gauge field $A_\mu$ in the continuum limit. In this section we search for the corresponding terms in the lattice model which produce the spatial components $A_i$ of Eq. (\ref{eq:HamChiral}). Appendix \ref{app:A0} provides a way to generate the temporal component $A_0$ in continuum limit of the KHLM, although this term commutes with the rest of the Hamiltonian and cannot affect any physical observables.

\subsection{Continuum limit of the $\mathbb{Z}_2$ gauge field}
\label{sec:Z2_continuum}
Consider coupling the KHLM to a homogeneous $\mathbb{Z}_2$ gauge field $u_{ij}$. The many-body Hamiltonian for $K=0$ is given by
\begin{equation}
H = \frac{i}{4} \sum_{\langle i , j \rangle} 2 J_{ij} u_{ij} c_i c_j,
\end{equation}
where $\langle i , j \rangle$ is a sum over nearest neighbour pairs (links), cf. Eq. (\ref{eq:KHLM_many_body_ham}). We focus on the isotropic case, $J_x=J_y=J_z=1$, and introduce a gauge field $u_{ij}$ taking values $- 1$ on all $z$ links and $+1$ on all $x$ and $y$ links. Equivalently, this gauge field can be simply encoded on the values of the couplings themselves by setting $u_{ij} = +1$ for all links, then taking $J_x = J_y = 1$ and allowing $J_z$ to take a value of $-1$~\cite{Ville4}, which is the method we use in this section. 

\begin{figure}[t]
\center
\includegraphics[width=\linewidth ]{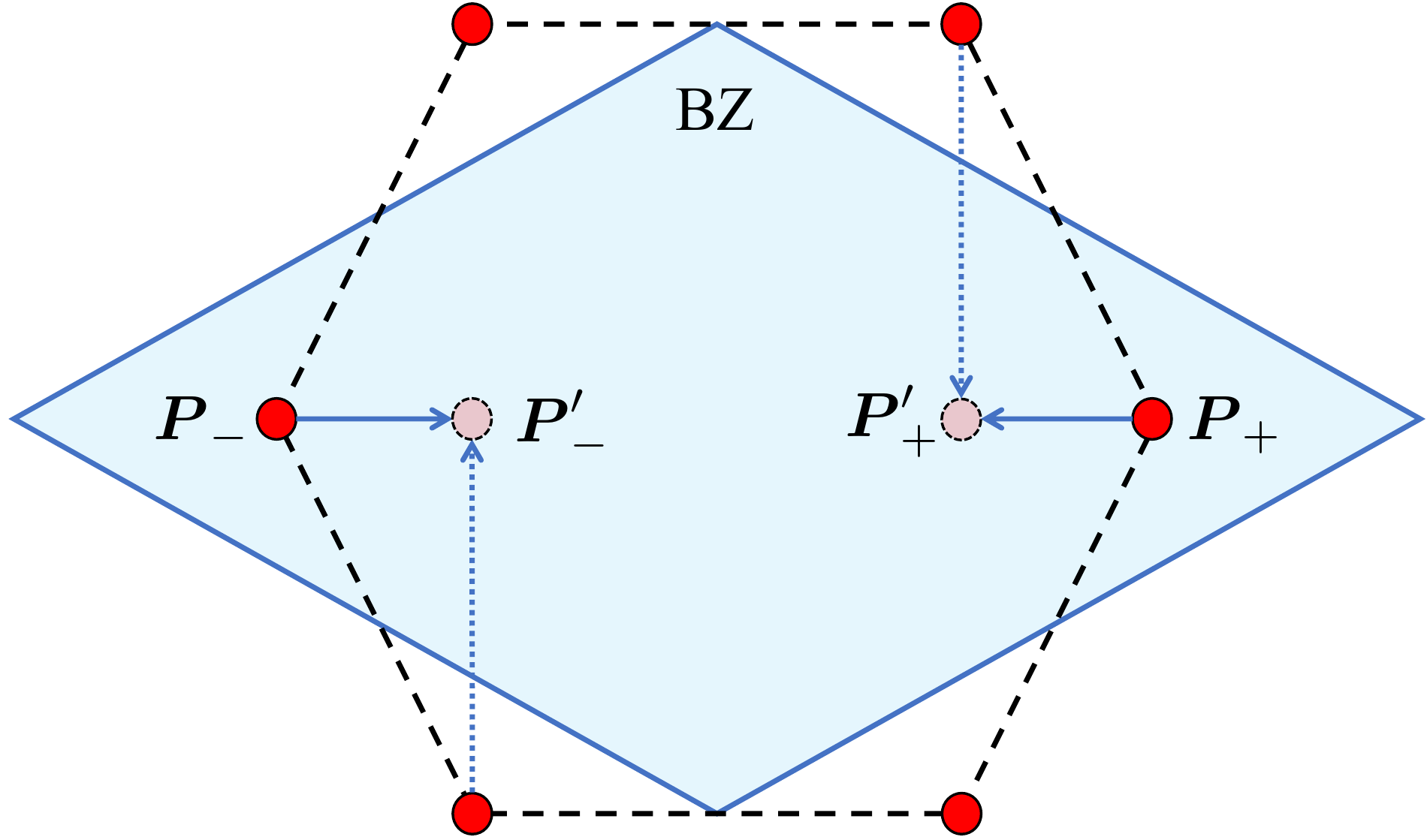}
\caption{The Brillouin zone (BZ) of the honeycomb lattice with two Fermi points, $\boldsymbol{P}_+$ and $\boldsymbol{P}_-$ corresponding to the isotropic couplings $J_x=J_y=J_z=1$. Continuously changing the coupling $J_z$ from $+1$ to $-1$ everywhere on the lattice shifts the Fermi points along the $x$-direction to the positions $\boldsymbol{P}'_+$ and $\boldsymbol{P}'_-$, as shown by the horizontal solid arrow. Due to the parity symmetry of the model, the shift is anti-parallel so $\Delta\boldsymbol{P}_+=-\Delta\boldsymbol{P}_-$, which gives rise to the chiral gauge field $\boldsymbol{A} =\left(2\pi/(3\sqrt{3}),0 \right)$. The final configuration of the Fermi points can also be viewed as an anti-parallel shift of the Fermi points from outside the Brillouin zone in the $y$-direction as shown by the vertical dashed arrows. This shift yields the chiral gauge field $\boldsymbol{A} =\left(0, 2\pi/3 \right)$.}
\label{fig:BZ}
\end{figure}

We can generate the change in sign of $J_z$ with a continuous transformation by allowing $J_z$ to take values in the interval $J_z\in[-1,1]$ across all $z$ links. Using the general result Eq.  (\ref{eq:fermi_points}), the Fermi points of this model are given by 
\begin{equation}
\boldsymbol{P}_\pm(J_z) = \pm \frac{2}{\sqrt{3}} \left( \arccos \left( - \frac{J_z}{2} \right), 0 \right) .
\end{equation}
From this formula, we see that when we switch on the $\mathbb{Z}_2$ gauge field by changing $J_z$ from $+1$ to $-1$, the Fermi points transform as
\begin{equation}
\boldsymbol{P}_\pm = \pm \left( \frac{ 4 \pi }{3 \sqrt{3}} , 0 \right) \  \mapsto \ \boldsymbol{P}'_\pm = \pm \left( \frac{ 2 \pi }{3 \sqrt{3}} , 0 \right). \label{eqn:FPtransx}
\end{equation}
Therefore, upon interpreting the gauge field as the shift of the Fermi points, we conclude from the general formula Eq. (\ref{eq:gauge_field}) that this corresponds to the chiral gauge field $\boldsymbol{A}\gamma_5$ with $\boldsymbol{A} = \left(2\pi/(3\sqrt{3}),0 \right)$. This $x$-direction gauge field corresponds to the anti-parallel displacement of the Fermi points $\Delta \boldsymbol{P}_\pm$ horizontally in the $x$-direction as shown in Fig.~\ref{fig:BZ}.

The particular transformation of the Fermi points given by Eq. (\ref{eqn:FPtransx}), corresponding to $J_z$ changing from $+1$ to $-1$, can have an alternative representation. One can obtain the same final configuration of the Fermi points from the initial configuration with an anti-parallel transportation vertically in the $y$-direction. If we shift $\boldsymbol{P}_+$ up by $\left(0, 2\pi/3 \right)$ and shift $\boldsymbol{P}_-$ down by $\left(0, - 2 \pi/3 \right)$, the Fermi points shift into neighbouring Brillouin zones and we arrive at the final configuration, as can be seen in Fig.~\ref{fig:BZ}. Note that under this transformation the initial points $\boldsymbol{P}_\pm$ from neighbouring Brillouin zones are mapped to the final points $\boldsymbol{P}'_\mp$, therefore our shift is given by $\Delta \boldsymbol{P}_\pm = \boldsymbol{P}'_\pm - \boldsymbol{P}_\mp = \mp \left(0 , 2\pi / 3 \right)$. Using the general formula Eq. (\ref{eq:gauge_field}), this interpretation corresponds to a chiral gauge field pointing in the $y$-direction given by 
\begin{equation}
\boldsymbol{A} =  \left( 0 ,  \frac{2 \pi}{3} \right). \label{eq:gauge_field_zflip}
\end{equation}
In other words, for the transformation of the Fermi points given by Eq. (\ref{eqn:FPtransx}), one can equivalently interpret it as an anti-parallel shift of the Fermi points in the $x$ direction or as an anti-parallel shift of the Fermi points in the $y$ direction. The possibility to interpret the final configuration of the Fermi points in these two equivalent ways is due to the periodicity of momentum space.  

The corresponding $4 \times 4$ continuum limit Hamiltonian, with the interpretation that the Fermi points have shifted anti-parallel in the $y$-direction, is given by
\begin{equation}
h_z(\boldsymbol{p}) = 3 \gamma^0 \left[ \gamma^1 p_x + \gamma^2 \left( p_y + \frac{2 \pi}{3} \gamma^5 \right) \right], \label{eq:z_flip}
\end{equation}
which is the original isotropic case Eq. (\ref{eq:DiracKit}) coupled to a chiral gauge field with a non-zero $y$ component. The sign of the $y$ component kinetic term has flipped relative to Eq. (\ref{eq:DiracKit}), which can be attributed to a non-trivial dreibein. These sign flips will not alter the continuum limit geometry of the model because the dreibein are only defined up to a Lorentz transformation, as $g_{\mu \nu} = e^a_{\ \mu} e^b_{\ \nu} \eta_{ab}$. This is discussed further in Appendix \ref{app:contlimit} and Ref. [\onlinecite{Farjami}].

The representation of the Fermi point transformation in terms of a chiral gauge field in the $y$ direction will help the interpretation of the transformation in terms of a generated flux in the continuum representation of the model, which will be presented in the next section. This latter interpretation follows the equivalence between Peierls substitution and minimal coupling of lattice gauge theories. A detailed discussion of this point is given in Appendix \ref{app:z2_cont}.

Note that one might be tempted to interpret the displacement of the Fermi points due to the change of $J_z$ couplings from $+1$ to $-1$ as a $U(1)$ gauge field. Indeed, the final position of the Fermi points can be obtained from the initial by a parallel shift in the $y$ direction, i.e. where both Fermi points shift in the \textit{same} direction, which is how a $U(1)$ phase would shift the dispersion for graphene. However, we discard this possibility as the resulting $4 \times 4$ Hamiltonian density in the continuum limit would have a term of the form $\mathcal{H}_\text{int}= A_\mu \bar{\Psi} \gamma^\mu \Psi$, which vanishes for the case of Majorana spinors $\Psi$. This is because $j^\mu = \bar{\Psi} \gamma^\mu \Psi$ is the electric current density due to $U(1)$ symmetry and under charge conjugation $\Psi \rightarrow \Psi^{(c)}$ this quantity changes sign. Therefore, for a Majorana spinor, where $\Psi = \Psi^{(c)}$, this quantity vanishes. On the other hand, the $U(1)_A$ interpretation would yield the term $\mathcal{H}_\text{int} = A_\mu \bar{\Psi} \gamma^\mu \gamma^5 \Psi$, where $j_A^\mu = \bar{\Psi} \gamma^\mu \gamma^5 \Psi$ is the axial vector current. This does not vanish for Majorana spinors and is explicitly given by $j^\mu_A = 2 e^{\ \mu}_a \psi_+^\dagger \sigma^a \psi_+$, where $\psi_+$ is the Weyl spinor about the Fermi point $\boldsymbol{P}_+$ and $\sigma^a = (\mathbb{I},\sigma^x,\sigma^y)$.

\subsection{Twists in the lattice}

In this section we modify the couplings of the isotropic model by adding and removing links on the honeycomb lattice. We consider two particular lattice deformations. First, we consider a lattice deformation  that has an equivalent representation in the continuum limit as the $\mathbb{Z}_2$ gauge field. Second, we employ a lattice twist similar to twists that have been considered in the literature in the context of KHLM~\cite{Brennan,Petrova}. Twist defects are of interest as they have been shown to support Majorana modes~\cite{Zheng,Bombin}.

\subsubsection{Twists of Type I}
\label{sec:twist_1}
First, we modify the isotropic model by removing \textit{all} $z$ links and adding two diagonal links across each plaquette of the honeycomb lattice. The corresponding Hamiltonian for $K=0$ is given by
\begin{eqnarray}
H_\text{I} = \frac{i}{4} \sum_{\boldsymbol{r} \in B} && 2 c^b_{\boldsymbol{r}} \left(c^a_{\boldsymbol{r} + \boldsymbol{n}_1} + c^a_{\boldsymbol{r} + \boldsymbol{n}_2} + c^a_{\boldsymbol{r} + \boldsymbol{n}_1 - \boldsymbol{n}_2} \right) \nonumber \\
&&+ 2c^b_{\boldsymbol{r} + \boldsymbol{n}_1 - \boldsymbol{n}_2 } c^a_{\boldsymbol{r}}  + \text{H.c.},
\label{eq:two_cross_ham}
\end{eqnarray}
The red links of the top right honeycomb of Fig. \ref{fig:zero-mode-comparison-A1-z} show an example of these modified couplings inserted \textit{locally}. This lattice modification does not change the Brillouin zone as the lattice retains its periodicity. This modifies $f(\boldsymbol{q}) \rightarrow f_\text{I}(\boldsymbol{q})$ of the single-particle Hamiltonian Eq. (\ref{eq:KHLM_sp_ham}), where 
\begin{equation}
f_\text{I}(\boldsymbol{q}) = 2\left[ e^{i \boldsymbol{q} \cdot \boldsymbol{n}_1} + e^{i \boldsymbol{q} \cdot \boldsymbol{n}_2 } + 2 \cos( \boldsymbol{q} \cdot (\boldsymbol{n}_1 - \boldsymbol{n}_2)) \right].
\end{equation}
The Fermi points of this model are given by
\begin{equation}
\boldsymbol{P}^\text{I}_\pm = \pm \left( \frac{2 \pi}{3 \sqrt{3}} , 0 \right), \label{eq:twist_fermi_point}
\end{equation}
which are the same Fermi points as the ones obtained from a global $J_z$ sign change given by Eq. (\ref{eqn:FPtransx}). We again interpret the shift in the Fermi points relative to the isotropic case as a displacement in the $y$ direction which therefore yields the same chiral gauge field Eq. (\ref{eq:gauge_field_zflip}) in the continuum limit. The corresponding Hamiltonian is given by
\begin{equation}
h_\text{I}(\boldsymbol{p}) = 3 \gamma^0 \left[ 3 \gamma^1 p_x  + \gamma^2 \left( p_y  + \frac{2 \pi}{3} \gamma^5 \right) \right]. \label{eq:cross_ham_cont}
\end{equation}
If we compare Eq. (\ref{eq:cross_ham_cont}) to Eq. (\ref{eq:z_flip}), we see that the continuum limits look identical, apart from a factor of $3$ in front of the $x$ component kinetic term. The emergent chiral gauge fields are the same as the Fermi points of both models have shifted by the same amount relative to the isotropic case. The factor of $3$ is the result of the additional next-to-next-to-nearest-neighbour couplings that changed the geometry of the lattice. Its effect is to scale the $x$ direction of the continuum limit and can be absorbed in the dreibein of the continuum limit. For this reason, we conclude that both lattice models are equivalent as they yield the same continuum limits up to a smooth deformation of the dreibein, so correspond to the same phase. 

\subsubsection{Twists of Type II}
\label{sec:twist_2}
Now consider the case where we modify the isotropic model by removing \textit{all} $z$ links and inserting a single new link across each plaquette, which is similar to what has been used in the literature~\cite{Brennan,Petrova}. The Hamiltonian then becomes
\begin{equation}
H_\text{II} = \frac{i}{4} \sum_{ \boldsymbol{r} \in B} 2 c^b_{\boldsymbol{r}} \left( c^a_{\boldsymbol{r} + \boldsymbol{n}_1} + c^a_{\boldsymbol{r} + \boldsymbol{n}_2} +  c^a_{\boldsymbol{r} + \boldsymbol{n}_2 - \boldsymbol{n}_1 }\right).
\end{equation}
The red links of the top right honeycomb of Fig. \ref{fig:zero-mode-Type2} show an example of these modified couplings inserted \textit{locally}. This modifies $f(\boldsymbol{q}) \rightarrow f_\text{II}(\boldsymbol{q})$ of the single-particle Hamiltonian Eq. (\ref{eq:KHLM_sp_ham}), where 
\begin{equation}
f_\text{II}(\boldsymbol{q}) = 2\left(e^{i \boldsymbol{q} \cdot \boldsymbol{n}_1} + e^{i \boldsymbol{q} \cdot \boldsymbol{n}_2} + e^{i \boldsymbol{q} \cdot ( \boldsymbol{n}_2 - \boldsymbol{n}_1)} \right).
\end{equation}
The Fermi points of this model are given by
\begin{equation}
\boldsymbol{P}^\text{II}_\pm = \pm \left( \frac{2 \pi}{3 \sqrt{3}} , \frac{2 \pi }{9} \right).
\end{equation}
which yields a shift in the Fermi points of $\Delta \boldsymbol{P}_\pm = \pm \left( - 2 \pi/(3 \sqrt{3}) , 2\pi / 9  \right)$, however there is no alternative interpretation of this shift as we had before. The corresponding $4 \times 4$ continuum limit is given by
\begin{equation}
h_\text{II}(\boldsymbol{p}) = \gamma^0 \gamma^x \left( p_x + A_x \gamma^5 \right) + \gamma^0 \gamma^y \left( p_y + A_y \gamma^5 \right),
\end{equation}
which is in Riemann-Cartan form with, using formula Eq. (\ref{eq:gauge_field}), a chiral gauge field $\boldsymbol{A} \gamma^5$, where $\boldsymbol{A} =  \left( 2 \pi/(3 \sqrt{3}) , -2\pi / 9  \right)$. The curved space gamma matrices are given by
\begin{align}
\gamma^x & = e_a^{\ x} \gamma^a = \frac{1}{2}\left( 9 \gamma^1 - \sqrt{3} \gamma^2 \right),
\label{eq:type_II_x_gamma} \\
\gamma^y & = e_a^{\ y} \gamma^a = \frac{1}{2} \left( 3 \sqrt{3} \gamma^1 + 3 \gamma^2 \right),
\label{eq:type_II_y_gamma}
\end{align}
which signifies a non-trivial dreibein $e_a^{\ \mu}$. This non-trivial dreibein corresponds to a non-trivial metric in the continuum limit. This is to be expected, as the twists have changed the geometry of the honeycomb lattice.

\subsection{Transforming between $\mathbb{Z}_2$ gauge field and twists}

We now consider the continuous transformation between the two modified Hamiltonians, $H_z$ with a global $\mathbb{Z}_2$ gauge field manifested by $J_z=-1$ [see Eq. (\ref{eq:H_J})] and $H_\text{I}$ with type I twists as defined in Eq. (\ref{eq:two_cross_ham}), and trace the motion of the Fermi points. We define the Hamiltonian
\begin{equation}
H(\lambda) = (1- \lambda) H_z + \lambda H_\text{I}, \quad \lambda \in [0,1],
\end{equation}
such that when we change $\lambda$ from $0$ to $1$, we transform the Hamiltonian from $H_z$ to $H_\text{I}$. The single-particle Hamiltonian corresponding to $H(\lambda)$ is given by Eq. (\ref{eq:KHLM_sp_ham}), where $f(\boldsymbol{q})$ is now given by
\begin{equation}
f(\boldsymbol{q},\lambda) = 2\big[e^{i \boldsymbol{q} \cdot \boldsymbol{n}_1 } + e^{i \boldsymbol{q} \cdot \boldsymbol{n}_2} + 2 \lambda \cos ( \boldsymbol{q} \cdot (\boldsymbol{n}_1 - \boldsymbol{n}_2 )) + (\lambda - 1)\big].
\end{equation}
and $\Delta(\boldsymbol{q})= 0$ as we keep for convenience $K=0$. The corresponding dispersion relation is given by $E(\boldsymbol{p},\lambda) = \pm |f(\boldsymbol{q},\lambda)|$ and has the Fermi points given by
\begin{equation}
\boldsymbol{P}_\pm(\lambda) = \pm \left( \frac{2 \pi}{3 \sqrt{3}} , 0 \right).
\end{equation} 
We observe that the Fermi points are independent of the value of $\lambda$, remaining fixed at their corresponding values, cf. Eqs. (\ref{eq:twist_fermi_point}) and (\ref{eqn:FPtransx}).
As a result, the global $\mathbb{Z}_2$ gauge field can be continuously deformed to a global lattice modification of the form given by Eq. (\ref{eq:two_cross_ham}) without changing the corresponding chiral gauge fields.

A natural question to ask is whether these two modifications are equivalent locally. In the continuum limit, such local modifications are expected to correspond to locally varying chiral gauge fields giving rise to non-trivial chiral fluxes. From the lattice description, we know that local $\mathbb{Z}_2$ transformations give rise to Majorana bounding vortices, while local lattice deformations of the form Eq. (\ref{eq:two_cross_ham}) can also trap Majorana zero modes. As they both correspond to the same chiral gauge field, we expect them to give rise to the same Majorana zero modes. This will be explicitly verified in the following.

\section{Chiral gauge fields and Majorana zero modes}
\label{sec:zero_modes}
In this section we investigate the relation between local chiral gauge fields and Majorana zero modes. We have seen that homogeneous $\mathbb{Z}_2$ gauge fields on the links of the honeycomb lattice and homogeneous lattice deformations of the form Eq. (\ref{eq:two_cross_ham}) give rise to the same continuum limit up to a rescaling of the dreibein. This result suggests that both models are equivalent at the \textit{lattice} level too. In this section we test this numerically by introducing the deformations and $\mathbb{Z}_2$ gauge fields \textit{locally} along a finite path through the lattice. We demonstrate both models are adiabatically connected and produce non-trivial fluxes at the endpoints of the path which trap zero modes in the same way, allowing us to conclude that both models are equivalent at the lattice level too.


\subsection{Flux of chiral gauge fields}

If a $\mathbb{Z}_2$ gauge field is inserted on the lattice of the KHLM by flipping the gauge field from $+1$ to $-1$ locally, one can produce $\pi$-vortices which trap Majorana zero modes~\cite{Kitaev}. For example, if we inserted a gauge field taking values $-1$ on all $z$ links along a path $P$ and $+1$ on all other links, then one finds vortices localised at each end of the path. 
A natural question to ask is whether such vortices appear in the continuous representation of the model. In particular, we want to investigate whether the chiral gauge field associated with local configurations of the $\mathbb{Z}_2$ gauge field can give rise to the same $\pi$ fluxes that trap Majorana zero modes in the continuum~\cite{Jackiw,Chamon}.

In Sec.~\ref{sec:gauge}, we deduced that a \textit{global} $\mathbb{Z}_2$ gauge field taking values $-1$ on all $z$ links and $+1$ on all $x$ and $y$ links yields a chiral gauge field in the continuum limit of the form $\boldsymbol{A} \gamma^5$, where $\boldsymbol{A} = \left( 0 , \frac{2 \pi}{3} \right)$. If we were to perform the same calculation on the brick wall lattice representation of the honeycomb lattice, the resulting chiral gauge field is given by $\boldsymbol{A} = (0, \pi)$, as shown in Appendix \ref{app:z2_cont}. This is in agreement with the equivalence between Peierls substitution and minimal coupling. Nevertheless, this is not the case in the honeycomb lattice model. The discrepancy is due to the fact that $x$ and $y$ links of the honeycomb lattice have a spatial $y$ component when oriented in the honeycomb lattice configuration, yet they receive no contribution from the gauge field. Hence, the value $2\pi/3$ is obtained from an average along strips in the $y$-direction of length $1$ with phase $\pi$ ($z$ links) and of length $1/2$ with phase $0$ ($x$ and $y$ links). As the argument below is concerned with horizontal paths $P$, which are well-localised in the $y$ direction crossing $z$ links that contribute a $\pi$ phase, we will take the corresponding chiral gauge field to be $\boldsymbol{A} = (0, \pi)$.

\begin{figure}[t]
\center
\includegraphics[width=\linewidth ]{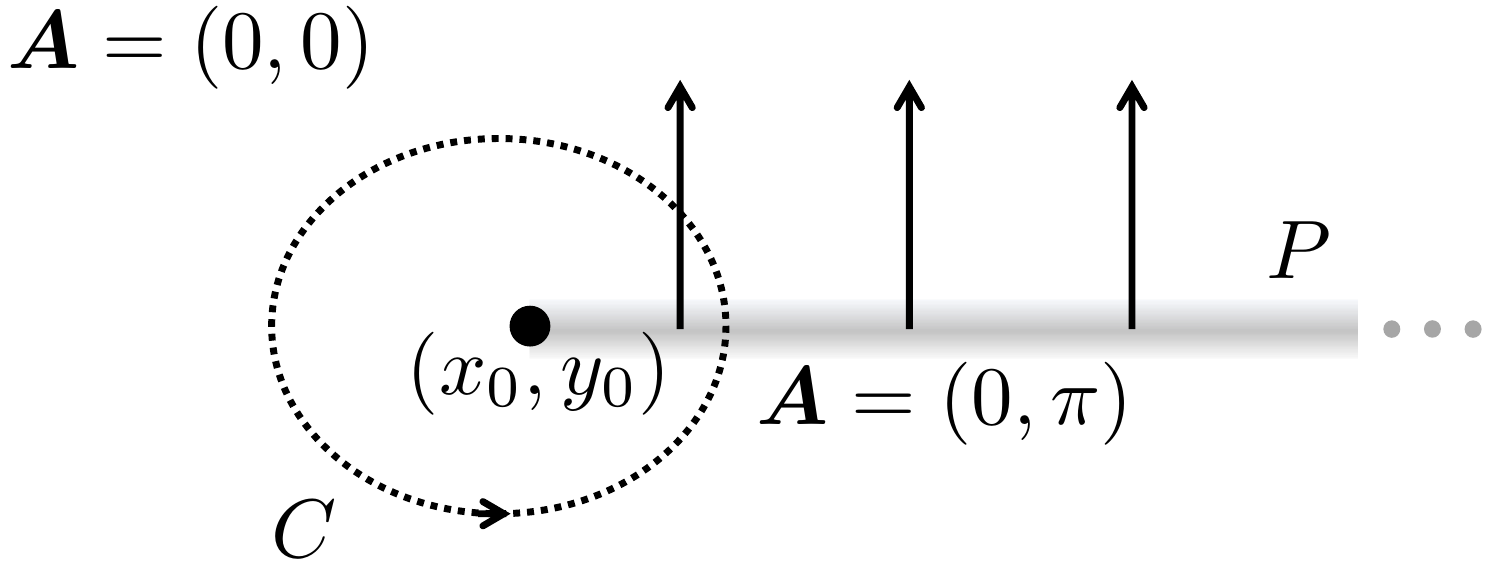}
\caption{The configuration of the chiral gauge field of the form $\boldsymbol{A}(\boldsymbol{r})\gamma^5 = \boldsymbol{A} \theta(x-x_0) \delta(y-y_0) \gamma^5$ confined in the $y$ direction along the path $P$ that starts at the point $\boldsymbol{r}_0=(x_0,y_0)$ and extends to infinity in the $x$ direction. Along the path, the gauge field takes value $\boldsymbol{A}=(0,\pi)$, while it takes the value $\boldsymbol{A}=(0,0)$ outside the path. This configuration of $\boldsymbol{A}\gamma^5$ gives rise to a flux $\Phi = \oint_C \mathrm{d} \boldsymbol{l} \cdot \boldsymbol{A}=\pi$ going through the loop $C$ that encloses $\boldsymbol{r}_0$.}
\label{fig:Flux}
\end{figure}

Suppose we insert the $\mathbb{Z}_2$ gauge field  locally along a horizontal straight path $P$ starting at the point $\boldsymbol{r}_0 = (x_0,y_0)$ heading off to infinity in the $x$ direction, as shown in Fig.~\ref{fig:Flux}. In the continuum limit, this would be described by a chiral gauge field 
\begin{equation}
\boldsymbol{A}(\boldsymbol{r})\gamma^5 = \boldsymbol{A} \theta(x-x_0) \delta(y-y_0) \gamma^5,
\label{eqn:pathchiral}
\end{equation}
where $\boldsymbol{A} = (0,\pi)$. The ``magnetic field" of this gauge field configuration is given by
\begin{equation}
\boldsymbol{B} \gamma^5 = \nabla \times \boldsymbol{A}(\boldsymbol{r})\gamma^5 = \pi \delta(x-x_0)\delta(y-y_0) \gamma^5 \hat{\boldsymbol{z}}.
\end{equation}
The phase along a loop $C$ that surrounds the endpoint $\boldsymbol{r}_0$ of $P$ is given by
\begin{equation}
\Phi = \oint_C \mathrm{d} \boldsymbol{l} \cdot \boldsymbol{A} = \int_S \mathrm{d} \boldsymbol{S} \cdot \boldsymbol{B} = \pi,
\label{eqn:piflux}
\end{equation}
where $S$ is the surface enclosed by the path $C$. Hence, the configuration Eq. (\ref{eqn:pathchiral}) of the chiral gauge field gives rise to a chiral $\pi$ flux. Similarly, if we insert the twists of type I from the previous section locally, along the same path $P$, we achieve the same gauge field Eq. (\ref{eqn:pathchiral}) and $\pi$ flux Eq. (\ref{eqn:piflux}). This suggests that the Majorana zero modes produced by the twists are equivalent to the Majorana zero modes trapped by $\mathbb{Z}_2$ vortices. Indeed, when inserting this gauge field into the Dirac Eq. (\ref{eq:HamChiral}), it is known that vortex profiles will trap zero modes~\cite{Jackiw}.

In the following, we first consider the generation of Majorana zero modes when local $\mathbb{Z}_2$ gauge fields or local twists are created. Then we adiabatically connect these zero modes, thus demonstrating that they are equivalent.

\subsection{Majorana zero modes}

\begin{figure}[t]
\center
\includegraphics[width=\linewidth ]{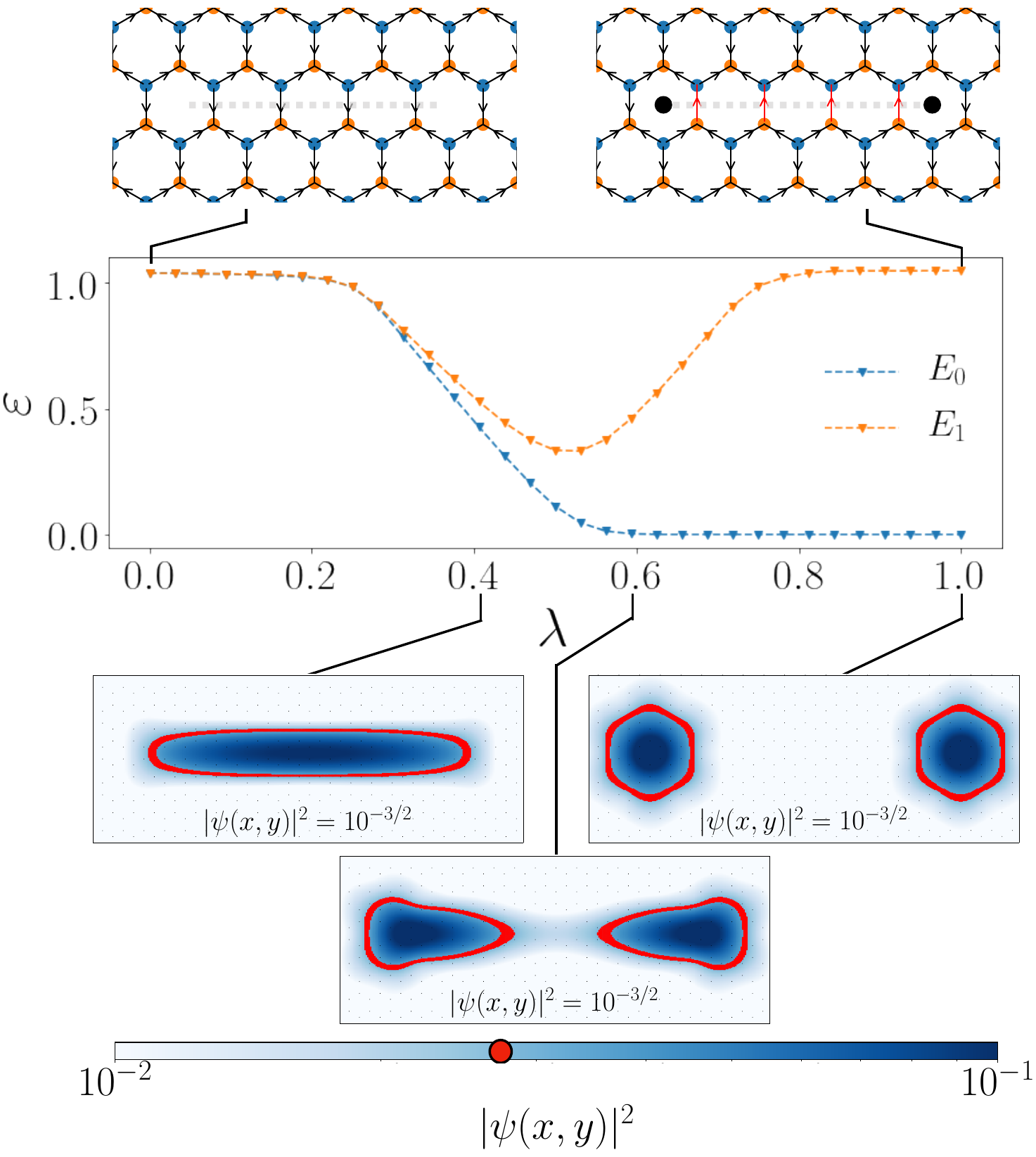}
\caption{The formation of zero modes in $H^\text{P}_\text{v}(\lambda)$. Top: A sketch of $H^\text{P}_\text{v}(\lambda)$ at $\lambda=0$ and $\lambda=1$ for a smaller system size. The path $P$, indicated by a dashed grey line, runs perpendicular to the $z$-links of the lattice and has a length $L/2$. Black links take the value $u_{ij}=+1$, red links take the value $u_{ij}=-1$. The black dots in the centre of plaquettes indicate the approximate position of vortices. Middle: The energy gap of $H^\text{P}_\text{v}(\lambda)$ as a function of $\lambda$ for a system with linear dimension $L=30$, isotropic $J=1$, and $K=0.1$. Zero modes are created with an energy gap above them as the sign of $u_z$ flips, i.e, at $\lambda\approx 0.5$. Bottom: The continuous profile of the wave function $|\psi(\boldsymbol{r})|^2$ of the gradually generated localised zero modes at $\lambda\approx 0.4,0.6,1$. The size and shape of the vortices are characterised by finding the set of points where $|\psi(\boldsymbol{r})|^2 = 10^{-3/2}$, as illustrated by the red boundary line.}
\label{fig:z2-gap}
\end{figure}

While the $\mathbb{Z}_2$ values of the links can change through a discrete process, it is possible to implement it in a continuous way. We observe the formation of zero modes throughout this continuous process by studying the behaviour of the energy spectrum and wave functions. For example, consider an initial Hamiltonian $H_0$, where all the gauge degrees of freedom have value $u_{ij}=+1$. Consider also a final Hamiltonian $H^\text{P}_\text{v}$, where the vertical $z$ links along a local path $P$ in the $x$-direction take the opposite sign $u_{ij}=-1$, as shown in Fig.~\ref{fig:z2-gap}. We label the links along this path as $u_z$. To shift from one Hamiltonian to the other we consider the interpolating Hamiltonian
\begin{equation}
H^\text{P}_\text{v}({\lambda})=(1-\lambda)H_0+\lambda H^\text{P}_\text{v}, \quad \lambda \in [0,1].
\end{equation}
The result is a continuous change in the value of $u_z$ from $u_z=1$ for $\lambda=0$ to $u_z=-1$ for $\lambda=1$. 
Thus, we expect to see Majorana zero modes appearing at the end points of $P$ as $\lambda$ approaches $1$. All numerical simulations presented in this section are for models with periodic boundary conditions, system size $L=30$, isotropic $J=1$ and $K=0.1$.

\begin{figure}[t]
\center
\includegraphics[width=\linewidth ]{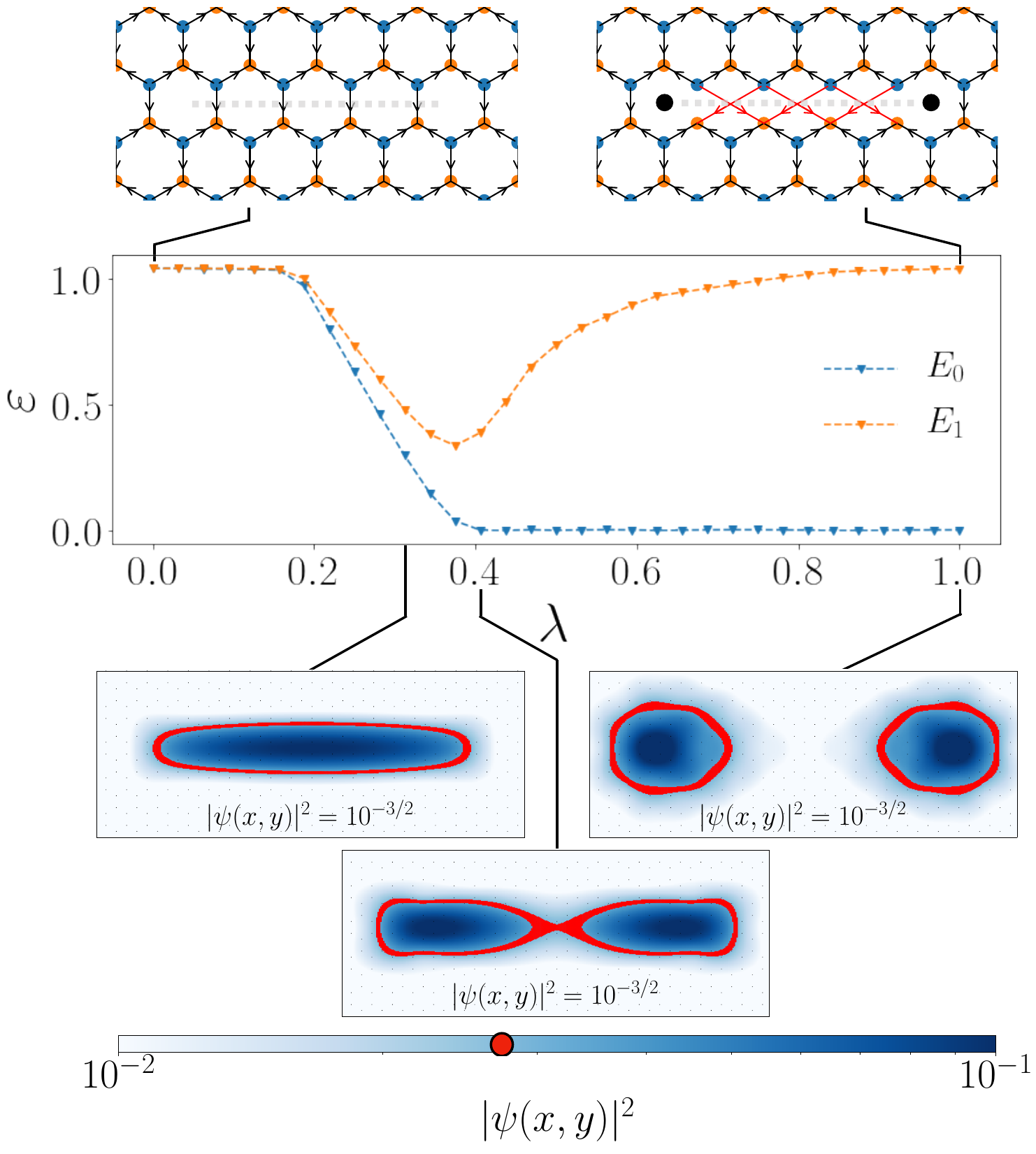}
\caption{The formation of zero modes in $H^\text{P}_\text{I}(\lambda)$.  Top: A sketch of $H^\text{P}_\text{I}(\lambda)$ at $\lambda=0$ and $\lambda=1$ for a smaller system. The path $P$, indicated by a dashed grey line, runs perpendicular to the $z$ links of the lattice and has a length $L/2$. The new next-to-next-to-nearest neighbour couplings are highlighted in red. The black dots in the centre of plaquettes indicate the approximate position of vortices. Middle: The energy gap of $H^\text{P}_\text{I}(\lambda)$ as a function of $\lambda$ for a system with linear dimension $L=30$, isotropic $J$, and $K=0.1$. Zero modes are created with an energy gap above them at $\lambda\approx 0.4$. The behaviour of the gap is similar to the gap observed as the sign of the $\mathbb{Z}_2$ gauge field flips in Fig.~\ref{fig:z2-gap}. Bottom: The continuous profile of the wave function $|\psi(\boldsymbol{r})|^2$ of the gradually generated localised zero modes at $\lambda\approx 0.4,0.6,1$. The size and shape of the vortices are characterised by finding the set of points where $|\psi(\boldsymbol{r})|^2 = 10^{-3/2}$, as illustrated by the red boundary line.}
\label{fig:zero-mode-comparison-A1-z}
\end{figure}

The generation of localised Majorana zero modes is shown in Fig.~\ref{fig:z2-gap} as $\lambda$ increases in discrete steps demonstrating that the local $\mathbb{Z}_2$ gauge field creates $\pi$-vortices. The single particle Hamiltonian $H_\text{vortex}(\lambda)$ is diagonalised for each discrete value of $\lambda$ and the energies $E_0$ and $E_1$ of the two lowest eigenstates are plotted in Fig.~\ref{fig:z2-gap}. At $\lambda=0$ the model is clearly gapped with no zero energy modes, while at $\lambda=1$ there is a clear zero energy mode with a gap above it. The gap between $E_0$ and $E_1$ forms at a transition point around $\lambda\approx 0.5$. From the diagonalisation of $H_\text{vortex}(\lambda)$, we also obtain the probability density at each lattice site $|\psi_i|^2$ for the lowest energy eigenstate. We call this the spatial wave function of the vortices. To visualise the shape of the zero modes, we approximate them with a continuous function as shown in Fig.~\ref{fig:z2-gap} (bottom) [see Appendix \ref{App:shape}]. As we approach the transition point $\lambda\approx 0.5$ a single fermion mode appears over the length of the path $P$. This mode splits into two Majorana zero modes as $\lambda$ increases, becoming exponentially localised at the end points of $P$ as we approach $\lambda=1$.

\begin{figure}[t]
\center
\includegraphics[width=\linewidth]{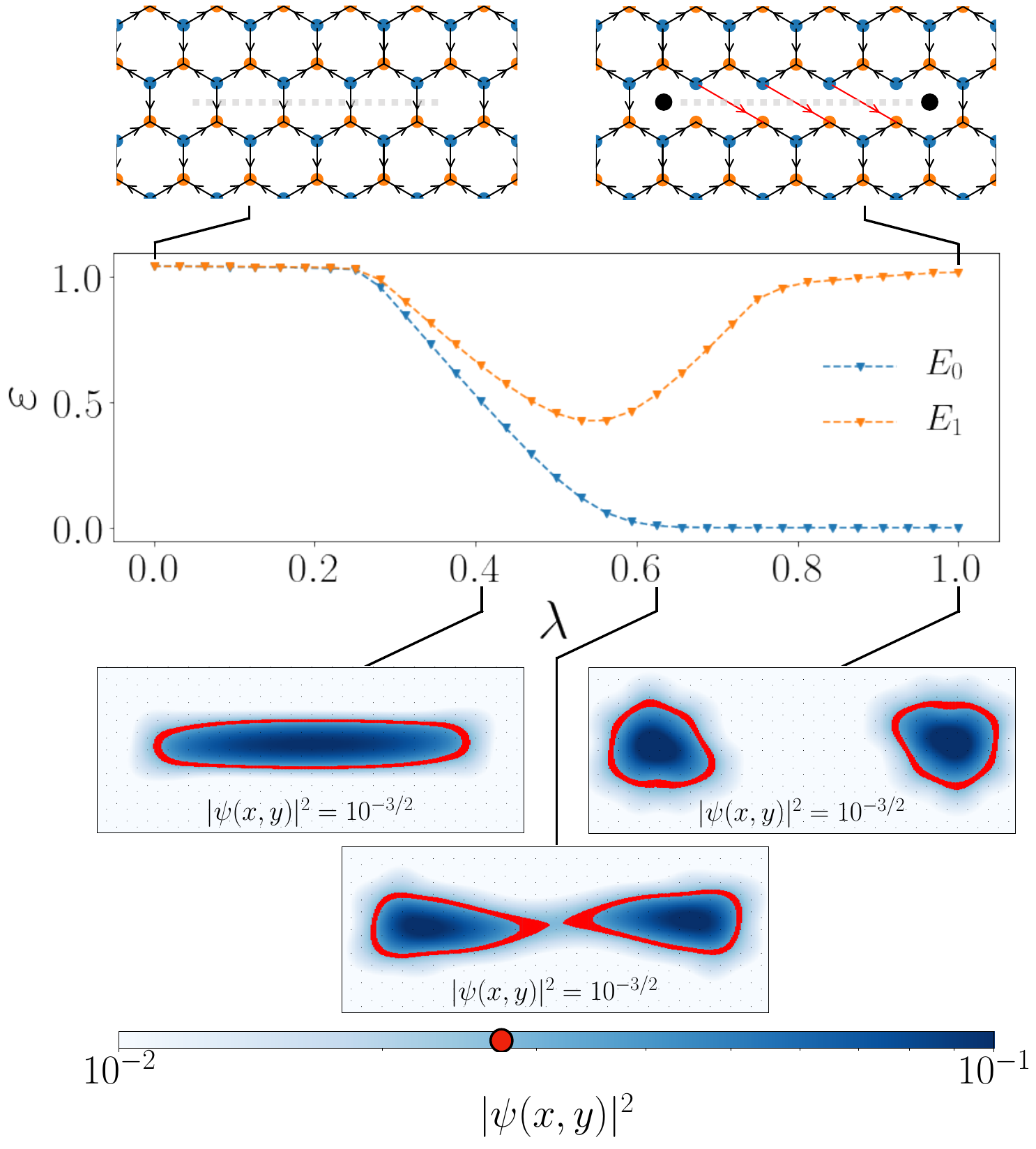}
\centering
\caption{The formation of zero modes in $H^\text{P}_\text{II}(\lambda)$. Top: A sketch of $H^\text{P}_\text{II}(\lambda)$ at $\lambda=0$ and $\lambda=1$ for a smaller system. The path $P$, indicated by a dashed grey line, runs perpendicular to the $z$-links of the lattice and has a length $L/2$. The new next-to-next-to-nearest neighbour couplings are highlighted in red. The black dots in the centre of plaquettes indicate the approximate position of vortices. Middle: The energy gap of $H^\text{P}_\text{II}(\lambda)$ as a function of $\lambda$ for a system with linear dimension $L=30$, isotropic $J$ and $K=0.1$. Zero modes are created with an energy gap above them at $\lambda\approx 0.5$. The behaviour of the gap is similar to the gap observed as the sign of the $\mathbb{Z}_2$ gauge field flips in Fig.~\ref{fig:z2-gap}. Bottom: The continuous profile of the wave function $|\psi(\boldsymbol{r})|^2$ of the gradually generated localised zero modes at $\lambda\approx 0.4,0.6,1$. The size and shape of the vortices are characterised by finding the set of points where $|\psi(\boldsymbol{r})|^2 = 10^{-3/2}$, as illustrated by the red boundary line.}
\label{fig:zero-mode-Type2}
\end{figure}

We now consider the isotropic vortex-free KHLM Hamiltonian $H_0$ and we create a non-zero chiral gauge field by introducing lattice deformations of type I as the ones in Hamiltonian (\ref{eq:two_cross_ham}). We consider these deformations along a horizontal path $P$ that result in the creation of twists at the endpoints of the path, as shown in Fig.~\ref{fig:zero-mode-comparison-A1-z}. We denote the resulting Hamiltonian as $H^\text{P}_\text{I}$. We use the same method as above to continuously shift between these two Hamiltonians:
\begin{equation}
H^\text{P}_\text{I}({\lambda})=(1-\lambda)H_0+\lambda H^\text{P}_\text{I}, \quad \lambda \in [0,1].
\end{equation}
Figure \ref{fig:zero-mode-comparison-A1-z} shows the energies of the two lowest eigenstates of the single particle Hamiltonian produced by varying $\lambda$ as well as the continuous approximations of the spatial wave function as vortices are produced. Similar to the vortex creation, we observe that the formation of twists give rise of stable Majorana zero modes as $\lambda$ increases and the gap begins to open. Hence, type I twists bound Majorana zero modes much like the $\mathbb{Z}_2$ vortices do.

Finally, we consider the equivalent generation of twists of type II along a horizontal path $P$. The resulting energies and wave functions are depicted in Fig.~\ref{fig:zero-mode-Type2}, demonstrating that type II twists bound Majorana zero modes in much the same way as $\mathbb{Z}_2$ vortices and type I twists.

\subsection{Adiabatic equivalence between lattice twists and vortices}

\begin{figure}[t]
\center
\includegraphics[width=\linewidth ]{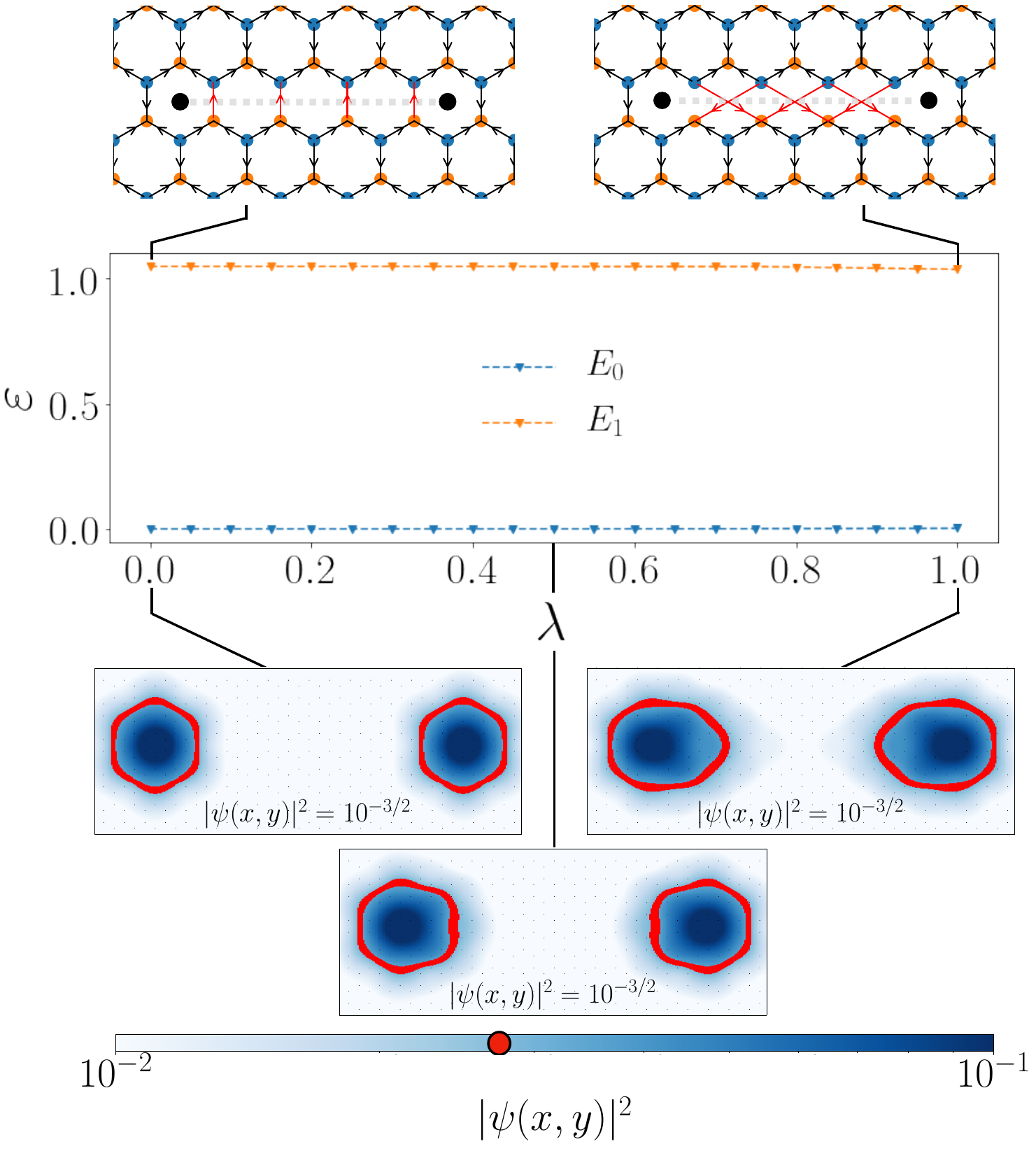}
\caption{The adiabatic equivalence of zero modes in $H^\text{P}_\text{v-I}(\lambda)$. Top: A sketch of $H^\text{P}_\text{v-I}(\lambda)$ at $\lambda=0$ and $\lambda=1$ for a smaller system size. The path $P$, indicated by a dashed grey line, remains constant, runs perpendicular to the $z$ links of the lattice and has a length $L/2$. The modified links along the path $P$ are highlighted in red. The black dots in the centre of plaquettes indicate the approximate position of vortices. Middle: The energy gap of $H^\text{P}_\text{v-I}(\lambda)$ as a function of $\lambda$ that interpolates between the two $\mathbb{Z}_2$ vortex configuration and lattice twists configuration of type I, for a system with linear dimension $L=30$, isotropic $J=1$, and $K=0.1$. The gap remains almost constant for all values of $\lambda$, indicating stable zero modes throughout the transition. Bottom: The continuous profile zero modes at $\lambda\approx 0,0.5,1$ shows they remain fixed in place and well-localised throughout the adiabatic transition. The shape of the zero modes at $\lambda=1$ appear stretched in the $x$-direction compared to $\lambda=0$ due to the change in the dreibein in Eq. (\ref{eq:cross_ham_cont}). The size and shape of the vortices are characterised by finding the set of points where $|\psi(\boldsymbol{r})|^2 = 10^{-3/2}$, as illustrated by the red boundary line.}
\label{fig:adiabatic1}
\end{figure}

We established in the previous section that string-like configurations of twists in the lattice give rise to Majorana zero modes at the end points of the string. This is very similar to the zero modes trapped by string-like configurations of the $\mathbb{Z}_2$ gauge field that creates $\pi$ flux vortices at its end-points. Here we demonstrate that these two apparently different ways of realising Majorana zero modes, i.e., by changing the sign of certain links or by modifying the connectivity of the lattice, are actually physically equivalent. We demonstrate this by adiabatically transforming between these two configurations and considering both the behaviour of the energy spectrum as well as the wave function of the zero modes. 

\begin{figure}[t]
\center
\includegraphics[width=\linewidth ]{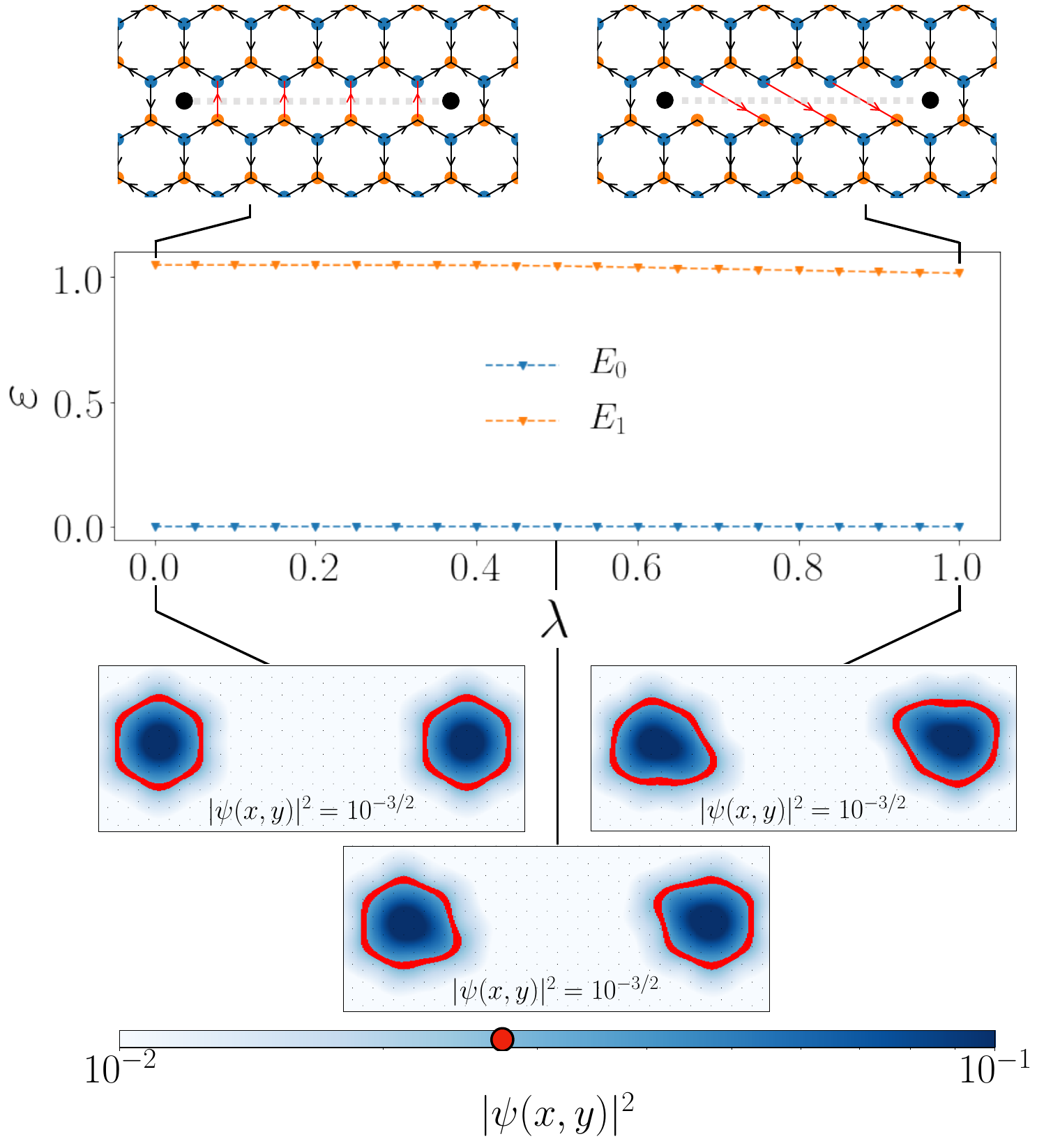}
\caption{The adiabatic equivalence of zero modes in $H^\text{P}_\text{v-II}(\lambda)$. Top: A sketch of $H^\text{P}_\text{v-II}(\lambda)$ at $\lambda=0$ and $\lambda=1$ for a smaller system size. The path $P$, indicated by a dashed grey line, remains constant, runs perpendicular to the $z$ links of the lattice and has a length $L/2$. The modified links along the path $P$ are highlighted in red. The black dots in the centre of plaquettes indicate the approximate position of vortices. Middle: The energy gap of $H^\text{P}_\text{v-II}(\lambda)$ as a function of $\lambda$ that interpolates between the two $\mathbb{Z}_2$ vortex configuration and lattice twists configuration of type II, for a system with linear dimension $L=30$, isotropic $J=1$, and $K=0.1$. The gap remains almost constant for all values of $\lambda$, indicating stable zero modes throughout the transition. Bottom: The continuous profile zero modes at $\lambda\approx 0,0.5,1$ shows they remain fixed in place and well-localised throughout the adiabatic transition. The asymmetry in the shape of the zero modes at $\lambda=1$ compared to $\lambda=0$ is reflected in the asymmetry of the dreibein in Eqs. (\ref{eq:type_II_x_gamma}) and (\ref{eq:type_II_y_gamma}). The size and shape of the vortices are characterised by finding the set of points where $|\psi(\boldsymbol{r})|^2 = 10^{-3/2}$, as illustrated by the red boundary line.}
\label{fig:adiabatic2}
\end{figure}

We take the Hamiltonians $H^\text{P}_\text{v}$ and $H^\text{P}_\text{I}$, defined in the previous section and depicted in Fig.~\ref{fig:z2-gap} (top, right) and \ref{fig:zero-mode-comparison-A1-z} (top, right), respectively. We define the Hamiltonian
\begin{equation}
H^\text{P}_\text{v-I}(\lambda) = (1-\lambda)H^\text{P}_\text{v} + \lambda H^\text{P}_\text{I}, \quad \lambda \in [0,1].
\label{eqn:adia}
\end{equation}
This allows us to adiabatically transition between the two Hamiltonians by varying $\lambda$. The path $P$ remains fixed throughout this transition. Figure \ref{fig:adiabatic1} shows the energy gap of the system and the continuous approximation of the wave function of a pair of zero modes as we adiabatically transition between $H^\text{P}_\text{v}$ and $H^\text{P}_\text{I}$. We observe that  the zero modes remain energetically separated from the rest of the states for all $\lambda$ with an energy gap that remains more or less constant throughout the process.  Moreover, the zero modes of the model remain fixed in place and well-localised throughout the adiabatic transition. Hence, the two ways of generating vortices are physically equivalent. The shape of the zero modes of $H^\text{P}_\text{I}$ appear stretched in the $x$ direction compared to $H^\text{P}_\text{v}$. This is due to the change in the dreibein in  Eq.(\ref{eq:cross_ham_cont}). This adiabatic process also demonstrates that there is a continuous family of lattice configurations given by $H^\text{P}_\text{v-I}(\lambda)$ for $\lambda\in [0,1]$ that give rise to the same localised Majorana zero modes. 

Similarly, Majorana zero modes produced by twists of type II are also adiabatically connected to zero modes produced by $\mathbb{Z}_2$ vortices. This is shown explicitly in Fig.~\ref{fig:adiabatic2}. The asymmetry in the shape of the zero modes for $H^\text{P}_\text{II}$ is reflected in the asymmetry of the dreibein in Eqs. (\ref{eq:type_II_x_gamma}) and (\ref{eq:type_II_y_gamma}), which demonstrates that the continuum limit geometry is scaled unevenly along each axis. However, the analysis of Sec. \ref{sec:twist_2} concluded that twists of type II do not yield a gauge field with exactly a $\pi$ flux. This is because the Fermi points of this model do not shift in the same way as they did for the case of implementing a $\mathbb{Z}_2$ gauge field. Therefore, Fig.~\ref{fig:adiabatic2} also demonstrates that the zero modes are stable as the flux of the underlying gauge field changes adiabatically as we transition between the two models.

\section{Conclusion}

The generation and manipulation Majorana fermions is one of the central problems in the current effort to understand the physics of non-Abelian anyons and employ them for quantum technologies. Here we demonstrated that two of the leading ways of trapping Majorana zero modes, employing vortices and employing lattice twists, are physically equivalent. We demonstrated this equivalence by finding the appropriate representation of these lattice defects in the continuum limit in terms of chiral gauge fields. We showed analytically that both $\mathbb{Z}_2$ gauge fields and lattice deformations have an equivalent representation in the low-energy spectrum of the system in terms of chiral gauge field coupled to the Majorana version of the Dirac equation. As the two continuum limits differed only by a smooth transformation of the dreibein, this suggested that the lattice level Hamiltonians must also be equivalent which we investigated numerically by simulation local configurations of the $\mathbb{Z}_2$ gauge field and lattice deformations. 

We observed numerically that local configurations of this chiral gauge field can create $\pi$ flux vortices. Motivated by this equivalence we investigated the possibility of Majorana bounding twists being physically equivalent to Majorana bounding vortices. We performed an adiabatic transformation between Hamiltonians that encode twists and vortices and showed that both the structure of the energy spectrum as well as the localisation properties of the Majorana zero modes remain invariant during the adiabatic transformation.

Our investigation demonstrates that Majorana bounding twists are physically equivalent to vortices even though they do not have a gauge field representation in the lattice level. Nevertheless, they give rise to a chiral gauge field with configurations that in the continuum limit are equivalent to the $\mathbb{Z}_2$ gauge configurations. This opens up a variety of possible investigations. First, it is possible to realise gauge theories that do not necessarily have a traditional interpretation in the lattice level in terms of Wilson lines. This can give wider flexibility for the realisation of gauge theories in the laboratory, e.g. with optical lattices~\cite{Alba}. Second, the adiabatic transformation between vortices and twists created a continuous spectrum of defects that can support Majorana zero modes beyond the two limiting cases. The possibility of having a wider range of Majorana bounding defects can facilitate their experimental generation and detection.

\acknowledgements

We would like to thank Omri Golan, Jaakko Nissinen, Zlatko Papi\'c, and Chris Self for inspiring conversations. We would like to thank Chris Self for providing the simulation code for the KHLM~\cite{Self_code}. This work was supported by EPSRC Grant No. EP/R020612/1. Statement of compliance with EPSRC policy framework on research data: This publication is theoretical work that does not require supporting research data.

\bibliography{chiral_references}

\begin{thebibliography}{63}%
\makeatletter
\providecommand \@ifxundefined [1]{%
 \@ifx{#1\undefined}
}%
\providecommand \@ifnum [1]{%
 \ifnum #1\expandafter \@firstoftwo
 \else \expandafter \@secondoftwo
 \fi
}%
\providecommand \@ifx [1]{%
 \ifx #1\expandafter \@firstoftwo
 \else \expandafter \@secondoftwo
 \fi
}%
\providecommand \natexlab [1]{#1}%
\providecommand \enquote  [1]{``#1''}%
\providecommand \bibnamefont  [1]{#1}%
\providecommand \bibfnamefont [1]{#1}%
\providecommand \citenamefont [1]{#1}%
\providecommand \href@noop [0]{\@secondoftwo}%
\providecommand \href [0]{\begingroup \@sanitize@url \@href}%
\providecommand \@href[1]{\@@startlink{#1}\@@href}%
\providecommand \@@href[1]{\endgroup#1\@@endlink}%
\providecommand \@sanitize@url [0]{\catcode `\\12\catcode `\$12\catcode
  `\&12\catcode `\#12\catcode `\^12\catcode `\_12\catcode `\%12\relax}%
\providecommand \@@startlink[1]{}%
\providecommand \@@endlink[0]{}%
\providecommand \url  [0]{\begingroup\@sanitize@url \@url }%
\providecommand \@url [1]{\endgroup\@href {#1}{\urlprefix }}%
\providecommand \urlprefix  [0]{URL }%
\providecommand \Eprint [0]{\href }%
\providecommand \doibase [0]{http://dx.doi.org/}%
\providecommand \selectlanguage [0]{\@gobble}%
\providecommand \bibinfo  [0]{\@secondoftwo}%
\providecommand \bibfield  [0]{\@secondoftwo}%
\providecommand \translation [1]{[#1]}%
\providecommand \BibitemOpen [0]{}%
\providecommand \bibitemStop [0]{}%
\providecommand \bibitemNoStop [0]{.\EOS\space}%
\providecommand \EOS [0]{\spacefactor3000\relax}%
\providecommand \BibitemShut  [1]{\csname bibitem#1\endcsname}%
\let\auto@bib@innerbib\@empty
\bibitem [{\citenamefont {Castro~Neto}\ \emph {et~al.}(2009)\citenamefont
  {Castro~Neto}, \citenamefont {Guinea}, \citenamefont {Peres}, \citenamefont
  {Novoselov},\ and\ \citenamefont {Geim}}]{Neto}%
  \BibitemOpen
  \bibfield  {author} {\bibinfo {author} {\bibfnamefont {A.~H.}\ \bibnamefont
  {Castro~Neto}}, \bibinfo {author} {\bibfnamefont {F.}~\bibnamefont {Guinea}},
  \bibinfo {author} {\bibfnamefont {N.~M.~R.}\ \bibnamefont {Peres}}, \bibinfo
  {author} {\bibfnamefont {K.~S.}\ \bibnamefont {Novoselov}}, \ and\ \bibinfo
  {author} {\bibfnamefont {A.~K.}\ \bibnamefont {Geim}},\ }\bibfield  {title}
  {\enquote {\bibinfo {title} {The electronic properties of graphene},}\ }\href
  {\doibase 10.1103/RevModPhys.81.109} {\bibfield  {journal} {\bibinfo
  {journal} {Rev. Mod. Phys.}\ }\textbf {\bibinfo {volume} {81}},\ \bibinfo
  {pages} {109--162} (\bibinfo {year} {2009})}\BibitemShut {NoStop}%
\bibitem [{\citenamefont {DiVincenzo}\ and\ \citenamefont
  {Mele}(1984)}]{DiVincenzo}%
  \BibitemOpen
  \bibfield  {author} {\bibinfo {author} {\bibfnamefont {D.~P.}\ \bibnamefont
  {DiVincenzo}}\ and\ \bibinfo {author} {\bibfnamefont {E.~J.}\ \bibnamefont
  {Mele}},\ }\bibfield  {title} {\enquote {\bibinfo {title} {Self-consistent
  effective-mass theory for intralayer screening in graphite intercalation
  compounds},}\ }\href {\doibase 10.1103/PhysRevB.29.1685} {\bibfield
  {journal} {\bibinfo  {journal} {Phys. Rev. B}\ }\textbf {\bibinfo {volume}
  {29}},\ \bibinfo {pages} {1685--1694} (\bibinfo {year} {1984})}\BibitemShut
  {NoStop}%
\bibitem [{\citenamefont {Semenoff}(1984)}]{Semenoff}%
  \BibitemOpen
  \bibfield  {author} {\bibinfo {author} {\bibfnamefont {Gordon~W.}\
  \bibnamefont {Semenoff}},\ }\bibfield  {title} {\enquote {\bibinfo {title}
  {Condensed-matter simulation of a three-dimensional anomaly},}\ }\href
  {\doibase 10.1103/PhysRevLett.53.2449} {\bibfield  {journal} {\bibinfo
  {journal} {Phys. Rev. Lett.}\ }\textbf {\bibinfo {volume} {53}},\ \bibinfo
  {pages} {2449--2452} (\bibinfo {year} {1984})}\BibitemShut {NoStop}%
\bibitem [{\citenamefont {Kitaev}(2006)}]{Kitaev}%
  \BibitemOpen
  \bibfield  {author} {\bibinfo {author} {\bibfnamefont {Alexei}\ \bibnamefont
  {Kitaev}},\ }\bibfield  {title} {\enquote {\bibinfo {title} {Anyons in an
  exactly solved model and beyond},}\ }\href {\doibase
  https://doi.org/10.1016/j.aop.2005.10.005} {\bibfield  {journal} {\bibinfo
  {journal} {Annals of Physics}\ }\textbf {\bibinfo {volume} {321}},\ \bibinfo
  {pages} {2 -- 111} (\bibinfo {year} {2006})},\ \bibinfo {note} {january
  Special Issue}\BibitemShut {NoStop}%
\bibitem [{\citenamefont {Farjami}\ \emph {et~al.}(2020)\citenamefont
  {Farjami}, \citenamefont {Horner}, \citenamefont {Self}, \citenamefont
  {Papi\ifmmode~\acute{c}\else \'{c}\fi{}},\ and\ \citenamefont
  {Pachos}}]{Farjami}%
  \BibitemOpen
  \bibfield  {author} {\bibinfo {author} {\bibfnamefont {Ashk}\ \bibnamefont
  {Farjami}}, \bibinfo {author} {\bibfnamefont {Matthew~D.}\ \bibnamefont
  {Horner}}, \bibinfo {author} {\bibfnamefont {Chris~N.}\ \bibnamefont {Self}},
  \bibinfo {author} {\bibfnamefont {Zlatko}\ \bibnamefont
  {Papi\ifmmode~\acute{c}\else \'{c}\fi{}}}, \ and\ \bibinfo {author}
  {\bibfnamefont {Jiannis~K.}\ \bibnamefont {Pachos}},\ }\bibfield  {title}
  {\enquote {\bibinfo {title} {Geometric description of the kitaev honeycomb
  lattice model},}\ }\href {\doibase 10.1103/PhysRevB.101.245116} {\bibfield
  {journal} {\bibinfo  {journal} {Phys. Rev. B}\ }\textbf {\bibinfo {volume}
  {101}},\ \bibinfo {pages} {245116} (\bibinfo {year} {2020})}\BibitemShut
  {NoStop}%
\bibitem [{\citenamefont {Kitaev}(2003)}]{Kitaev2}%
  \BibitemOpen
  \bibfield  {author} {\bibinfo {author} {\bibfnamefont {A.Yu.}\ \bibnamefont
  {Kitaev}},\ }\bibfield  {title} {\enquote {\bibinfo {title} {Fault-tolerant
  quantum computation by anyons},}\ }\href {\doibase
  https://doi.org/10.1016/S0003-4916(02)00018-0} {\bibfield  {journal}
  {\bibinfo  {journal} {Annals of Physics}\ }\textbf {\bibinfo {volume}
  {303}},\ \bibinfo {pages} {2 -- 30} (\bibinfo {year} {2003})}\BibitemShut
  {NoStop}%
\bibitem [{\citenamefont {Pachos}(2007)}]{Pachos}%
  \BibitemOpen
  \bibfield  {author} {\bibinfo {author} {\bibfnamefont {Jiannis~K.}\
  \bibnamefont {Pachos}},\ }\bibfield  {title} {\enquote {\bibinfo {title} {The
  wavefunction of an anyon},}\ }\href {\doibase
  https://doi.org/10.1016/j.aop.2006.05.007} {\bibfield  {journal} {\bibinfo
  {journal} {Annals of Physics}\ }\textbf {\bibinfo {volume} {322}},\ \bibinfo
  {pages} {1254 -- 1264} (\bibinfo {year} {2007})}\BibitemShut {NoStop}%
\bibitem [{\citenamefont {Schmidt}\ \emph {et~al.}(2008)\citenamefont
  {Schmidt}, \citenamefont {Dusuel},\ and\ \citenamefont {Vidal}}]{Vidal3}%
  \BibitemOpen
  \bibfield  {author} {\bibinfo {author} {\bibfnamefont {Kai~Phillip}\
  \bibnamefont {Schmidt}}, \bibinfo {author} {\bibfnamefont {S\'ebastien}\
  \bibnamefont {Dusuel}}, \ and\ \bibinfo {author} {\bibfnamefont {Julien}\
  \bibnamefont {Vidal}},\ }\bibfield  {title} {\enquote {\bibinfo {title}
  {Emergent fermions and anyons in the kitaev model},}\ }\href {\doibase
  10.1103/PhysRevLett.100.057208} {\bibfield  {journal} {\bibinfo  {journal}
  {Phys. Rev. Lett.}\ }\textbf {\bibinfo {volume} {100}},\ \bibinfo {pages}
  {057208} (\bibinfo {year} {2008})}\BibitemShut {NoStop}%
\bibitem [{\citenamefont {Lahtinen}\ \emph {et~al.}(2008)\citenamefont
  {Lahtinen}, \citenamefont {Kells}, \citenamefont {Carollo}, \citenamefont
  {Stitt}, \citenamefont {Vala},\ and\ \citenamefont {Pachos}}]{Ville1}%
  \BibitemOpen
  \bibfield  {author} {\bibinfo {author} {\bibfnamefont {V.}~\bibnamefont
  {Lahtinen}}, \bibinfo {author} {\bibfnamefont {G.}~\bibnamefont {Kells}},
  \bibinfo {author} {\bibfnamefont {A.}~\bibnamefont {Carollo}}, \bibinfo
  {author} {\bibfnamefont {T.}~\bibnamefont {Stitt}}, \bibinfo {author}
  {\bibfnamefont {J.}~\bibnamefont {Vala}}, \ and\ \bibinfo {author}
  {\bibfnamefont {Jiannis~K.}\ \bibnamefont {Pachos}},\ }\bibfield  {title}
  {\enquote {\bibinfo {title} {Spectrum of the non-abelian phase in kitaev's
  honeycomb lattice model},}\ }\href {\doibase 10.1016/j.aop.2007.12.009}
  {\bibfield  {journal} {\bibinfo  {journal} {Annals of Physics}\ }\textbf
  {\bibinfo {volume} {323}},\ \bibinfo {pages} {2286} (\bibinfo {year}
  {2008})}\BibitemShut {NoStop}%
\bibitem [{\citenamefont {Self}\ \emph {et~al.}(2019)\citenamefont {Self},
  \citenamefont {Knolle}, \citenamefont {Iblisdir},\ and\ \citenamefont
  {Pachos}}]{Self}%
  \BibitemOpen
  \bibfield  {author} {\bibinfo {author} {\bibfnamefont {Chris~N.}\
  \bibnamefont {Self}}, \bibinfo {author} {\bibfnamefont {Johannes}\
  \bibnamefont {Knolle}}, \bibinfo {author} {\bibfnamefont {Sofyan}\
  \bibnamefont {Iblisdir}}, \ and\ \bibinfo {author} {\bibfnamefont
  {Jiannis~K.}\ \bibnamefont {Pachos}},\ }\bibfield  {title} {\enquote
  {\bibinfo {title} {Thermally induced metallic phase in a gapped quantum spin
  liquid: Monte carlo study of the kitaev model with parity projection},}\
  }\href {\doibase 10.1103/PhysRevB.99.045142} {\bibfield  {journal} {\bibinfo
  {journal} {Phys. Rev. B}\ }\textbf {\bibinfo {volume} {99}},\ \bibinfo
  {pages} {045142} (\bibinfo {year} {2019})}\BibitemShut {NoStop}%
\bibitem [{\citenamefont {Otten}\ \emph {et~al.}(2019)\citenamefont {Otten},
  \citenamefont {Roy},\ and\ \citenamefont {Hassler}}]{Otten}%
  \BibitemOpen
  \bibfield  {author} {\bibinfo {author} {\bibfnamefont {Daniel}\ \bibnamefont
  {Otten}}, \bibinfo {author} {\bibfnamefont {Ananda}\ \bibnamefont {Roy}}, \
  and\ \bibinfo {author} {\bibfnamefont {Fabian}\ \bibnamefont {Hassler}},\
  }\bibfield  {title} {\enquote {\bibinfo {title} {Dynamical structure factor
  in the non-abelian phase of the kitaev honeycomb model in the presence of
  quenched disorder},}\ }\href {\doibase 10.1103/PhysRevB.99.035137} {\bibfield
   {journal} {\bibinfo  {journal} {Phys. Rev. B}\ }\textbf {\bibinfo {volume}
  {99}},\ \bibinfo {pages} {035137} (\bibinfo {year} {2019})}\BibitemShut
  {NoStop}%
\bibitem [{\citenamefont {Dusuel}\ \emph {et~al.}(2008)\citenamefont {Dusuel},
  \citenamefont {Schmidt},\ and\ \citenamefont {Vidal}}]{Vidal4}%
  \BibitemOpen
  \bibfield  {author} {\bibinfo {author} {\bibfnamefont {S\'ebastien}\
  \bibnamefont {Dusuel}}, \bibinfo {author} {\bibfnamefont {Kai~Phillip}\
  \bibnamefont {Schmidt}}, \ and\ \bibinfo {author} {\bibfnamefont {Julien}\
  \bibnamefont {Vidal}},\ }\bibfield  {title} {\enquote {\bibinfo {title}
  {Creation and manipulation of anyons in the kitaev model},}\ }\href {\doibase
  10.1103/PhysRevLett.100.177204} {\bibfield  {journal} {\bibinfo  {journal}
  {Phys. Rev. Lett.}\ }\textbf {\bibinfo {volume} {100}},\ \bibinfo {pages}
  {177204} (\bibinfo {year} {2008})}\BibitemShut {NoStop}%
\bibitem [{\citenamefont {Jackiw}\ and\ \citenamefont {Rossi}(1981)}]{Jackiw}%
  \BibitemOpen
  \bibfield  {author} {\bibinfo {author} {\bibfnamefont {R.}~\bibnamefont
  {Jackiw}}\ and\ \bibinfo {author} {\bibfnamefont {P.}~\bibnamefont {Rossi}},\
  }\bibfield  {title} {\enquote {\bibinfo {title} {Zero modes of the
  vortex-fermion system},}\ }\href {\doibase
  https://doi.org/10.1016/0550-3213(81)90044-4} {\bibfield  {journal} {\bibinfo
   {journal} {Nuclear Physics B}\ }\textbf {\bibinfo {volume} {190}},\ \bibinfo
  {pages} {681 -- 691} (\bibinfo {year} {1981})}\BibitemShut {NoStop}%
\bibitem [{\citenamefont {Hou}\ \emph {et~al.}(2007)\citenamefont {Hou},
  \citenamefont {Chamon},\ and\ \citenamefont {Mudry}}]{Hou}%
  \BibitemOpen
  \bibfield  {author} {\bibinfo {author} {\bibfnamefont {Chang-Yu}\
  \bibnamefont {Hou}}, \bibinfo {author} {\bibfnamefont {Claudio}\ \bibnamefont
  {Chamon}}, \ and\ \bibinfo {author} {\bibfnamefont {Christopher}\
  \bibnamefont {Mudry}},\ }\bibfield  {title} {\enquote {\bibinfo {title}
  {Electron fractionalization in two-dimensional graphenelike structures},}\
  }\href {\doibase 10.1103/PhysRevLett.98.186809} {\bibfield  {journal}
  {\bibinfo  {journal} {Phys. Rev. Lett.}\ }\textbf {\bibinfo {volume} {98}},\
  \bibinfo {pages} {186809} (\bibinfo {year} {2007})}\BibitemShut {NoStop}%
\bibitem [{\citenamefont {Lahtinen}(2011)}]{Ville4}%
  \BibitemOpen
  \bibfield  {author} {\bibinfo {author} {\bibfnamefont {Ville}\ \bibnamefont
  {Lahtinen}},\ }\bibfield  {title} {\enquote {\bibinfo {title} {Interacting
  non-abelian anyons as majorana fermions in the honeycomb lattice model},}\
  }\href {\doibase 10.1088/1367-2630/13/7/075009} {\bibfield  {journal}
  {\bibinfo  {journal} {New Journal of Physics}\ }\textbf {\bibinfo {volume}
  {13}},\ \bibinfo {pages} {075009} (\bibinfo {year} {2011})}\BibitemShut
  {NoStop}%
\bibitem [{\citenamefont {Lahtinen}\ and\ \citenamefont
  {Pachos}(2009)}]{Ville5}%
  \BibitemOpen
  \bibfield  {author} {\bibinfo {author} {\bibfnamefont {Ville}\ \bibnamefont
  {Lahtinen}}\ and\ \bibinfo {author} {\bibfnamefont {Jiannis~K}\ \bibnamefont
  {Pachos}},\ }\bibfield  {title} {\enquote {\bibinfo {title} {Non-abelian
  statistics as a berry phase in exactly solvable models},}\ }\href {\doibase
  10.1088/1367-2630/11/9/093027} {\bibfield  {journal} {\bibinfo  {journal}
  {New Journal of Physics}\ }\textbf {\bibinfo {volume} {11}},\ \bibinfo
  {pages} {093027} (\bibinfo {year} {2009})}\BibitemShut {NoStop}%
\bibitem [{\citenamefont {Chaloupka}\ \emph {et~al.}(2010)\citenamefont
  {Chaloupka}, \citenamefont {Jackeli},\ and\ \citenamefont
  {Khaliullin}}]{Chaloupka}%
  \BibitemOpen
  \bibfield  {author} {\bibinfo {author} {\bibfnamefont {Jiri}\ \bibnamefont
  {Chaloupka}}, \bibinfo {author} {\bibfnamefont {George}\ \bibnamefont
  {Jackeli}}, \ and\ \bibinfo {author} {\bibfnamefont {Giniyat}\ \bibnamefont
  {Khaliullin}},\ }\bibfield  {title} {\enquote {\bibinfo {title}
  {Kitaev-heisenberg model on a honeycomb lattice: Possible exotic phases in
  iridium oxides ${A}_{2}{\mathrm{iro}}_{3}$},}\ }\href {\doibase
  10.1103/PhysRevLett.105.027204} {\bibfield  {journal} {\bibinfo  {journal}
  {Phys. Rev. Lett.}\ }\textbf {\bibinfo {volume} {105}},\ \bibinfo {pages}
  {027204} (\bibinfo {year} {2010})}\BibitemShut {NoStop}%
\bibitem [{\citenamefont {Choi}\ \emph {et~al.}(2012)\citenamefont {Choi},
  \citenamefont {Coldea}, \citenamefont {Kolmogorov}, \citenamefont
  {Lancaster}, \citenamefont {Mazin}, \citenamefont {Blundell}, \citenamefont
  {Radaelli}, \citenamefont {Singh}, \citenamefont {Gegenwart}, \citenamefont
  {Choi}, \citenamefont {Cheong}, \citenamefont {Baker}, \citenamefont
  {Stock},\ and\ \citenamefont {Taylor}}]{Choi}%
  \BibitemOpen
  \bibfield  {author} {\bibinfo {author} {\bibfnamefont {S.~K.}\ \bibnamefont
  {Choi}}, \bibinfo {author} {\bibfnamefont {R.}~\bibnamefont {Coldea}},
  \bibinfo {author} {\bibfnamefont {A.~N.}\ \bibnamefont {Kolmogorov}},
  \bibinfo {author} {\bibfnamefont {T.}~\bibnamefont {Lancaster}}, \bibinfo
  {author} {\bibfnamefont {I.~I.}\ \bibnamefont {Mazin}}, \bibinfo {author}
  {\bibfnamefont {S.~J.}\ \bibnamefont {Blundell}}, \bibinfo {author}
  {\bibfnamefont {P.~G.}\ \bibnamefont {Radaelli}}, \bibinfo {author}
  {\bibfnamefont {Yogesh}\ \bibnamefont {Singh}}, \bibinfo {author}
  {\bibfnamefont {P.}~\bibnamefont {Gegenwart}}, \bibinfo {author}
  {\bibfnamefont {K.~R.}\ \bibnamefont {Choi}}, \bibinfo {author}
  {\bibfnamefont {S.-W.}\ \bibnamefont {Cheong}}, \bibinfo {author}
  {\bibfnamefont {P.~J.}\ \bibnamefont {Baker}}, \bibinfo {author}
  {\bibfnamefont {C.}~\bibnamefont {Stock}}, \ and\ \bibinfo {author}
  {\bibfnamefont {J.}~\bibnamefont {Taylor}},\ }\bibfield  {title} {\enquote
  {\bibinfo {title} {Spin waves and revised crystal structure of honeycomb
  iridate ${\mathrm{na}}_{2}{\mathrm{iro}}_{3}$},}\ }\href {\doibase
  10.1103/PhysRevLett.108.127204} {\bibfield  {journal} {\bibinfo  {journal}
  {Phys. Rev. Lett.}\ }\textbf {\bibinfo {volume} {108}},\ \bibinfo {pages}
  {127204} (\bibinfo {year} {2012})}\BibitemShut {NoStop}%
\bibitem [{\citenamefont {Jackeli}\ and\ \citenamefont
  {Khaliullin}(2009)}]{Jackeli}%
  \BibitemOpen
  \bibfield  {author} {\bibinfo {author} {\bibfnamefont {G.}~\bibnamefont
  {Jackeli}}\ and\ \bibinfo {author} {\bibfnamefont {G.}~\bibnamefont
  {Khaliullin}},\ }\bibfield  {title} {\enquote {\bibinfo {title} {Mott
  insulators in the strong spin-orbit coupling limit: From heisenberg to a
  quantum compass and kitaev models},}\ }\href {\doibase
  10.1103/PhysRevLett.102.017205} {\bibfield  {journal} {\bibinfo  {journal}
  {Phys. Rev. Lett.}\ }\textbf {\bibinfo {volume} {102}},\ \bibinfo {pages}
  {017205} (\bibinfo {year} {2009})}\BibitemShut {NoStop}%
\bibitem [{\citenamefont {Banerjee}\ \emph {et~al.}(2016)\citenamefont
  {Banerjee}, \citenamefont {Bridges}, \citenamefont {Yan}, \citenamefont
  {Aczel}, \citenamefont {Li}, \citenamefont {Stone}, \citenamefont {Granroth},
  \citenamefont {Lumsden}, \citenamefont {Yiu}, \citenamefont {Knolle},
  \citenamefont {Bhattacharjee}, \citenamefont {Kovrizhin}, \citenamefont
  {Moessner}, \citenamefont {Tennant}, \citenamefont {Mandrus},\ and\
  \citenamefont {Nagler}}]{Banerjee}%
  \BibitemOpen
  \bibfield  {author} {\bibinfo {author} {\bibfnamefont {A.}~\bibnamefont
  {Banerjee}}, \bibinfo {author} {\bibfnamefont {C.}~\bibnamefont {Bridges}},
  \bibinfo {author} {\bibfnamefont {J.-Q.}\ \bibnamefont {Yan}}, \bibinfo
  {author} {\bibfnamefont {A.A.}\ \bibnamefont {Aczel}}, \bibinfo {author}
  {\bibfnamefont {L.}~\bibnamefont {Li}}, \bibinfo {author} {\bibfnamefont
  {B.}~\bibnamefont {Stone}}, \bibinfo {author} {\bibfnamefont {G.E.}\
  \bibnamefont {Granroth}}, \bibinfo {author} {\bibfnamefont {M.D.}\
  \bibnamefont {Lumsden}}, \bibinfo {author} {\bibfnamefont {Y.}~\bibnamefont
  {Yiu}}, \bibinfo {author} {\bibfnamefont {J.}~\bibnamefont {Knolle}},
  \bibinfo {author} {\bibfnamefont {S.}~\bibnamefont {Bhattacharjee}}, \bibinfo
  {author} {\bibfnamefont {D.L.}\ \bibnamefont {Kovrizhin}}, \bibinfo {author}
  {\bibfnamefont {R.}~\bibnamefont {Moessner}}, \bibinfo {author}
  {\bibfnamefont {D.A.}\ \bibnamefont {Tennant}}, \bibinfo {author}
  {\bibfnamefont {D.G.}\ \bibnamefont {Mandrus}}, \ and\ \bibinfo {author}
  {\bibfnamefont {S.E.}\ \bibnamefont {Nagler}},\ }\bibfield  {title} {\enquote
  {\bibinfo {title} {Proximate kitaev quantum spin liquid behaviour in a
  honeycomb magnet},}\ }\href {\doibase 10.1038/nmat4604} {\bibfield  {journal}
  {\bibinfo  {journal} {Nature Materials}\ }\textbf {\bibinfo {volume} {15}},\
  \bibinfo {pages} {733--740} (\bibinfo {year} {2016})}\BibitemShut {NoStop}%
\bibitem [{\citenamefont {Das~Sarma}\ and\ \citenamefont
  {Nayak}(2015)}]{DasSarma}%
  \BibitemOpen
  \bibfield  {author} {\bibinfo {author} {\bibfnamefont {M.}~\bibnamefont
  {Das~Sarma}, \bibfnamefont {S.~Freedman}}\ and\ \bibinfo {author}
  {\bibfnamefont {C.}~\bibnamefont {Nayak}},\ }\bibfield  {title} {\enquote
  {\bibinfo {title} {Majorana zero modes and topological quantum
  computation},}\ }\href {\doibase 10.1038/npjqi.2015.1} {\bibfield  {journal}
  {\bibinfo  {journal} {npj Quantum Information}\ }\textbf {\bibinfo {volume}
  {1}},\ \bibinfo {pages} {15001} (\bibinfo {year} {2015})}\BibitemShut
  {NoStop}%
\bibitem [{\citenamefont {Nayak}\ \emph {et~al.}(2008)\citenamefont {Nayak},
  \citenamefont {Simon}, \citenamefont {Stern}, \citenamefont {Freedman},\ and\
  \citenamefont {Das~Sarma}}]{Nayak}%
  \BibitemOpen
  \bibfield  {author} {\bibinfo {author} {\bibfnamefont {Chetan}\ \bibnamefont
  {Nayak}}, \bibinfo {author} {\bibfnamefont {Steven~H.}\ \bibnamefont
  {Simon}}, \bibinfo {author} {\bibfnamefont {Ady}\ \bibnamefont {Stern}},
  \bibinfo {author} {\bibfnamefont {Michael}\ \bibnamefont {Freedman}}, \ and\
  \bibinfo {author} {\bibfnamefont {Sankar}\ \bibnamefont {Das~Sarma}},\
  }\bibfield  {title} {\enquote {\bibinfo {title} {Non-abelian anyons and
  topological quantum computation},}\ }\href {\doibase
  10.1103/RevModPhys.80.1083} {\bibfield  {journal} {\bibinfo  {journal} {Rev.
  Mod. Phys.}\ }\textbf {\bibinfo {volume} {80}},\ \bibinfo {pages}
  {1083--1159} (\bibinfo {year} {2008})}\BibitemShut {NoStop}%
\bibitem [{\citenamefont {Lampen-Kelley}\ \emph {et~al.}(2017)\citenamefont
  {Lampen-Kelley}, \citenamefont {Banerjee}, \citenamefont {Aczel},
  \citenamefont {Cao}, \citenamefont {Stone}, \citenamefont {Bridges},
  \citenamefont {Yan}, \citenamefont {Nagler},\ and\ \citenamefont
  {Mandrus}}]{Lampen-Kelley}%
  \BibitemOpen
  \bibfield  {author} {\bibinfo {author} {\bibfnamefont {P.}~\bibnamefont
  {Lampen-Kelley}}, \bibinfo {author} {\bibfnamefont {A.}~\bibnamefont
  {Banerjee}}, \bibinfo {author} {\bibfnamefont {A.~A.}\ \bibnamefont {Aczel}},
  \bibinfo {author} {\bibfnamefont {H.~B.}\ \bibnamefont {Cao}}, \bibinfo
  {author} {\bibfnamefont {M.~B.}\ \bibnamefont {Stone}}, \bibinfo {author}
  {\bibfnamefont {C.~A.}\ \bibnamefont {Bridges}}, \bibinfo {author}
  {\bibfnamefont {J.-Q.}\ \bibnamefont {Yan}}, \bibinfo {author} {\bibfnamefont
  {S.~E.}\ \bibnamefont {Nagler}}, \ and\ \bibinfo {author} {\bibfnamefont
  {D.}~\bibnamefont {Mandrus}},\ }\bibfield  {title} {\enquote {\bibinfo
  {title} {Destabilization of magnetic order in a dilute kitaev spin liquid
  candidate},}\ }\href {\doibase 10.1103/PhysRevLett.119.237203} {\bibfield
  {journal} {\bibinfo  {journal} {Phys. Rev. Lett.}\ }\textbf {\bibinfo
  {volume} {119}},\ \bibinfo {pages} {237203} (\bibinfo {year}
  {2017})}\BibitemShut {NoStop}%
\bibitem [{\citenamefont {Do}\ \emph {et~al.}(2020)\citenamefont {Do},
  \citenamefont {Lee}, \citenamefont {Kihara}, \citenamefont {Choi},
  \citenamefont {Yoon}, \citenamefont {Kim}, \citenamefont {Cheong},
  \citenamefont {Chen}, \citenamefont {Chou}, \citenamefont {Nojiri},\ and\
  \citenamefont {Choi}}]{Do}%
  \BibitemOpen
  \bibfield  {author} {\bibinfo {author} {\bibfnamefont {Seung-Hwan}\
  \bibnamefont {Do}}, \bibinfo {author} {\bibfnamefont {C.~H.}\ \bibnamefont
  {Lee}}, \bibinfo {author} {\bibfnamefont {T.}~\bibnamefont {Kihara}},
  \bibinfo {author} {\bibfnamefont {Y.~S.}\ \bibnamefont {Choi}}, \bibinfo
  {author} {\bibfnamefont {Sungwon}\ \bibnamefont {Yoon}}, \bibinfo {author}
  {\bibfnamefont {Kangwon}\ \bibnamefont {Kim}}, \bibinfo {author}
  {\bibfnamefont {Hyeonsik}\ \bibnamefont {Cheong}}, \bibinfo {author}
  {\bibfnamefont {Wei-Tin}\ \bibnamefont {Chen}}, \bibinfo {author}
  {\bibfnamefont {Fangcheng}\ \bibnamefont {Chou}}, \bibinfo {author}
  {\bibfnamefont {H.}~\bibnamefont {Nojiri}}, \ and\ \bibinfo {author}
  {\bibfnamefont {Kwang-Yong}\ \bibnamefont {Choi}},\ }\bibfield  {title}
  {\enquote {\bibinfo {title} {Randomly hopping majorana fermions in the
  diluted kitaev system
  $\ensuremath{\alpha}$-${\mathrm{ru}}_{0.8}{\mathrm{ir}}_{0.2}{\mathrm{cl}}_{3}$},}\
  }\href {\doibase 10.1103/PhysRevLett.124.047204} {\bibfield  {journal}
  {\bibinfo  {journal} {Phys. Rev. Lett.}\ }\textbf {\bibinfo {volume} {124}},\
  \bibinfo {pages} {047204} (\bibinfo {year} {2020})}\BibitemShut {NoStop}%
\bibitem [{\citenamefont {Winter}\ \emph {et~al.}(2017)\citenamefont {Winter},
  \citenamefont {Tsirlin}, \citenamefont {Daghofer}, \citenamefont {van~den
  Brink}, \citenamefont {Singh}, \citenamefont {Gegenwart},\ and\ \citenamefont
  {Valent{\'{\i}}}}]{Winter_2017}%
  \BibitemOpen
  \bibfield  {author} {\bibinfo {author} {\bibfnamefont {Stephen~M}\
  \bibnamefont {Winter}}, \bibinfo {author} {\bibfnamefont {Alexander~A}\
  \bibnamefont {Tsirlin}}, \bibinfo {author} {\bibfnamefont {Maria}\
  \bibnamefont {Daghofer}}, \bibinfo {author} {\bibfnamefont {Jeroen}\
  \bibnamefont {van~den Brink}}, \bibinfo {author} {\bibfnamefont {Yogesh}\
  \bibnamefont {Singh}}, \bibinfo {author} {\bibfnamefont {Philipp}\
  \bibnamefont {Gegenwart}}, \ and\ \bibinfo {author} {\bibfnamefont {Roser}\
  \bibnamefont {Valent{\'{\i}}}},\ }\bibfield  {title} {\enquote {\bibinfo
  {title} {Models and materials for generalized kitaev magnetism},}\ }\href
  {\doibase 10.1088/1361-648x/aa8cf5} {\bibfield  {journal} {\bibinfo
  {journal} {Journal of Physics: Condensed Matter}\ }\textbf {\bibinfo {volume}
  {29}},\ \bibinfo {pages} {493002} (\bibinfo {year} {2017})}\BibitemShut
  {NoStop}%
\bibitem [{\citenamefont {Kim}\ and\ \citenamefont {Kee}(2016)}]{Kim}%
  \BibitemOpen
  \bibfield  {author} {\bibinfo {author} {\bibfnamefont {Heung-Sik}\
  \bibnamefont {Kim}}\ and\ \bibinfo {author} {\bibfnamefont {Hae-Young}\
  \bibnamefont {Kee}},\ }\bibfield  {title} {\enquote {\bibinfo {title}
  {Crystal structure and magnetism in
  $\ensuremath{\alpha}\ensuremath{-}{\mathrm{rucl}}_{3}$: An ab initio
  study},}\ }\href {\doibase 10.1103/PhysRevB.93.155143} {\bibfield  {journal}
  {\bibinfo  {journal} {Phys. Rev. B}\ }\textbf {\bibinfo {volume} {93}},\
  \bibinfo {pages} {155143} (\bibinfo {year} {2016})}\BibitemShut {NoStop}%
\bibitem [{\citenamefont {Petrova}\ \emph {et~al.}(2013)\citenamefont
  {Petrova}, \citenamefont {Mellado},\ and\ \citenamefont
  {Tchernyshyov}}]{Petrova}%
  \BibitemOpen
  \bibfield  {author} {\bibinfo {author} {\bibfnamefont {Olga}\ \bibnamefont
  {Petrova}}, \bibinfo {author} {\bibfnamefont {Paula}\ \bibnamefont
  {Mellado}}, \ and\ \bibinfo {author} {\bibfnamefont {Oleg}\ \bibnamefont
  {Tchernyshyov}},\ }\bibfield  {title} {\enquote {\bibinfo {title} {Unpaired
  majorana modes in the gapped phase of kitaev's honeycomb model},}\ }\href
  {\doibase 10.1103/PhysRevB.88.140405} {\bibfield  {journal} {\bibinfo
  {journal} {Phys. Rev. B}\ }\textbf {\bibinfo {volume} {88}},\ \bibinfo
  {pages} {140405} (\bibinfo {year} {2013})}\BibitemShut {NoStop}%
\bibitem [{\citenamefont {Petrova}\ \emph {et~al.}(2014)\citenamefont
  {Petrova}, \citenamefont {Mellado},\ and\ \citenamefont
  {Tchernyshyov}}]{Petrova2}%
  \BibitemOpen
  \bibfield  {author} {\bibinfo {author} {\bibfnamefont {Olga}\ \bibnamefont
  {Petrova}}, \bibinfo {author} {\bibfnamefont {Paula}\ \bibnamefont
  {Mellado}}, \ and\ \bibinfo {author} {\bibfnamefont {Oleg}\ \bibnamefont
  {Tchernyshyov}},\ }\bibfield  {title} {\enquote {\bibinfo {title} {Unpaired
  majorana modes on dislocations and string defects in kitaev's honeycomb
  model},}\ }\href {\doibase 10.1103/PhysRevB.90.134404} {\bibfield  {journal}
  {\bibinfo  {journal} {Phys. Rev. B}\ }\textbf {\bibinfo {volume} {90}},\
  \bibinfo {pages} {134404} (\bibinfo {year} {2014})}\BibitemShut {NoStop}%
\bibitem [{\citenamefont {Willans}\ \emph {et~al.}(2011)\citenamefont
  {Willans}, \citenamefont {Chalker},\ and\ \citenamefont
  {Moessner}}]{Willans}%
  \BibitemOpen
  \bibfield  {author} {\bibinfo {author} {\bibfnamefont {A.~J.}\ \bibnamefont
  {Willans}}, \bibinfo {author} {\bibfnamefont {J.~T.}\ \bibnamefont
  {Chalker}}, \ and\ \bibinfo {author} {\bibfnamefont {R.}~\bibnamefont
  {Moessner}},\ }\bibfield  {title} {\enquote {\bibinfo {title} {Site dilution
  in the kitaev honeycomb model},}\ }\href {\doibase
  10.1103/PhysRevB.84.115146} {\bibfield  {journal} {\bibinfo  {journal} {Phys.
  Rev. B}\ }\textbf {\bibinfo {volume} {84}},\ \bibinfo {pages} {115146}
  (\bibinfo {year} {2011})}\BibitemShut {NoStop}%
\bibitem [{\citenamefont {Golan}\ and\ \citenamefont {Stern}(2018)}]{Golan}%
  \BibitemOpen
  \bibfield  {author} {\bibinfo {author} {\bibfnamefont {Omri}\ \bibnamefont
  {Golan}}\ and\ \bibinfo {author} {\bibfnamefont {Ady}\ \bibnamefont
  {Stern}},\ }\bibfield  {title} {\enquote {\bibinfo {title} {Probing
  topological superconductors with emergent gravity},}\ }\href {\doibase
  10.1103/PhysRevB.98.064503} {\bibfield  {journal} {\bibinfo  {journal} {Phys.
  Rev. B}\ }\textbf {\bibinfo {volume} {98}},\ \bibinfo {pages} {064503}
  (\bibinfo {year} {2018})}\BibitemShut {NoStop}%
\bibitem [{\citenamefont {Nissinen}(2020)}]{Nissinen_2020}%
  \BibitemOpen
  \bibfield  {author} {\bibinfo {author} {\bibfnamefont {Jaakko}\ \bibnamefont
  {Nissinen}},\ }\bibfield  {title} {\enquote {\bibinfo {title} {Emergent
  spacetime and gravitational nieh-yan anomaly in chiral $p+ip$ weyl
  superfluids and superconductors},}\ }\href {\doibase
  10.1103/PhysRevLett.124.117002} {\bibfield  {journal} {\bibinfo  {journal}
  {Phys. Rev. Lett.}\ }\textbf {\bibinfo {volume} {124}},\ \bibinfo {pages}
  {117002} (\bibinfo {year} {2020})}\BibitemShut {NoStop}%
\bibitem [{\citenamefont {Hehl}\ and\ \citenamefont {Datta}(1971)}]{Hehl}%
  \BibitemOpen
  \bibfield  {author} {\bibinfo {author} {\bibfnamefont {F.W.}\ \bibnamefont
  {Hehl}}\ and\ \bibinfo {author} {\bibfnamefont {B.K.}\ \bibnamefont
  {Datta}},\ }\bibfield  {title} {\enquote {\bibinfo {title} {Nonlinear spinor
  equation and asymmetric connection in general relativity},}\ }\href {\doibase
  10.1063/1.1665738} {\bibfield  {journal} {\bibinfo  {journal} {Journal of
  Mathematical Physics}\ }\textbf {\bibinfo {volume} {12}},\ \bibinfo {pages}
  {1334} (\bibinfo {year} {1971})}\BibitemShut {NoStop}%
\bibitem [{\citenamefont {Maggiore}(2005)}]{Maggiore}%
  \BibitemOpen
  \bibfield  {author} {\bibinfo {author} {\bibfnamefont {Michele}\ \bibnamefont
  {Maggiore}},\ }\href@noop {} {\emph {\bibinfo {title} {A Modern Introduction
  to Quantum Field Theory}}}\ (\bibinfo  {publisher} {Oxford University
  Press},\ \bibinfo {address} {Oxford},\ \bibinfo {year} {2005})\BibitemShut
  {NoStop}%
\bibitem [{\citenamefont {Creutz}(1994)}]{Creutz}%
  \BibitemOpen
  \bibfield  {author} {\bibinfo {author} {\bibfnamefont {Michael}\ \bibnamefont
  {Creutz}},\ }\href@noop {} {\enquote {\bibinfo {title} {Chiral symmetry and
  lattice gauge theory},}\ } (\bibinfo {year} {1994}),\ \Eprint
  {http://arxiv.org/abs/hep-lat/9410008} {arXiv:hep-lat/9410008 [hep-lat]}
  \BibitemShut {NoStop}%
\bibitem [{\citenamefont {Laurila}\ and\ \citenamefont
  {Nissinen}(2020)}]{laurila2020torsional}%
  \BibitemOpen
  \bibfield  {author} {\bibinfo {author} {\bibfnamefont {Sara}\ \bibnamefont
  {Laurila}}\ and\ \bibinfo {author} {\bibfnamefont {Jaakko}\ \bibnamefont
  {Nissinen}},\ }\href@noop {} {\enquote {\bibinfo {title} {Torsional landau
  levels and geometric anomalies in condensed matter weyl systems},}\ }
  (\bibinfo {year} {2020}),\ \Eprint {http://arxiv.org/abs/2007.10682}
  {arXiv:2007.10682 [cond-mat.str-el]} \BibitemShut {NoStop}%
\bibitem [{\citenamefont {Cortijo}\ \emph {et~al.}(2015)\citenamefont
  {Cortijo}, \citenamefont {Ferreir\'os}, \citenamefont {Landsteiner},\ and\
  \citenamefont {Vozmediano}}]{Cortijo}%
  \BibitemOpen
  \bibfield  {author} {\bibinfo {author} {\bibfnamefont {Alberto}\ \bibnamefont
  {Cortijo}}, \bibinfo {author} {\bibfnamefont {Yago}\ \bibnamefont
  {Ferreir\'os}}, \bibinfo {author} {\bibfnamefont {Karl}\ \bibnamefont
  {Landsteiner}}, \ and\ \bibinfo {author} {\bibfnamefont {Mar\'{\i}a A.~H.}\
  \bibnamefont {Vozmediano}},\ }\bibfield  {title} {\enquote {\bibinfo {title}
  {Elastic gauge fields in weyl semimetals},}\ }\href {\doibase
  10.1103/PhysRevLett.115.177202} {\bibfield  {journal} {\bibinfo  {journal}
  {Phys. Rev. Lett.}\ }\textbf {\bibinfo {volume} {115}},\ \bibinfo {pages}
  {177202} (\bibinfo {year} {2015})}\BibitemShut {NoStop}%
\bibitem [{\citenamefont {Sumiyoshi}\ and\ \citenamefont
  {Fujimoto}(2016)}]{Sumiyoshi}%
  \BibitemOpen
  \bibfield  {author} {\bibinfo {author} {\bibfnamefont {Hiroaki}\ \bibnamefont
  {Sumiyoshi}}\ and\ \bibinfo {author} {\bibfnamefont {Satoshi}\ \bibnamefont
  {Fujimoto}},\ }\bibfield  {title} {\enquote {\bibinfo {title} {Torsional
  chiral magnetic effect in a weyl semimetal with a topological defect},}\
  }\href {\doibase 10.1103/PhysRevLett.116.166601} {\bibfield  {journal}
  {\bibinfo  {journal} {Phys. Rev. Lett.}\ }\textbf {\bibinfo {volume} {116}},\
  \bibinfo {pages} {166601} (\bibinfo {year} {2016})}\BibitemShut {NoStop}%
\bibitem [{\citenamefont {Landsteiner}(2016)}]{Landsteiner}%
  \BibitemOpen
  \bibfield  {author} {\bibinfo {author} {\bibfnamefont {K.}~\bibnamefont
  {Landsteiner}},\ }\bibfield  {title} {\enquote {\bibinfo {title} {Notes on
  anomaly induced transport},}\ }\href {\doibase 10.5506/aphyspolb.47.2617}
  {\bibfield  {journal} {\bibinfo  {journal} {Acta Physica Polonica B}\
  }\textbf {\bibinfo {volume} {47}},\ \bibinfo {pages} {2617} (\bibinfo {year}
  {2016})}\BibitemShut {NoStop}%
\bibitem [{\citenamefont {Grushin}\ \emph {et~al.}(2016)\citenamefont
  {Grushin}, \citenamefont {Venderbos}, \citenamefont {Vishwanath},\ and\
  \citenamefont {Ilan}}]{Grushin_2016}%
  \BibitemOpen
  \bibfield  {author} {\bibinfo {author} {\bibfnamefont {Adolfo~G.}\
  \bibnamefont {Grushin}}, \bibinfo {author} {\bibfnamefont {J\"orn W.~F.}\
  \bibnamefont {Venderbos}}, \bibinfo {author} {\bibfnamefont {Ashvin}\
  \bibnamefont {Vishwanath}}, \ and\ \bibinfo {author} {\bibfnamefont {Roni}\
  \bibnamefont {Ilan}},\ }\bibfield  {title} {\enquote {\bibinfo {title}
  {Inhomogeneous weyl and dirac semimetals: Transport in axial magnetic fields
  and fermi arc surface states from pseudo-landau levels},}\ }\href {\doibase
  10.1103/PhysRevX.6.041046} {\bibfield  {journal} {\bibinfo  {journal} {Phys.
  Rev. X}\ }\textbf {\bibinfo {volume} {6}},\ \bibinfo {pages} {041046}
  (\bibinfo {year} {2016})}\BibitemShut {NoStop}%
\bibitem [{\citenamefont {Pikulin}\ \emph {et~al.}(2016)\citenamefont
  {Pikulin}, \citenamefont {Chen},\ and\ \citenamefont {Franz}}]{Pikulin_2016}%
  \BibitemOpen
  \bibfield  {author} {\bibinfo {author} {\bibfnamefont {D.~I.}\ \bibnamefont
  {Pikulin}}, \bibinfo {author} {\bibfnamefont {Anffany}\ \bibnamefont {Chen}},
  \ and\ \bibinfo {author} {\bibfnamefont {M.}~\bibnamefont {Franz}},\
  }\bibfield  {title} {\enquote {\bibinfo {title} {Chiral anomaly from
  strain-induced gauge fields in dirac and weyl semimetals},}\ }\href {\doibase
  10.1103/PhysRevX.6.041021} {\bibfield  {journal} {\bibinfo  {journal} {Phys.
  Rev. X}\ }\textbf {\bibinfo {volume} {6}},\ \bibinfo {pages} {041021}
  (\bibinfo {year} {2016})}\BibitemShut {NoStop}%
\bibitem [{\citenamefont {Gorbar}\ \emph {et~al.}(2017)\citenamefont {Gorbar},
  \citenamefont {Miransky}, \citenamefont {Shovkovy},\ and\ \citenamefont
  {Sukhachov}}]{Gorbar_2017}%
  \BibitemOpen
  \bibfield  {author} {\bibinfo {author} {\bibfnamefont {E.~V.}\ \bibnamefont
  {Gorbar}}, \bibinfo {author} {\bibfnamefont {V.~A.}\ \bibnamefont
  {Miransky}}, \bibinfo {author} {\bibfnamefont {I.~A.}\ \bibnamefont
  {Shovkovy}}, \ and\ \bibinfo {author} {\bibfnamefont {P.~O.}\ \bibnamefont
  {Sukhachov}},\ }\bibfield  {title} {\enquote {\bibinfo {title} {Consistent
  chiral kinetic theory in weyl materials: Chiral magnetic plasmons},}\ }\href
  {\doibase 10.1103/PhysRevLett.118.127601} {\bibfield  {journal} {\bibinfo
  {journal} {Phys. Rev. Lett.}\ }\textbf {\bibinfo {volume} {118}},\ \bibinfo
  {pages} {127601} (\bibinfo {year} {2017})}\BibitemShut {NoStop}%
\bibitem [{\citenamefont {Ferreiros}\ \emph {et~al.}(2019)\citenamefont
  {Ferreiros}, \citenamefont {Kedem}, \citenamefont {Bergholtz},\ and\
  \citenamefont {Bardarson}}]{Ferreiros_2019}%
  \BibitemOpen
  \bibfield  {author} {\bibinfo {author} {\bibfnamefont {Yago}\ \bibnamefont
  {Ferreiros}}, \bibinfo {author} {\bibfnamefont {Yaron}\ \bibnamefont
  {Kedem}}, \bibinfo {author} {\bibfnamefont {Emil~J.}\ \bibnamefont
  {Bergholtz}}, \ and\ \bibinfo {author} {\bibfnamefont {Jens~H.}\ \bibnamefont
  {Bardarson}},\ }\bibfield  {title} {\enquote {\bibinfo {title} {Mixed
  axial-torsional anomaly in weyl semimetals},}\ }\href {\doibase
  10.1103/PhysRevLett.122.056601} {\bibfield  {journal} {\bibinfo  {journal}
  {Phys. Rev. Lett.}\ }\textbf {\bibinfo {volume} {122}},\ \bibinfo {pages}
  {056601} (\bibinfo {year} {2019})}\BibitemShut {NoStop}%
\bibitem [{\citenamefont {Grushin}\ and\ \citenamefont {Palumbo}(2020)}]{Gian}%
  \BibitemOpen
  \bibfield  {author} {\bibinfo {author} {\bibfnamefont {Adolfo~G.}\
  \bibnamefont {Grushin}}\ and\ \bibinfo {author} {\bibfnamefont
  {Giandomenico}\ \bibnamefont {Palumbo}},\ }\href@noop {} {\enquote {\bibinfo
  {title} {Fermionic dualities with axial gauge fields},}\ } (\bibinfo {year}
  {2020}),\ \Eprint {http://arxiv.org/abs/2007.02944} {arXiv:2007.02944
  [cond-mat.mes-hall]} \BibitemShut {NoStop}%
\bibitem [{\citenamefont {Bertlmann}(2000)}]{Bertlmann}%
  \BibitemOpen
  \bibfield  {author} {\bibinfo {author} {\bibfnamefont {Reinhold~A.}\
  \bibnamefont {Bertlmann}},\ }\href@noop {} {\emph {\bibinfo {title}
  {Anomalies in Quantum Field Theory}}}\ (\bibinfo  {publisher} {Oxford
  University Press},\ \bibinfo {address} {Oxford},\ \bibinfo {year}
  {2000})\BibitemShut {NoStop}%
\bibitem [{\citenamefont {Liu}\ \emph {et~al.}(2013)\citenamefont {Liu},
  \citenamefont {Ye},\ and\ \citenamefont {Qi}}]{Liu}%
  \BibitemOpen
  \bibfield  {author} {\bibinfo {author} {\bibfnamefont {Chao-Xing}\
  \bibnamefont {Liu}}, \bibinfo {author} {\bibfnamefont {Peng}\ \bibnamefont
  {Ye}}, \ and\ \bibinfo {author} {\bibfnamefont {Xiao-Liang}\ \bibnamefont
  {Qi}},\ }\bibfield  {title} {\enquote {\bibinfo {title} {Chiral gauge field
  and axial anomaly in a weyl semimetal},}\ }\href {\doibase
  10.1103/PhysRevB.87.235306} {\bibfield  {journal} {\bibinfo  {journal} {Phys.
  Rev. B}\ }\textbf {\bibinfo {volume} {87}},\ \bibinfo {pages} {235306}
  (\bibinfo {year} {2013})}\BibitemShut {NoStop}%
\bibitem [{\citenamefont {Jackiw}\ and\ \citenamefont {Pi}(2007)}]{Jackiw2}%
  \BibitemOpen
  \bibfield  {author} {\bibinfo {author} {\bibfnamefont {R.}~\bibnamefont
  {Jackiw}}\ and\ \bibinfo {author} {\bibfnamefont {S.-Y.}\ \bibnamefont
  {Pi}},\ }\bibfield  {title} {\enquote {\bibinfo {title} {Chiral gauge theory
  for graphene},}\ }\href {\doibase 10.1103/PhysRevLett.98.266402} {\bibfield
  {journal} {\bibinfo  {journal} {Phys. Rev. Lett.}\ }\textbf {\bibinfo
  {volume} {98}},\ \bibinfo {pages} {266402} (\bibinfo {year}
  {2007})}\BibitemShut {NoStop}%
\bibitem [{\citenamefont {Lahtinen}\ and\ \citenamefont
  {Pachos}(2010)}]{Ville3}%
  \BibitemOpen
  \bibfield  {author} {\bibinfo {author} {\bibfnamefont {Ville}\ \bibnamefont
  {Lahtinen}}\ and\ \bibinfo {author} {\bibfnamefont {Jiannis~K.}\ \bibnamefont
  {Pachos}},\ }\bibfield  {title} {\enquote {\bibinfo {title} {Topological
  phase transitions driven by gauge fields in an exactly solvable model},}\
  }\href {\doibase 10.1103/PhysRevB.81.245132} {\bibfield  {journal} {\bibinfo
  {journal} {Phys. Rev. B}\ }\textbf {\bibinfo {volume} {81}},\ \bibinfo
  {pages} {245132} (\bibinfo {year} {2010})}\BibitemShut {NoStop}%
\bibitem [{\citenamefont {Rachel}\ \emph {et~al.}(2016)\citenamefont {Rachel},
  \citenamefont {Fritz},\ and\ \citenamefont {Vojta}}]{Rachel}%
  \BibitemOpen
  \bibfield  {author} {\bibinfo {author} {\bibfnamefont {Stephan}\ \bibnamefont
  {Rachel}}, \bibinfo {author} {\bibfnamefont {Lars}\ \bibnamefont {Fritz}}, \
  and\ \bibinfo {author} {\bibfnamefont {Matthias}\ \bibnamefont {Vojta}},\
  }\bibfield  {title} {\enquote {\bibinfo {title} {Landau levels of majorana
  fermions in a spin liquid},}\ }\href {\doibase
  10.1103/PhysRevLett.116.167201} {\bibfield  {journal} {\bibinfo  {journal}
  {Phys. Rev. Lett.}\ }\textbf {\bibinfo {volume} {116}},\ \bibinfo {pages}
  {167201} (\bibinfo {year} {2016})}\BibitemShut {NoStop}%
\bibitem [{\citenamefont {Giuliani}\ \emph {et~al.}(2010)\citenamefont
  {Giuliani}, \citenamefont {Mastropietro},\ and\ \citenamefont
  {Porta}}]{Giuliani}%
  \BibitemOpen
  \bibfield  {author} {\bibinfo {author} {\bibfnamefont {Alessandro}\
  \bibnamefont {Giuliani}}, \bibinfo {author} {\bibfnamefont {Vieri}\
  \bibnamefont {Mastropietro}}, \ and\ \bibinfo {author} {\bibfnamefont
  {Marcello}\ \bibnamefont {Porta}},\ }\bibfield  {title} {\enquote {\bibinfo
  {title} {Lattice gauge theory model for graphene},}\ }\href {\doibase
  10.1103/PhysRevB.82.121418} {\bibfield  {journal} {\bibinfo  {journal} {Phys.
  Rev. B}\ }\textbf {\bibinfo {volume} {82}},\ \bibinfo {pages} {121418}
  (\bibinfo {year} {2010})}\BibitemShut {NoStop}%
\bibitem [{\citenamefont {Gusynin}\ \emph {et~al.}(2007)\citenamefont
  {Gusynin}, \citenamefont {Sharapov},\ and\ \citenamefont
  {Carbotte}}]{Gusynin}%
  \BibitemOpen
  \bibfield  {author} {\bibinfo {author} {\bibfnamefont {V.~P.}\ \bibnamefont
  {Gusynin}}, \bibinfo {author} {\bibfnamefont {S.~G.}\ \bibnamefont
  {Sharapov}}, \ and\ \bibinfo {author} {\bibfnamefont {J.~P.}\ \bibnamefont
  {Carbotte}},\ }\bibfield  {title} {\enquote {\bibinfo {title} {Ac
  conductivity of graphene: from tight-binding model to 2+1-dimensional quantum
  electrodynamics},}\ }\href {\doibase 10.1142/S0217979207038022} {\bibfield
  {journal} {\bibinfo  {journal} {International Journal of Modern Physics B}\
  }\textbf {\bibinfo {volume} {21}},\ \bibinfo {pages} {4611} (\bibinfo {year}
  {2007})}\BibitemShut {NoStop}%
\bibitem [{\citenamefont {Aidelsburger}\ \emph {et~al.}(2018)\citenamefont
  {Aidelsburger}, \citenamefont {Nascimbene},\ and\ \citenamefont
  {Goldman}}]{Aidelsburger}%
  \BibitemOpen
  \bibfield  {author} {\bibinfo {author} {\bibfnamefont {Monika}\ \bibnamefont
  {Aidelsburger}}, \bibinfo {author} {\bibfnamefont {Sylvain}\ \bibnamefont
  {Nascimbene}}, \ and\ \bibinfo {author} {\bibfnamefont {Nathan}\ \bibnamefont
  {Goldman}},\ }\bibfield  {title} {\enquote {\bibinfo {title} {Artificial
  gauge fields in materials and engineered systems},}\ }\href {\doibase
  https://doi.org/10.1016/j.crhy.2018.03.002} {\bibfield  {journal} {\bibinfo
  {journal} {Comptes Rendus Physique}\ }\textbf {\bibinfo {volume} {19}},\
  \bibinfo {pages} {394 -- 432} (\bibinfo {year} {2018})},\ \bibinfo {note}
  {quantum simulation / Simulation quantique}\BibitemShut {NoStop}%
\bibitem [{\citenamefont {Rothe}(2012)}]{Rothe}%
  \BibitemOpen
  \bibfield  {author} {\bibinfo {author} {\bibfnamefont {Heinz~J}\ \bibnamefont
  {Rothe}},\ }\href {\doibase 10.1142/8229} {\emph {\bibinfo {title} {Lattice
  Gauge Theories}}},\ \bibinfo {edition} {4th}\ ed.\ (\bibinfo  {publisher}
  {World Scientific},\ \bibinfo {address} {Singapore},\ \bibinfo {year}
  {2012})\BibitemShut {NoStop}%
\bibitem [{\citenamefont {Münster}\ and\ \citenamefont
  {Walzl}(2000)}]{Munster}%
  \BibitemOpen
  \bibfield  {author} {\bibinfo {author} {\bibfnamefont {G.}~\bibnamefont
  {Münster}}\ and\ \bibinfo {author} {\bibfnamefont {M.}~\bibnamefont
  {Walzl}},\ }\href@noop {} {\enquote {\bibinfo {title} {Lattice gauge theory -
  a short primer},}\ } (\bibinfo {year} {2000}),\ \Eprint
  {http://arxiv.org/abs/hep-lat/0012005} {arXiv:hep-lat/0012005 [hep-lat]}
  \BibitemShut {NoStop}%
\bibitem [{\citenamefont {Volovik}(2003)}]{Volovik}%
  \BibitemOpen
  \bibfield  {author} {\bibinfo {author} {\bibfnamefont {Grigori~E.}\
  \bibnamefont {Volovik}},\ }\href@noop {} {\emph {\bibinfo {title} {The
  Universe in a Helium Droplet}}}\ (\bibinfo  {publisher} {Oxford University
  Press},\ \bibinfo {address} {Oxford},\ \bibinfo {year} {2003})\BibitemShut
  {NoStop}%
\bibitem [{\citenamefont {Nielsen}\ and\ \citenamefont
  {Ninomiya}(1981)}]{Nielsen}%
  \BibitemOpen
  \bibfield  {author} {\bibinfo {author} {\bibfnamefont {H.B.}\ \bibnamefont
  {Nielsen}}\ and\ \bibinfo {author} {\bibfnamefont {M.}~\bibnamefont
  {Ninomiya}},\ }\bibfield  {title} {\enquote {\bibinfo {title} {Absence of
  neutrinos on a lattice: (ii). intuitive topological proof},}\ }\href
  {\doibase https://doi.org/10.1016/0550-3213(81)90524-1} {\bibfield  {journal}
  {\bibinfo  {journal} {Nuclear Physics B}\ }\textbf {\bibinfo {volume}
  {193}},\ \bibinfo {pages} {173 -- 194} (\bibinfo {year} {1981})}\BibitemShut
  {NoStop}%
\bibitem [{\citenamefont {Brennan}\ and\ \citenamefont {Vala}(2016)}]{Brennan}%
  \BibitemOpen
  \bibfield  {author} {\bibinfo {author} {\bibfnamefont {John}\ \bibnamefont
  {Brennan}}\ and\ \bibinfo {author} {\bibfnamefont {Jiří}\ \bibnamefont
  {Vala}},\ }\bibfield  {title} {\enquote {\bibinfo {title} {Lattice defects in
  the kitaev honeycomb model},}\ }\href {\doibase 10.1021/acs.jpca.6b00149}
  {\bibfield  {journal} {\bibinfo  {journal} {The Journal of Physical Chemistry
  A}\ }\textbf {\bibinfo {volume} {120}},\ \bibinfo {pages} {3326--3334}
  (\bibinfo {year} {2016})},\ \bibinfo {note} {pMID: 26886150}\BibitemShut
  {NoStop}%
\bibitem [{\citenamefont {Zheng}\ \emph {et~al.}(2015)\citenamefont {Zheng},
  \citenamefont {Dua},\ and\ \citenamefont {Jiang}}]{Zheng}%
  \BibitemOpen
  \bibfield  {author} {\bibinfo {author} {\bibfnamefont {Huaixiu}\ \bibnamefont
  {Zheng}}, \bibinfo {author} {\bibfnamefont {Arpit}\ \bibnamefont {Dua}}, \
  and\ \bibinfo {author} {\bibfnamefont {Liang}\ \bibnamefont {Jiang}},\
  }\bibfield  {title} {\enquote {\bibinfo {title} {Demonstrating non-abelian
  statistics of majorana fermions using twist defects},}\ }\href {\doibase
  10.1103/PhysRevB.92.245139} {\bibfield  {journal} {\bibinfo  {journal} {Phys.
  Rev. B}\ }\textbf {\bibinfo {volume} {92}},\ \bibinfo {pages} {245139}
  (\bibinfo {year} {2015})}\BibitemShut {NoStop}%
\bibitem [{\citenamefont {Bombin}(2010)}]{Bombin}%
  \BibitemOpen
  \bibfield  {author} {\bibinfo {author} {\bibfnamefont {H.}~\bibnamefont
  {Bombin}},\ }\bibfield  {title} {\enquote {\bibinfo {title} {Topological
  order with a twist: Ising anyons from an abelian model},}\ }\href {\doibase
  10.1103/PhysRevLett.105.030403} {\bibfield  {journal} {\bibinfo  {journal}
  {Phys. Rev. Lett.}\ }\textbf {\bibinfo {volume} {105}},\ \bibinfo {pages}
  {030403} (\bibinfo {year} {2010})}\BibitemShut {NoStop}%
\bibitem [{\citenamefont {Chamon}\ \emph {et~al.}(2010)\citenamefont {Chamon},
  \citenamefont {Jackiw}, \citenamefont {Nishida}, \citenamefont {Pi},\ and\
  \citenamefont {Santos}}]{Chamon}%
  \BibitemOpen
  \bibfield  {author} {\bibinfo {author} {\bibfnamefont {C.}~\bibnamefont
  {Chamon}}, \bibinfo {author} {\bibfnamefont {R.}~\bibnamefont {Jackiw}},
  \bibinfo {author} {\bibfnamefont {Y.}~\bibnamefont {Nishida}}, \bibinfo
  {author} {\bibfnamefont {S.-Y.}\ \bibnamefont {Pi}}, \ and\ \bibinfo {author}
  {\bibfnamefont {L.}~\bibnamefont {Santos}},\ }\bibfield  {title} {\enquote
  {\bibinfo {title} {Quantizing majorana fermions in a superconductor},}\
  }\href {\doibase 10.1103/PhysRevB.81.224515} {\bibfield  {journal} {\bibinfo
  {journal} {Phys. Rev. B}\ }\textbf {\bibinfo {volume} {81}},\ \bibinfo
  {pages} {224515} (\bibinfo {year} {2010})}\BibitemShut {NoStop}%
\bibitem [{\citenamefont {Alba}\ \emph {et~al.}(2013)\citenamefont {Alba},
  \citenamefont {Fernandez-Gonzalvo}, \citenamefont {Mur-Petit}, \citenamefont
  {Garcia-Ripoll},\ and\ \citenamefont {Pachos}}]{Alba}%
  \BibitemOpen
  \bibfield  {author} {\bibinfo {author} {\bibfnamefont {E.}~\bibnamefont
  {Alba}}, \bibinfo {author} {\bibfnamefont {X.}~\bibnamefont
  {Fernandez-Gonzalvo}}, \bibinfo {author} {\bibfnamefont {J.}~\bibnamefont
  {Mur-Petit}}, \bibinfo {author} {\bibfnamefont {J.J.}\ \bibnamefont
  {Garcia-Ripoll}}, \ and\ \bibinfo {author} {\bibfnamefont {J.K.}\
  \bibnamefont {Pachos}},\ }\bibfield  {title} {\enquote {\bibinfo {title}
  {Simulating dirac fermions with abelian and non-abelian gauge fields in
  optical lattices},}\ }\href {\doibase
  https://doi.org/10.1016/j.aop.2012.10.005} {\bibfield  {journal} {\bibinfo
  {journal} {Annals of Physics}\ }\textbf {\bibinfo {volume} {328}},\ \bibinfo
  {pages} {64 -- 82} (\bibinfo {year} {2013})}\BibitemShut {NoStop}%
\bibitem [{\citenamefont {Self}(2019)}]{Self_code}%
  \BibitemOpen
  \bibfield  {author} {\bibinfo {author} {\bibfnamefont {Chris~N.•}\
  \bibnamefont {Self}},\ }\href@noop {} {\enquote {\bibinfo {title} {Kitaev
  honeycomb in python},}\ }\bibinfo {howpublished}
  {\url{https://github.com/chris-n-self/kitaev-honeycomb}} (\bibinfo {year}
  {2019})\BibitemShut {NoStop}%
\bibitem [{\citenamefont {Maraner}\ and\ \citenamefont
  {Pachos}(2009)}]{Maraner}%
  \BibitemOpen
  \bibfield  {author} {\bibinfo {author} {\bibfnamefont {Paolo}\ \bibnamefont
  {Maraner}}\ and\ \bibinfo {author} {\bibfnamefont {Jiannis~K.}\ \bibnamefont
  {Pachos}},\ }\bibfield  {title} {\enquote {\bibinfo {title} {Yang–mills
  gauge theories from simple fermionic lattice models},}\ }\href {\doibase
  https://doi.org/10.1016/j.physleta.2009.05.029} {\bibfield  {journal}
  {\bibinfo  {journal} {Physics Letters A}\ }\textbf {\bibinfo {volume}
  {373}},\ \bibinfo {pages} {2542 -- 2545} (\bibinfo {year}
  {2009})}\BibitemShut {NoStop}%
\bibitem [{\citenamefont {Yang}\ \emph {et~al.}(2019)\citenamefont {Yang},
  \citenamefont {Iadecola}, \citenamefont {Chamon},\ and\ \citenamefont
  {Mudry}}]{Yang}%
  \BibitemOpen
  \bibfield  {author} {\bibinfo {author} {\bibfnamefont {Zhi-Cheng}\
  \bibnamefont {Yang}}, \bibinfo {author} {\bibfnamefont {Thomas}\ \bibnamefont
  {Iadecola}}, \bibinfo {author} {\bibfnamefont {Claudio}\ \bibnamefont
  {Chamon}}, \ and\ \bibinfo {author} {\bibfnamefont {Christopher}\
  \bibnamefont {Mudry}},\ }\bibfield  {title} {\enquote {\bibinfo {title}
  {Hierarchical majoranas in a programmable nanowire network},}\ }\href
  {\doibase 10.1103/PhysRevB.99.155138} {\bibfield  {journal} {\bibinfo
  {journal} {Phys. Rev. B}\ }\textbf {\bibinfo {volume} {99}},\ \bibinfo
  {pages} {155138} (\bibinfo {year} {2019})}\BibitemShut {NoStop}%
\end{thebibliography}%

\appendix 

\section{The continuum limit of the most general KHLM}
\label{app:contlimit}
\subsection{The KHLM}
\begin{figure}[tp]
\center
\includegraphics[width=\columnwidth]{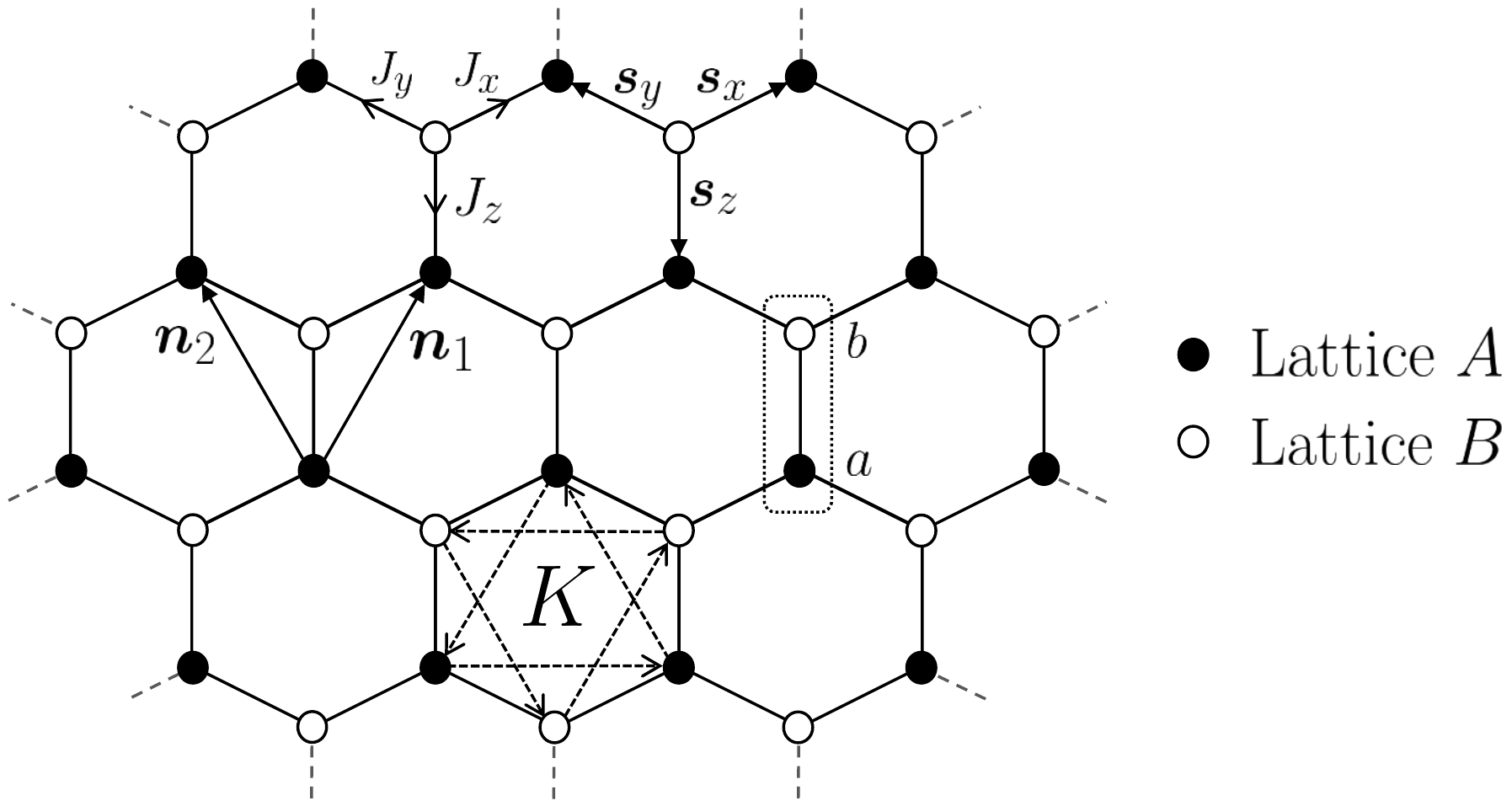}
\caption{The honeycomb lattice with Majorana fermions tunnelling between nearest neighbouring sites with couplings $J_x$, $J_y$, and $J_z$, depending on the direction of the link. Tunnelling between next-to-nearest-neighbouring sites with coupling $K$ is also indicated. The honeycomb lattice comprises two triangular sub-lattices, $A$ and $B$, denoted by full and empty circles, respectively. We take the unit cell along the $z$ links. The translation vectors between sites of the same sub-lattices are $\boldsymbol{n}_1=(\frac{\sqrt3}{2},\frac{3}{2})$ and $\boldsymbol{n}_2=(-\frac{\sqrt3}{2},\frac{3}{2})$. The orientations of the nearest tunnelings (from $A$ to $B$ sites) and next-to-nearest tunnelings (anticlockwise) are indicated.}
\label{fig:honeycomb_lattice}
\end{figure}

In this Appendix, we shall provide a derivation of the continuum limit of the KHLM. As shown in Ref. \onlinecite{Kitaev}, the KHLM Hamiltonian in the vortex-free sector can be brought into the Majorana form
\begin{equation}
H = \frac{i}{4}\left( \sum_{\langle i,j\rangle}  2J_{ij} u_{ij} c_i c_j + 2K \sum_{\langle\langle i,j\rangle\rangle} u_{ij} c_i c_j \right), \label{eq:KHLM_many_body_ham} 
\end{equation}
where $\{ c_i \}$ are Majorana modes, $u_{ij} \in \mathbb{Z}_2$ are the link operators, while $\langle i , j \rangle$ denotes a summation over pairs of nearest neighbours and similarly $\langle \langle i,j \rangle \rangle $ for next-to-nearest neighbours. The orientation of the links is shown in Fig. (\ref{fig:honeycomb_lattice}).

The honeycomb lattice can be generated by a unit cell consisting of the pair of lattice sites connected via a $z$ link, together with the basis vectors
\begin{equation}
\boldsymbol{n}_1 = \left( \frac{\sqrt{3}}{2},\frac{3}{2} \right), \quad \boldsymbol{n}_2 = \left( -\frac{\sqrt{3}}{2},\frac{3}{2} \right).
\end{equation}
As shown in Fig. (\ref{fig:honeycomb_lattice}), the honeycomb lattice also contains two triangular sub-lattices, $A$ and $B$. As each unit cell contains one site on $A$ and one on $B$, we label the sites of the honeycomb lattice by the pair $(\boldsymbol{r}, \alpha)$, where $\boldsymbol{r} \in B$ is the location of the site on sub-lattice $B$ that the unit cell overlaps and $\alpha \in (a,b)$ labels the site within the unit cell.

To reflect this symmetry of the lattice, we relabel our Majorana modes by defining $c^\alpha_{\boldsymbol{r}}$ as the mode of lattice site $(\boldsymbol{r}, \alpha)$. With this relabelling, the Hamiltonian in the \textit{vortex-free} sector, where all link operators are $u_{ij} = +1$, takes the form $H = H_J + H_K$, where
\begin{equation}
H_J  = \frac{i}{4} \sum_{\boldsymbol{r} \in B} 2c^b_{\boldsymbol{r}} \left( J^x c^a_{\boldsymbol{r} + \boldsymbol{n}_1} + J^y c^a_{\boldsymbol{r} + \boldsymbol{n}_2} + J^z c^a_{\boldsymbol{r}} \right) + \text{H.c.} \label{eq:H_J}
\end{equation}
and
\begin{equation}
\begin{aligned}
H_K   = \frac{iK}{4} & \sum_{\boldsymbol{r} \in B}  c^a_{\boldsymbol{r}} \left(-c^a_{\boldsymbol{r} + \boldsymbol{n}_1} + c^a_{\boldsymbol{r} + \boldsymbol{n}_2} + c^a_{\boldsymbol{r} + \boldsymbol{n}_1 - \boldsymbol{n}_2} \right) \\
& + c^b_{\boldsymbol{r}} \left(c^b_{\boldsymbol{r} + \boldsymbol{n}_1} - c^b_{\boldsymbol{r} + \boldsymbol{n}_2} - c^b_{\boldsymbol{r} + \boldsymbol{n}_1 - \boldsymbol{n}_2} \right) + \text{H.c.} \label{eq:H_K}
\end{aligned}
\end{equation}
We now Fourier transform the Hamiltonian with the definition $c^\alpha_{\boldsymbol{r}} = \int \mathrm{d}^2 q e^{- i \boldsymbol{q} \cdot \boldsymbol{r}} c^\alpha_{\boldsymbol{q}}$, which yields 
\begin{align}
H_J & = \frac{1}{4} \int \mathrm{d}^2 q \left( -i f(\boldsymbol{q}) c^{a \dagger}_{\boldsymbol{q}} c^b_{\boldsymbol{q}} + i f^*(\boldsymbol{q}) c^{b \dagger}_{\boldsymbol{q}} c^a_{\boldsymbol{q}} \right), \\
H_K & = \frac{1}{4} \int \mathrm{d}^2 q \Delta(\boldsymbol{q}) \left(c^{a \dagger}_{\boldsymbol{q}} c^a_{\boldsymbol{q}} - c^{b \dagger}_{\boldsymbol{q}} c^b_{\boldsymbol{q}} \right),
\end{align}
where $f(\boldsymbol{q}) = 2(J_x e^{i \boldsymbol{q} \cdot \boldsymbol{n}_1} + J_y e^{i \boldsymbol{q} \cdot \boldsymbol{n}_2} + J_z)$ and $\Delta(\boldsymbol{q})  = 2K [  - \sin(\boldsymbol{q} \cdot \boldsymbol{n}_1) + \sin(\boldsymbol{q} \cdot \boldsymbol{n}_2) + \sin( \boldsymbol{p} \cdot ( \boldsymbol{n}_1 - \boldsymbol{n}_2) ) ]$. If we define the two-component spinor $\Psi_{\boldsymbol{q}} = ( c^a_{\boldsymbol{q}} \ i c^b_{\boldsymbol{q}} )^\mathrm{T}$, we can write the total Hamiltonian $H$ as
\begin{equation}
H = \frac{1}{4} \int \mathrm{d}^2 q \Psi^\dagger_{\boldsymbol{q}} h(\boldsymbol{q}) \Psi_{\boldsymbol{q}},
\end{equation}
where the single-particle Hamiltonian $h(\boldsymbol{q})$ is given by
\begin{equation}
h(\boldsymbol{q}) = 
\begin{pmatrix}
\Delta(\boldsymbol{q}) & -f(\boldsymbol{q}) \\
-f^*(\boldsymbol{q}) & - \Delta(\boldsymbol{q})
\end{pmatrix}. \label{eq:sp_ham}
\end{equation}
\subsection{Fermi points}
From Eq. (\ref{eq:sp_ham}), we find that the single-particle dispersion relation is given by
\begin{equation}
E(\boldsymbol{q}) = \pm \sqrt{ \Delta^2(\boldsymbol{q}) + |f(\boldsymbol{q})|^2}.
\end{equation}
For now, we ignore the contribution of the $K$ term to the dispersion relation and first focus on the case where $E(\boldsymbol{q}) = \pm |f(\boldsymbol{q})|$. The Fermi points of the dispersion relation are defined as the points $\{ \boldsymbol{P}_i \}$ for which $E(\boldsymbol{P}_i)=0$. The Fermi points of the model therefore solve the equations
\begin{align}
J_x \cos(\boldsymbol{P}_i \cdot \boldsymbol{n}_1 ) + J_y \cos(\boldsymbol{P}_i \cdot \boldsymbol{n}_2) + J_z & = 0, \\
J_x \sin(\boldsymbol{P}_i \cdot \boldsymbol{n}_1 ) + J_y \sin(\boldsymbol{P}_i \cdot \boldsymbol{n}_2) & = 0.
\end{align}
The most general Fermi point was calculated in Ref. \onlinecite{Farjami}, however, it only applies for \textit{positive} values of the couplings $\{J_i\}$. A minor modification to the formula allows us to write down the Fermi point for the most general case which handles both positive and negative values. The Fermi point is given by
\begin{equation}
\boldsymbol{P}_\pm = \pm 
\begin{pmatrix}
\frac{1}{\sqrt{3}}( \mathrm{sgn}(J_y) \arccos(a) + \mathrm{sgn}(J_x) \arccos(b)) \\
\frac{1}{3}(\mathrm{sgn}(J_y) \arccos(a) - \mathrm{sgn}(J_x) \arccos(b))
\end{pmatrix}, \label{eq:fermi_points}
\end{equation}
where
\begin{equation}
a = \frac{J_y^2 - J_x^2 - J_z^2}{2J_x J_z}, \quad b = \frac{J_x^2 - J_y^2 - J_z^2}{2J_y J_z}
\end{equation}
When reinstating the $K$ term, the Fermi points are not shifted from these points if we take $K$ to be suitably small.
\subsection{The continuum limit}

\begin{table*}[tp]
\begin{center}
\begin{tabular}{|c | c | c | c | c |} 
\hline
\textbf{Type} & \multicolumn{1}{|p{2cm}|}{\centering \textbf{Spinor} \\ \textbf{bilinear} } & \multicolumn{1}{|p{3cm}|}{\centering \textbf{Single-particle} \\ \textbf{Hamiltonian form} } & \textbf{QFT Interpretation} & \textbf{Lattice Interpretation} \\ [0.5ex] 
\hline\hline
1 & $\bar{\psi}\psi$ & $\gamma^0$  & Mass & Kekul\'e distortion with real parameters \\
\hline
2 & $\bar{\psi} \gamma^a \psi$ & $\gamma^0 \gamma^a $  & $(2+1)$D kinetic terms & Nearest neighbour tunnelling $(J)$ \\
\hline
3 & $\bar{\psi}\gamma^3 \psi$ & $\gamma^0 \gamma^3 $  & $(3+1)$D $z$-direction kinetic term & None \\
\hline
4 & $\bar{\psi}\gamma^5 \psi$ & $\gamma^0 \gamma^5  $  & Pseudoscalar & Kekul\'e distortion with imaginary parameters \\
\hline
5 & $\bar{\psi} \gamma^a \gamma^5 \psi$  & $\gamma^0 \gamma^a \gamma^5 $  & $(2+1)$D chiral gauge field & A chiral shift of the Fermi points $\boldsymbol{P}_\pm$ \\
\hline 
6 & $\bar{\psi} \gamma^3 \gamma^5 \psi$ & $\gamma^0 \gamma^3 \gamma^5$  & Torsion & Next-to-nearest neighbour tunnelling $(K)$ \\
\hline
7 & $\bar{\psi} \gamma^a \gamma^b \psi$ & $\gamma^0 \gamma^a \gamma^b$  & Anti-symmetric rank 2 tensor & None \\
\hline
\end{tabular}
\end{center}
\caption{The 16 possible spinor bilinears produced from the five gamma matrices $\{ \gamma^a,\gamma^3,\gamma^5  \}$ obeying the $(3+1)$-dimensional Clifford algebra, where $a,b = 0,1,2$, which are split up into seven types. From left to right, we list the spinor bilinears, how they would appear in the single-particle Hamiltonian, their quantum field theory interpretation and finally the corresponding lattice terms that produces this bilinear in the continuum limit. Note that the interpretation of each term applies to a $(2+1)$-dimensional theory.}
\label{tab:bilinears}
\end{table*}

We define the continuum limit about each Fermi point $\boldsymbol{P}_\pm$ by restricting the Hamiltonian Eq. (\ref{eq:sp_ham}) to take values of momenta near the Fermi points as $\boldsymbol{q} = \boldsymbol{P}_\pm + \boldsymbol{p}$, for small $\boldsymbol{p}$. We define $h_\pm (\boldsymbol{p}) \equiv h(\boldsymbol{P}_\pm + \boldsymbol{p})$ as our continuum limit Hamiltonians about each Fermi point. We have
\begin{equation}
\begin{aligned}
f(\boldsymbol{P}_\pm + \boldsymbol{p})& = \boldsymbol{p} \cdot \nabla f(\boldsymbol{P}_\pm) + O(p^2) \\
& = ( \mp A + iB)p_x + iC p_y \label{eq:taylor_f}
\end{aligned}
\end{equation}
where the coefficients are given by
\begin{subequations}
\begin{align}
A & = \mathrm{sgn}(J_x)\mathrm{sgn}(J_y) \sqrt{ 12J_x^2 - 3 \frac{(J_y^2 - J_x^2 - J_z^2)^2}{J_z^2}} \\
B & = \sqrt{3} \frac{(J_y^2 - J_x^2)}{J_z} \\
C & = -3J_z 
\end{align}
\end{subequations}
Substituting Eq. (\ref{eq:taylor_f}) into Eq. (\ref{eq:sp_ham}) yields the two continuum limits
\begin{equation}
h_\pm(\boldsymbol{p}) = (\pm A \sigma^x + B \sigma^y)p_x + C \sigma^y p_y. \label{eq:h_gen_2x2}
\end{equation}
Now we consider the two Fermi points simultaneously by defining the four-component spinor $\Psi(\boldsymbol{p}) = (c^a_+ \ ic^b_+ \ i c^b_- \ c^a_-)$, where $c^{a/b}_\pm(\boldsymbol{p}) = c^{a/b}_{\boldsymbol{P}_\pm + \boldsymbol{p}}$. We combine the Hamiltonians $h_+(\boldsymbol{p}) $ and $h_-(\boldsymbol{p})$ by taking their direct sum with respect to the basis defined by $\Psi$. This yields the total $4 \times 4$ continuum limit Hamiltonian given by
\begin{equation}
\begin{aligned}
h_\text{KHLM}(\boldsymbol{p}) & = h_+ (\boldsymbol{p}) \oplus \sigma^x h_-(\boldsymbol{p}) \sigma^x \\
& = \left(A \sigma^z \otimes \sigma^x + B \sigma^z \otimes \sigma^y \right)p_x + C\sigma^z \otimes \sigma^y p_y
\end{aligned} \label{eq:4by4H_pauli}
\end{equation}
Note that we have rotated $h_-(\boldsymbol{p})$ with a $\sigma^x$ rotation before combining it with $h_+(\boldsymbol{p})$ due to our definition of $\Psi(\boldsymbol{p})$.

This low energy limit given by Eq. (\ref{eq:4by4H_pauli}) suggests that we use the Dirac $\boldsymbol{ \alpha}$ and $\beta$ matrices,
\begin{equation}
\boldsymbol{\alpha} = 
\begin{pmatrix}
\boldsymbol{\sigma} & 0 \\
0 & -\boldsymbol{\sigma}
\end{pmatrix}
= \sigma^z\otimes\boldsymbol{\sigma}, \quad
{\beta} = 
\begin{pmatrix}
0 &\mathbb{I} \\
\mathbb{I} & 0
\end{pmatrix}
= \sigma^x\otimes\mathbb{I},
\end{equation}
where $\boldsymbol{\sigma} =(\sigma^x,\sigma^y,\sigma^z)$ are the Pauli matrices and $\mathbb{I}$ is the two-dimensional identity. The corresponding Dirac gamma matrices are defined by $\gamma^0 = \beta$ and $\boldsymbol{\gamma} = \beta^{-1} \boldsymbol{\alpha}$ where
\begin{equation}
\gamma^0 = \begin{pmatrix}
0 &\mathbb{I} \\
\mathbb{I} & 0
\end{pmatrix}
= \sigma^x\otimes\mathbb{I}, \quad
\boldsymbol{\gamma}=
\begin{pmatrix}
0 & -\boldsymbol{\sigma} \\
\boldsymbol{\sigma} & 0
\end{pmatrix}
= -i\sigma^y\otimes\boldsymbol{\sigma}.
\end{equation}
These matrices satisfy the Clifford algebra $\{ \gamma^A, \gamma^B \} = 2\eta^{A B}$, where Latin indices $A,B \in (0,1,2,3)$ and $\eta_{AB} = \mathrm{diag}(1,-1,-1,-1)$ is the Minkowski metric. Despite working in $(2+1)$-dimensional space, the fact we are working with a $4 \times 4$ representation allows us to define $\gamma^3$, however, at this stage $\gamma^3$ is redundant. Using the gamma matrices, the Hamiltonian Eq. (\ref{eq:4by4H_pauli}) becomes
\begin{equation}
h_\text{KHLM}(\boldsymbol{p})  = \left( A \gamma^0 \gamma^1 + B \gamma^0 \gamma^2 \right)p_x + C\gamma^0 \gamma^2 p_y.  \label{eq:gen_con_lim}
\end{equation}
Comparison of this model to the Riemann-Cartan Hamiltonian (\ref{eq:RC Ham}), we can interpret (\ref{eq:gen_con_lim}) as a Dirac Hamiltonian defined on a Riemann-Cartan space-time with the dreibein and metric
\begin{equation}
e_a^{\ \mu}  = \begin{pmatrix}
1 & 0 & 0 \\
0 & A & 0 \\
0 & B & C
\end{pmatrix}, \quad g_{\mu \nu} = 
\begin{pmatrix}
1 & 0 & 0 \\
0 & - \frac{1}{A^2}-\frac{B^2}{A^2 C^2} & \frac{B}{AC^2} \\
0 &  \frac{B}{AC^2} & -\frac{1}{C^2}
\end{pmatrix}.
\end{equation}

\section{Generalised actions}
\label{app:gen_actions}

The usual action of a spin-$\frac{1}{2}$ particle $\psi$ of mass $m$ on a $(2+1)$-dimensional spacetime $M$ is given by
\begin{equation}
S_\text{RC}  = \frac{i}{2} \int_M \mathrm{d}^{2+1} x  |e| \left(  \bar{\psi} \gamma^\mu D_\mu \psi - \overline{D_\mu \psi} \gamma^\mu \psi + 2i m \bar{\psi}\psi \right), \label{eq:dirac action}
\end{equation} 
however is not the most general action that one could write down for a spinor field. As we have seen in the previous section, despite working in $(2+1)$-dimensional space, the continuum limit of the KHLM has provided us with a $4 \times 4$ representation of the gamma matrices obeying the $(3+1)$-dimensional Clifford algebra. It is known from the theory of spinors in $(3+1)$-dimensional space-times that the most general Lorentz invariant action one could write down is formed from $16$ spinor bilinears $\bar{\psi} \Gamma \psi$, where $\Gamma$ is a matrix constructed from products of gamma matrices~\cite{Maggiore}. We summarise the $16$ possibilities in Table \ref{tab:bilinears}.

Out of the 16 spinor bilinears, there are two types of bilinears we do not expect to see in any continuum limit of the KHLM: types $3$ and $7$ of Table \ref{tab:bilinears}. Coefficients of single gamma matrices are interpreted as momenta because they typically appear in the Hamiltonian as $\gamma^\mu p_\mu$. For this reason, a bilinear of type $3$ is interpreted as a $z$ component kinetic term. As we do not have access to the $z$ direction with our $(2+1)$-dimensional lattice, we do not expect to see this term. A bilinear of type $7$ is interpreted as an anti-symmetric rank-2 tensor because it transforms as one under Lorentz transformations. A bilinear of type $7$ \textit{could} arise in principle in our continuum limit, however, it would require us to introduce additional vector or tensor fields to the model to contract with the bilinear to produce a Lorentz invariant, e.g. $\gamma^\mu \gamma^\nu X_\mu Y_\nu$, $\gamma^\mu \gamma^\nu M_{\mu \nu}$, etc.~\cite{Maraner}. For this reason, we do not expect to see this term with only minor modifications to the lattice Hamiltonian.

The remaining bilinears listed in Table \ref{tab:bilinears} are possibilities in the continuum limit of the KHLM and the corresponding lattice interpretation is listed. Indeed, the kinetic terms of type $2$ correspond to the tunnelling coupling $J$ of Majorana between nearest neighbours, while the torsion term of type $6$ corresponds to the next-to-nearest-neighbour tunnelling coupling $K$~\cite{Farjami}. Moreover, the mass term of type $1$ and the pseudoscalar term of type $4$ can be generated by appropriately tuned Kekul\'e distortions of the nearest-neighbour tunnelling couplings~\cite{Hou,Yang}. 

The remaining bilinear of type $5$ has not been considered so far. This term correspond to a \textit{chiral gauge field} which is the focus of this paper.

\section{Continuum limit of the $\mathbb{Z}_2$ gauge field}
\label{app:z2_cont}
\begin{figure}[tp]
\center
\includegraphics[width=\columnwidth]{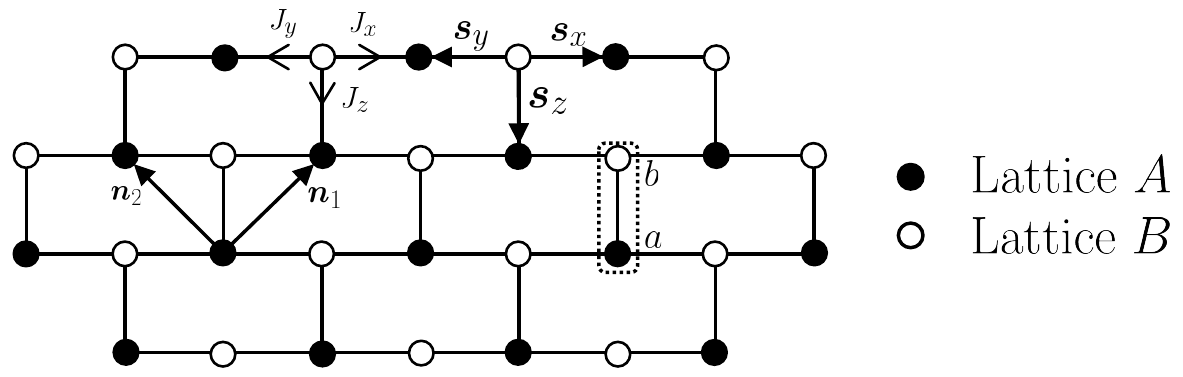}
\caption{The brick wall lattice with Majorana fermions tunnelling between nearest-neighbouring sites with couplings $J_x$, $J_y$, and $J_z$ depending on the direction of the link. The brick wall lattice has links of length $1$ and comprises two square sub-lattices, $A$ and $B$, denoted by full and empty circles, respectively. We take the unit cell along the $z$ links. The translation vectors between sites of the same sub-lattice are $\boldsymbol{n}_1=(1,1)$ and $\boldsymbol{n}_2=(-1,1)$, while the translation vectors between sites of different sub-lattice are $\boldsymbol{s}_x = (0,1), \boldsymbol{s}_y = (-1,0)$ and $(0,-1)$. The orientations of the nearest tunnellings (from $A$ to $B$ sites) are indicated.}
\label{fig:brickwall_lattice}
\end{figure}

In this Appendix we expand upon the analysis in Sec. \ref{sec:Z2_continuum} and provide a more detailed argument for how to take the continuum limit of the KHLM coupled to a $\mathbb{Z}_2$ gauge field. To make the continuum limit analysis simpler, we map the honeycomb lattice to a brick wall lattice, as shown in Fig. \ref{fig:brickwall_lattice}. This ensures that the links of the lattice align with the axes of the underlying Cartesian coordinate system. 

As discussed previously, we minimally couple a lattice theory to a gauge field by multiplying the hopping terms of the many-body Hamiltonian by link operators of the form $u_{ij} = \exp (i \int_i^j \mathrm{d} \boldsymbol{l} \cdot \boldsymbol{A})$, where $\boldsymbol{A}$ is an element of the Lie algebra corresponding to the gauge Lie group. For the KHLM, the many-body Hamiltonian coupled to a gauge field is given by Eq. (\ref{eq:KHLM_many_body_ham}). As $\mathbb{Z}_2$ is not a Lie group, it has no corresponding Lie algebra, however it is a subgroup of $U(1)$ so we are still able to express its link operators as $ u_\alpha = \exp( i \boldsymbol{A} \cdot \boldsymbol{s}_\alpha)$ for some suitable field $\boldsymbol{A}$, where $\alpha \in (x,y,z)$ labels the links of the lattice and $\boldsymbol{s}_x = (1,0)$, $\boldsymbol{s}_y = (-1,0)$ and $\boldsymbol{s}_z = (0,-1)$ are the three link vectors, see Fig. \ref{fig:brickwall_lattice}.

The many-body Hamiltonian Eq. (\ref{eq:KHLM_many_body_ham}) of the isotropic KHLM, where $J_x = J_y = J_z = 1$ and $K=0$, coupled to a $\mathbb{Z}_2$ gauge field is given by
\begin{equation}
H = \frac{i}{4} \sum_{\boldsymbol{r} \in B} \sum_{\alpha = x,y,z} 2 e^{i \boldsymbol{A}(\boldsymbol{r}) \cdot \boldsymbol{s}_\alpha} c^b_{\boldsymbol{r}} c^a_{\boldsymbol{r} + \boldsymbol{s}_\alpha} + \text{H.c.}.
\end{equation}
For the special case of constant $\boldsymbol{A}$, the corresponding single-particle Hamiltonian is given by Eq. (\ref{eq:sp_ham}), where $f(\boldsymbol{q})$ is substituted for
\begin{equation}
f_{\boldsymbol{A}} (\boldsymbol{q}) = 2 \sum_{\alpha=x,y,z} e^{i ( \boldsymbol{p} + \boldsymbol{A} ) \cdot \boldsymbol{s}_\alpha}.
\end{equation} 
We see that $f_{\boldsymbol{A}}(\boldsymbol{q}) = f(\boldsymbol{q} + \boldsymbol{A})$, where $f(\boldsymbol{q})$ is the function in the absence of a gauge field. It appears that the gauge field has the effect of translating the entire dispersion relation $E(\boldsymbol{q}) = \pm |f(\boldsymbol{q})|$ of the isotropic case by $-\boldsymbol{A}$. Consequently, one would conclude that both Fermi points $\boldsymbol{P}_\pm$ have shifted by $- \boldsymbol{A}$. Note that $\boldsymbol{A}$ cannot be arbitrary, but is heavily restricted to ensure that it exponentiates to an element of $\mathbb{Z}_2$. For this reason, these special values of $\boldsymbol{A}$ shift the Fermi points oppositely in such a way that it appears that there has been a global shift in one direction. 

\begin{figure}[tp]
\center
\includegraphics[width=0.75\columnwidth]{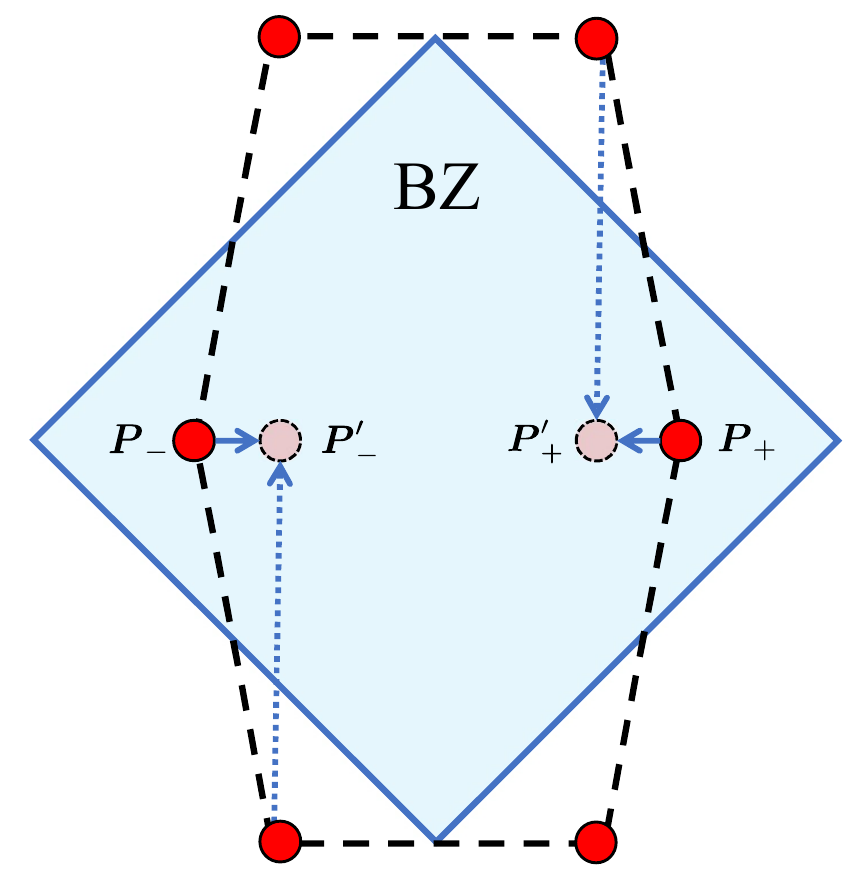}
\caption{The Brillouin zone (BZ) of the brick wall lattice with two Fermi points, $\boldsymbol{P}_+$ and $\boldsymbol{P}_-$ corresponding to the isotropic couplings $J_x=J_y=J_z=1$. Continuously changing the coupling $J_z$ from $+1$ to $-1$ everywhere on the lattice shifts the Fermi points along the $x$-direction to the positions $\boldsymbol{P}'_+$ and $\boldsymbol{P}'_-$, as shown by the horizontal solid arrow. Due to the parity symmetry of the model, the shift is anti-parallel so $\Delta\boldsymbol{P}_+=-\Delta\boldsymbol{P}_-$, which gives rise to the chiral gauge field $\boldsymbol{A} =\left(\pi/3,0 \right)$. The final configuration of the Fermi points can also be viewed as an anti-parallel shift of the Fermi points from outside the Brillouin zone in the $y$-direction, as shown by the vertical dashed arrows. This shift yields the chiral gauge field $\boldsymbol{A} =\left(0, \pi \right)$.}
\label{fig:BZ_brickwall_lattice}
\end{figure}

Consider the case of a global $\mathbb{Z}_2$ gauge field for which $u_x = u_y = +1$ and $u_z = -1$ everywhere. Solving for the Fermi points before and after switching on the gauge field, we find the Fermi points transform as
\begin{equation}
\boldsymbol{P}_\pm = \pm \left( \frac{2 \pi}{3} , 0 \right) \ \mapsto \ \boldsymbol{P}'_\pm = \pm \left( \frac{ \pi}{3} , 0 \right) \label{eq:z2_fp_transformation}
\end{equation}
so, looking at Fig. \ref{fig:BZ_brickwall_lattice}, this corresponds to a chiral shift of $\pi/3$ in the $x$-direction. Using the formula $\boldsymbol{A} = -\Delta \boldsymbol{P}_+$, the corresponding chiral gauge field of the continuum limit is given by $\boldsymbol{A} \gamma^5$, where $\boldsymbol{A} = ( \pi/3  , 0 )$.

However, there is an alternative interpretation. If we look at Fig. \ref{fig:BZ_brickwall_lattice}, we can interpret the transformation Eq. (\ref{eq:z2_fp_transformation}) as shifting $\boldsymbol{P}_+$ up by $(0, \pi)$ and shifting $\boldsymbol{P}_-$ down by $-(0,\pi)$ into neighbouring Brillouin zones. Under this transformation, the $\pm$ Fermi points are swapped as $\boldsymbol{P}_\pm$ of neighbouring Brillouin zones are mapped to $\boldsymbol{P}'_\mp$, therefore we take our shift to be $\Delta \boldsymbol{P}_\pm = \boldsymbol{P}_\pm - \boldsymbol{P}_\mp = \mp (0, \pi)$ and the corresponding gauge field is given by $\boldsymbol{A} = (0, \pi)$. Working backwards, we see that upon exponentiation $u_\alpha = \exp(i \boldsymbol{A} \cdot \boldsymbol{s}_\alpha)$ does indeed give us the correct link operators of $u_x = u_y = 1$ and $u_z = -1$.
 
The corresponding continuum limit Hamiltonians about each Fermi point, taking into account the shift in the $y$ direction, is given by
\begin{equation}
h_\pm(\boldsymbol{p})  = 2\left[ \pm \sqrt{3} \sigma^x p_x + \sigma^y (p_y \pm \pi) \right].
\end{equation}
Combining these two Hamiltonians into a single $4 \times 4$ Hamiltonian yields
\begin{equation}
h_z(\boldsymbol{p}) = 2 \left[ \sqrt{3} \gamma^0 \gamma^1 p_x + \gamma^0 \gamma^2 ( p_y + \pi \gamma^5 ) \right],
\end{equation}
so we see that the $\mathbb{Z}_2$ gauge field arises as a chiral gauge field in the continuum limit as expected.

\section{Generating the time component $A_0$ of a chiral gauge field}
\label{app:A0}

To obtain $A_0$ in the continuum limit one must modify the $K$ term of the original the KHLM. Note that this term couples sites that live on the same sub-lattice, either $A$ or $B$, with the same tunnelling amplitude for both sub-lattices. We modify this term so that there are different tunnelling amplitudes $K_a$ and $K_b$ for each sub-lattice. In this case, the contribution of the $K$ term to the single-particle Hamiltonian in momentum space becomes
\begin{equation}
h_K({\bs q}) = \begin{pmatrix} \Delta_a({\bs q}) & 0 \\ 0 & -\Delta_b({\bs q}) \end{pmatrix},
\end{equation}
where 
\begin{equation}
\begin{aligned}
\Delta_{a/b}({\bs q})  = 2K_{a / b} [& -\sin({\bs q} \cdot {\bs n}_1)  + \sin({\bs q} \cdot {\bs n}_2) \\
& + \sin({\bs q} \cdot ({\bs n}_1 - {\bs n}_2) )].
\end{aligned}
\end{equation}
These couplings do not shift the Fermi points so the analysis is straightforward.

We repeat the usual procedure by expanding the Hamiltonian about the two Fermi points by defining $(h_K)_\pm \equiv h_K(\boldsymbol{P}_\pm + \boldsymbol{p})$ to first order in $\boldsymbol{p}$. As $\Delta_{a/b}({\bs P}_\pm + {\bs p}) = \mp 3 \sqrt{3} K_{a/b} + O(p^2)$ we can combine the Hamiltonians of the two Fermi points into a single Hamiltonian as before, which yields the total Hamiltonian
\begin{equation}
\begin{aligned}
h_{K,\text{total}} = 3 \sqrt{3} \left(\frac{K_a - K_b}{2} \sigma^z \otimes \mathbb{I} -  \frac{K_a + K_b}{2} \mathbb{I} \otimes \sigma^z  \right).
\end{aligned}
\end{equation}
By a direct comparison with Eq. (\ref{eq:HamChiral}) and noting that $\gamma^5 = \sigma^z \otimes \mathbb{I}$, we have
\begin{equation}
A_0 = 3 \sqrt{3} \left(\frac{K_a - K_b}{2}\right).
\end{equation}
Moreover, the second part proportional to $\mathbb{I} \otimes \sigma^z$ corresponds to the torsion term of the Hamiltonian Eq. (\ref{eq:HamChiral}).

\section{The shape of Majorana zero modes}
\label{App:shape}

\begin{figure}[h]
\center
\includegraphics[width=\columnwidth]{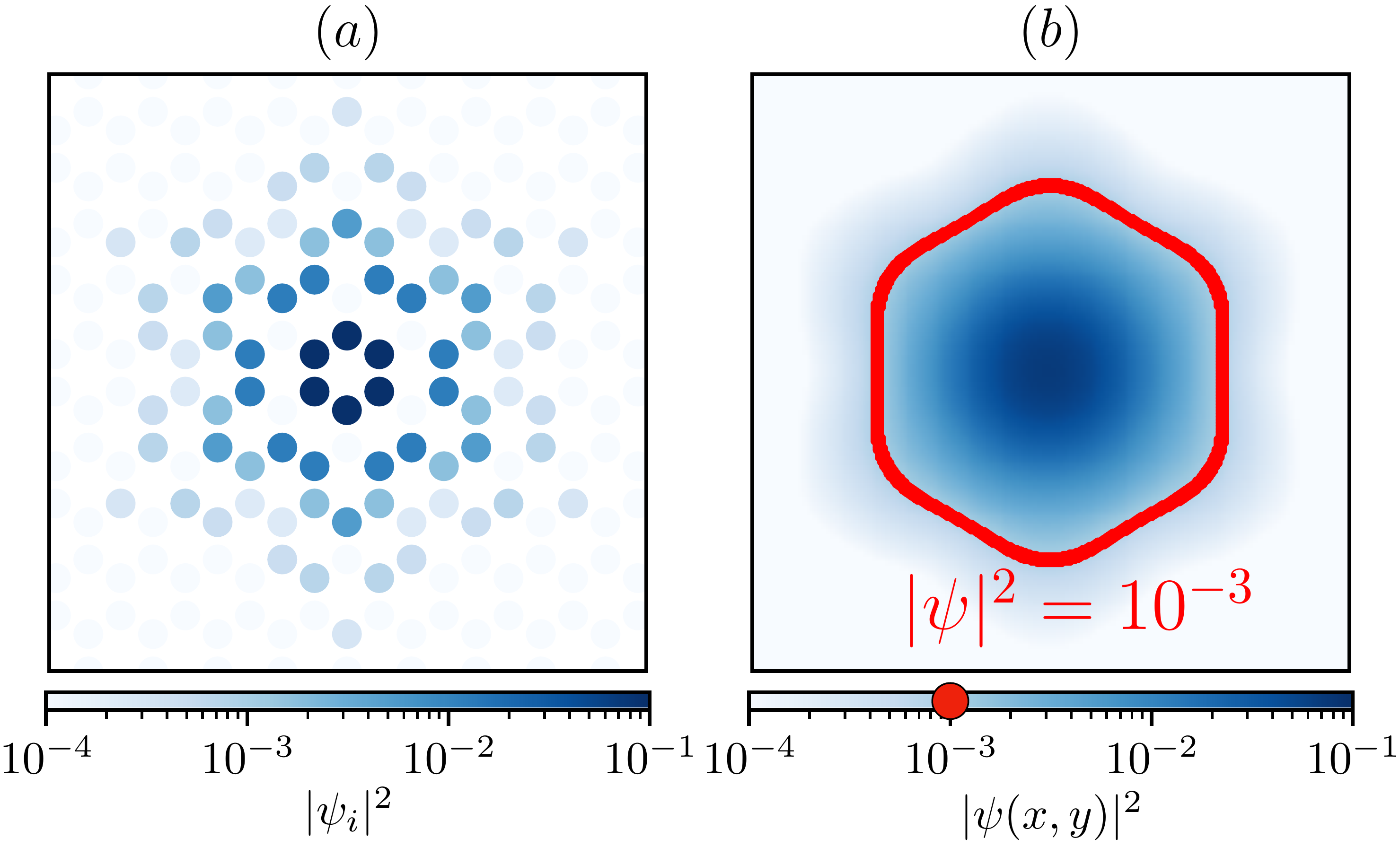}
\caption{Obtaining a continuous profile for the vortex and extracting its dimensions. (a) The lattice probability density $|\psi_i|^2$ of the wave function for a vortex, located on the plaquette in the centre. (b) A continuous approximation of the vortex probability distribution is constructed using two-dimensional Gaussians centred on each lattice site, as described in the text. The size and shape of the vortex are characterised by finding the set of points where $|\psi(\boldsymbol{r})|^2 = 10^{-3}$, as illustrated. Here we used $L=36$, $K=0.125$, and $\epsilon = 1$.}
\label{fig:lattice-to-bounding-box}
\end{figure}

For the purposes of visualising the localisation of zero modes we approximate their profile on the lattice with a continuous distribution by replacing each lattice point with two-dimensional Gaussians centred on each site,
\begin{equation*}
|\psi(\boldsymbol{r})|^2 = \sum_i |\psi_i|^2 \, \delta(\boldsymbol{r}-\boldsymbol{r}_i) \rightarrow \sum_i \frac{|\psi_i|^2}{2\pi\epsilon} e^{-{|\boldsymbol{r}-\boldsymbol{r}_i|^2\over 2\epsilon}} \, ,
\end{equation*}
where $\epsilon$ is taken to be similar to the lattice spacing so that the Gaussians of neighbouring sites overlap. Figure \ref{fig:lattice-to-bounding-box} illustrates this substitution. In the continuum, we expect a single wave function exponentially localised at the position of the vortex. This continuous profile reduces the discrete lattice effects allowing us to clearly observe the localisation or delocalisation of zero mode excitations.

\end{document}